\documentclass[aps,prb,reprint,onecolumn,11pt,superscriptaddress]{revtex4-2}
\usepackage[colorlinks=true,linkcolor=blue,citecolor=blue,urlcolor=blue]{hyperref}
\usepackage{amsmath}
\usepackage{amssymb} 
\usepackage[pdftex]{graphicx}
\usepackage{braket} 
\usepackage{orcidlink}

\newcounter{firstbib}

\begin{document}

\newcommand{\RedText}[1]{\textcolor{red}{#1}}
\newcommand{\GreenText}[1]{\textcolor{green}{#1}}
\newcommand{\BlueText}[1]{\textcolor{blue}{#1}}

\title{Few-electron spin qubits in optically active GaAs quantum dots}



\author{Peter Millington-Hotze \orcidlink{0000-0002-0861-7457}}
\affiliation{School of Mathematical and Physical Sciences, University of
Sheffield, Sheffield S3 7RH, United Kingdom}
\author{Petr Klenovsky}
\email[]{klenovsky@physics.muni.cz}
\affiliation{Department of Condensed Matter Physics, Faculty of Science, Masaryk University, Kotl\'a\v{r}sk\'a~267/2, 61137~Brno, Czech~Republic}
\affiliation{Czech Metrology Institute, Okru\v{z}n\'i 31, 63800~Brno, Czech~Republic}
\author{Harry E. Dyte}
\affiliation{School of Mathematical and Physical Sciences, University of
Sheffield, Sheffield S3 7RH, United Kingdom}
\author{George Gillard}
\affiliation{Department of Physics and Astronomy, University of
Sheffield, Sheffield S3 7RH, United Kingdom}
\author{Santanu Manna}
\affiliation{Institute of Semiconductor and Solid State Physics,
Johannes Kepler University Linz, Altenberger Str. 69, 4040 Linz,
Austria}
\author{Saimon F. Covre da Silva}
\affiliation{Institute of Semiconductor and Solid State Physics,
Johannes Kepler University Linz, Altenberger Str. 69, 4040 Linz,
Austria}
\author{Armando Rastelli}
\email[]{Armando.Rastelli@jku.at}
\affiliation{Institute of Semiconductor and Solid State Physics,
Johannes Kepler University Linz, Altenberger Str. 69, 4040 Linz,
Austria}
\author{Evgeny A. Chekhovich \orcidlink{0000-0003-1626-9015}}
\email[]{E.Chekhovich@sussex.ac.uk} 
\affiliation{Department of
Physics and Astronomy, University of Sussex, Brighton BN1 9QH,
United Kingdom}

\date{\today}

\begin{abstract}
The knowledge of the energy spectrum completely defines the dynamics of a quantum system for a given initial state. This makes spectroscopy a key characterization technique when studying or designing qubits and complex quantum systems. In semiconductor quantum dots, the electronic quantum states can be probed through charge transport spectroscopy, but the electric current itself disrupts the fragile quantum system, and the technique is practically limited to gate-defined quantum dots. Epitaxial quantum dots benefit from excellent optical properties, but are usually incompatible with charge transport, while alternative spectroscopy techniques provide only limited information. Here we demonstrate a spectroscopy technique which utilizes nuclear spins as a non-invasive probe. By using spin currents instead of the charge currents we achieve near-equilibrium probing. Experiments are conducted on low-strain GaAs/AlGaAs epitaxial dots, revealing energy spectra for charge configurations with up to seven electrons and the subtle properties of the multi-electron states. The rich variety of observations includes long-lived spin-qubit states in $s$ and $p$ shells, ground-state phase transitions, strong spin-orbit coupling regimes, and anomalously fast nuclear spin diffusion. Experiments are backed up by good agreement with the first-principles configuration-interaction numerical modelling. Our work uncovers few-electron states as a new operating regime for optically active quantum dots. Accurate control and probing of many-body states offers a test-bed system for fundamental physics studies, while prospective technological applications include electron spin qubits with extended coherence and scalable electrical control.   
\end{abstract}


\maketitle

\begin{figure*}
\includegraphics[width=0.99\linewidth]{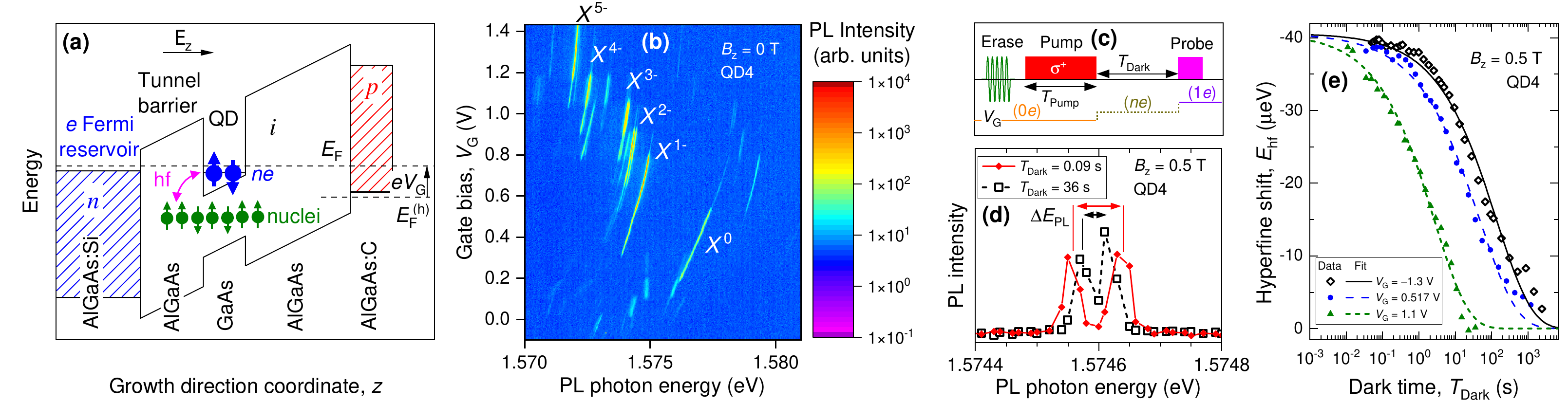}
\caption{\label{Fig:Intro} {Optical measurement of nuclear spin relaxation dynamics in GaAs QDs.} (a) Simplified schematic of the band
edge energies in a $p-i-n$ diode structure. The electron Fermi
reservoir formed by a Si doped $n$-type layer is separated from
the GaAs QDs by an AlGaAs tunnel barrier layer. The C doped
$p$-type layer is separated by a thick AlGaAs layer which inhibits
hole tunneling. The gate bias $V_{\rm{G}}$ is applied to shift the electron Fermi energy
$E_{\rm{F}}$ with respect to the hole Fermi energy
$E_{\rm{F}}^{(\rm{h})}$ and tune the QD potential energy. A QD can trap an integer number $n$ of
electrons which interact with the nuclear spins of the crystal
lattice via hyperfine (hf) coupling. (b) Photoluminescence (PL)
spectra of a single QD excited with a $\approx 690$~nm laser. Variation of the bias $V_{\rm{G}}$ results
in PL dominated by a neutral exciton $X^0$ or excitons $X^{n-}$
charged with $n$ electrons. (c) Timing diagram of the nuclear spin
relaxation (NSR) measurement. The cycle starts with a radio
frequency (rf) burst to erase any remnant nuclear spin polarization,
followed by nuclear spin polarization with a circularly polarized
pump laser, next followed by the dark evolution delay $T_{\rm{Dark}}$ under arbitrary gate bias
$V_{\rm{G}}$, and completed by a short optical probe used to measure the PL spectrum. (d) Example probe PL spectra. Decay of nuclear spin polarization under increasing $T_{\rm{Dark}}$ results in reduction of the
hyperfine shift $E_{\rm{hf}}$, observed as a decrease of the
$X^{1-}$ PL doublet splitting $\Delta E_{\rm{PL}}$. (e) The changes in $E_{\rm{hf}}$ are
measured as a function of $T_{\rm{Dark}}$ (symbols) under
different biases $V_{\rm{G}}$. Lines show stretched exponential
fitting used to derive the nuclear spin half-decay times
$T_{1,\rm{N}}$, for which the $E_{\rm{hf}}$ reduces to $1/2$ of
its initial level.}
\end{figure*}

\newcommand{\FigBand}{Fig.~\ref{Fig:Intro}(a)}
\newcommand{\FigBiasPL}{Fig.~\ref{Fig:Intro}(b)}
\newcommand{\FigTDiagNSR}{Fig.~\ref{Fig:Intro}(c)}
\newcommand{\FigPL}{Fig.~\ref{Fig:Intro}(d)}
\newcommand{\FigNucDec}{Fig.~\ref{Fig:Intro}(e)}

\begin{figure*}
\includegraphics[width=0.99\linewidth]{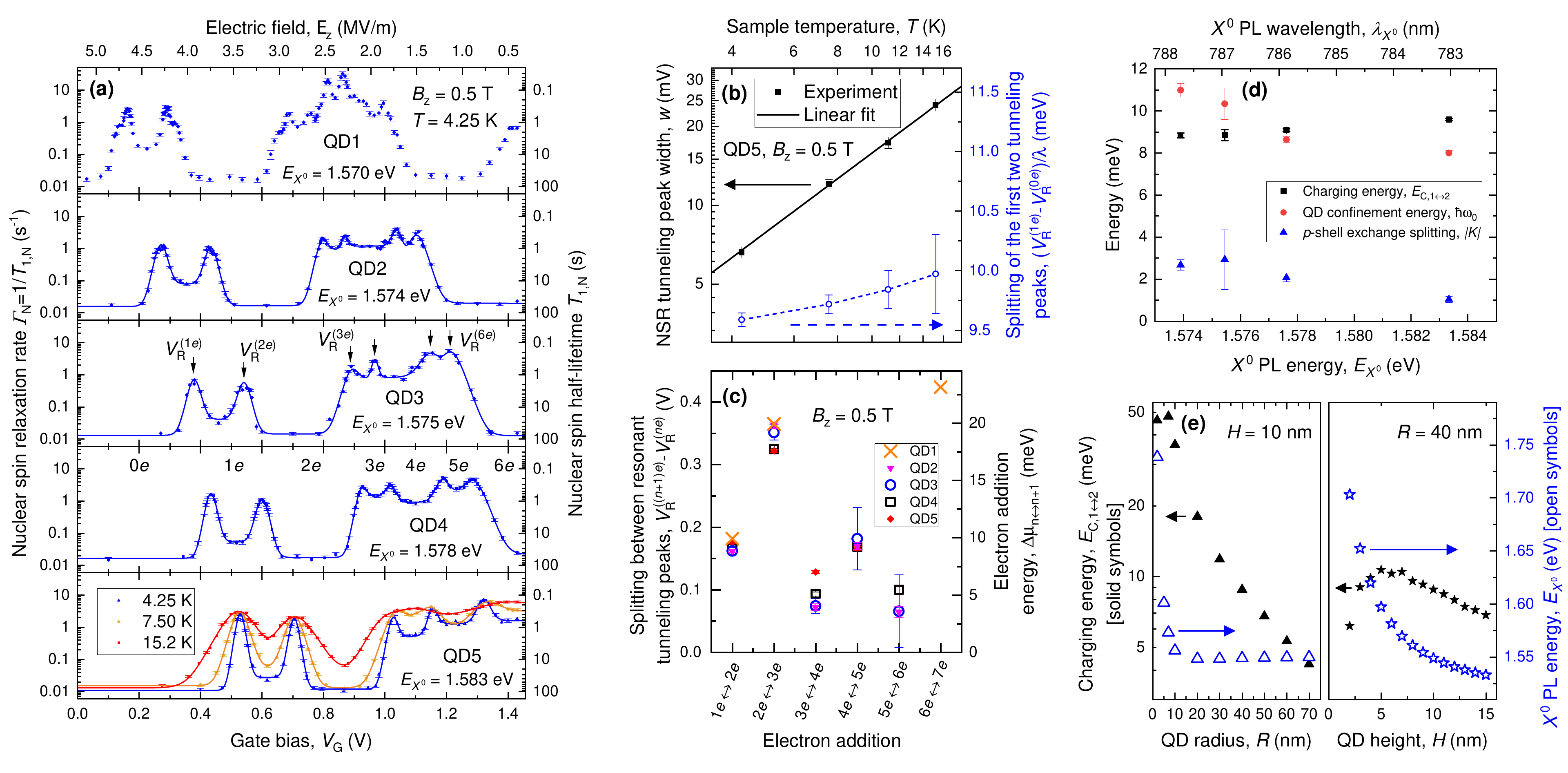}
\caption{\label{Fig:G1NBias} {Nuclear spin relaxation
spectroscopy of the few-electron states in GaAs epitaxial QDs.} (a)
Bias dependence (symbols) of the nuclear spin half-decay time
$T_{1,\rm{N}}$ (right scale) and the corresponding nuclear spin
relaxation rate $\varGamma_{\rm{N}}=1/T_{1,\rm{N}}$ (left scale)
measured in the dark. Results are obtained at sample
temperature $T=4.25$~K and are shown for individual dots
QD1$--$QD5 (from top to bottom) with different ground state exciton
emission energies $E_{X^{0}}$. Additional measurements on QD5 are shown
for $T=7.50$~K (diamonds) and $T=15.2$~K (squares). Lines show
model fitting. Top scale shows the calculated electric field $\mathcal{E}_{\rm{z}}$ as a
function of $V_{\rm{G}}$. Vertical arrows highlight the resonant biases $V_{\rm{R}}^{ne}$ of the first six peaks ($n\leq 6$) of QD3. (b) Temperature dependence of the
tunneling peak width (solid squares, left scale) and splitting of
the first two peaks (open circles, right scale) for QD5. Solid
line shows linear fitting of the peak widths. (c) Splittings of the
adjacent peaks $n$ and $n+1$ (left scale) and the corresponding
electron addition energies $\Delta\mu_{n\leftrightarrow n+1}$
(right scale) for different quantum dots as a function of the
electron number $n$. (d) Electron interaction energies measured for
different QDs and plotted as a function of their ground state PL energy
$E_{X^{0}}$. (e) QD energies calculated for a lens-shaped QD with different radii $R$ at fixed full height $H=10$~nm (left panel) or different $H$ at fixed $R=40$~nm (right panel). Solid black symbols (left scale) show the effective charging energy $E_{\rm{C},1\leftrightarrow2}=\Delta\mu_{1\leftrightarrow 2}$ for adding the second electron to the dot. Open blue symbols (right scale) show the corresponding ground state PL energy $E_{X^{0}}$. All error bars in this figure are
95\% confidence intervals.}
\end{figure*}

\newcommand{\FigNSRBias}{Fig.~\ref{Fig:G1NBias}(a)}
\newcommand{\FigwTemp}{Fig.~\ref{Fig:G1NBias}(b)}
\newcommand{\FigEnAdd}{Fig.~\ref{Fig:G1NBias}(c)}
\newcommand{\FigEnvsEnPL}{Fig.~\ref{Fig:G1NBias}(d)}
\newcommand{\FigEnCharge}{Fig.~\ref{Fig:G1NBias}(e)}

\begin{figure*}
\includegraphics[width=0.95\linewidth]{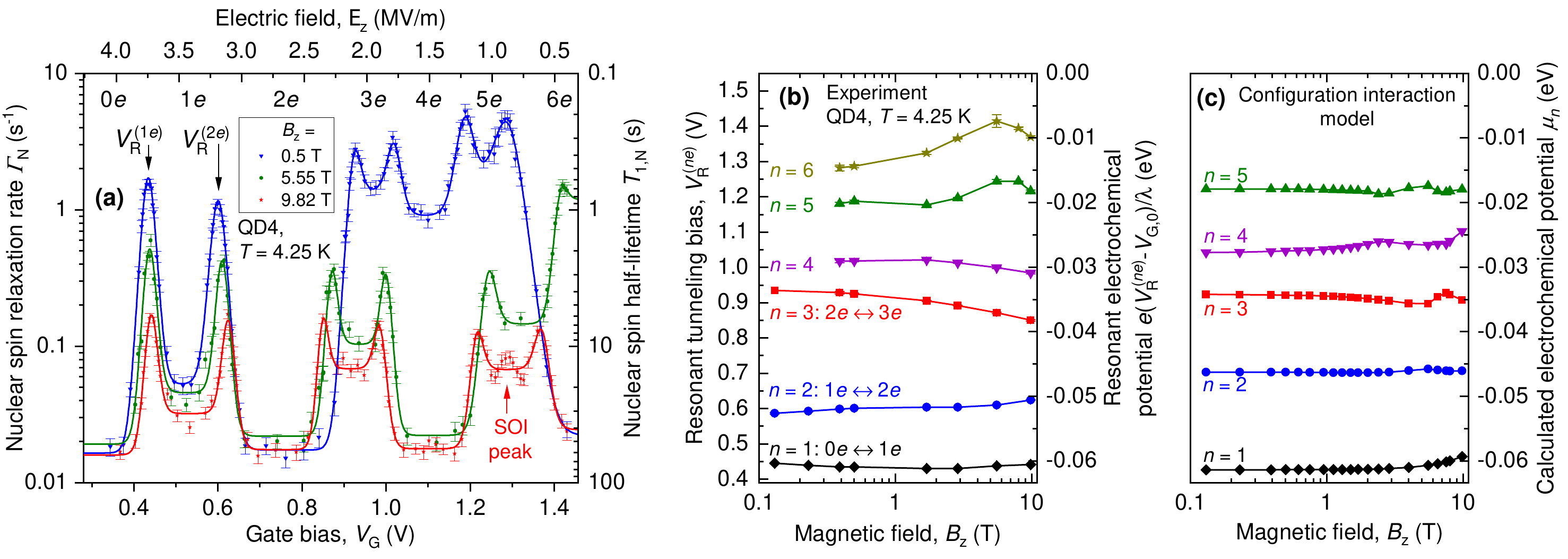}
\caption{\label{Fig:G1NBz} {Few-electron quantum dot states in
external magnetic field.} (a) Bias dependence (symbols) of the
nuclear spin half-decay time $T_{1,\rm{N}}$ (right scale) and the
corresponding nuclear spin relaxation rate
$\varGamma_{\rm{N}}=1/T_{1,\rm{N}}$ (left scale) measured on QD4
at different magnetic fields. Vertical arrows highlight the resonant biases $V_{\rm{R}}^{ne}$ of the first two peaks ($n=1,2$). Lines show model fitting.
Top scale shows the calculated electric field as a function of
$V_{\rm{G}}$. An extra peak in the 5$e$ valley, attributed to
spin-orbit interaction (SOI), is highlighted by the arrow. (b)
Resonant biases $V_{\rm{R}}^{(ne)}$ derived from the measured positions of
the cotunneling peaks in (a) and plotted as a function of magnetic
field $B_{\rm{z}}$. The right scale shows the corresponding QD
electrochemical potentials derived as
$e(V_{\rm{R}}^{(ne)}-V_{\rm{G,0}})/\lambda$, where
$V_{\rm{G,0}}=1.55$~V is the gate bias corresponding to zero
electric field at the QD. (c) Numerically calculated
electrochemical potentials for a model QD based on a shape derived from atomic force microscopy (AFM).  All
error bars are 95\% confidence intervals.}
\end{figure*}

\newcommand{\FigNSRBz}{Fig.~\ref{Fig:G1NBz}(a)}
\newcommand{\FigEnChargeBz}{Fig.~\ref{Fig:G1NBz}(b)}
\newcommand{\FigEnChargeBzCI}{Fig.~\ref{Fig:G1NBz}(c)}

\begin{figure*}
\includegraphics[width=0.95\linewidth]{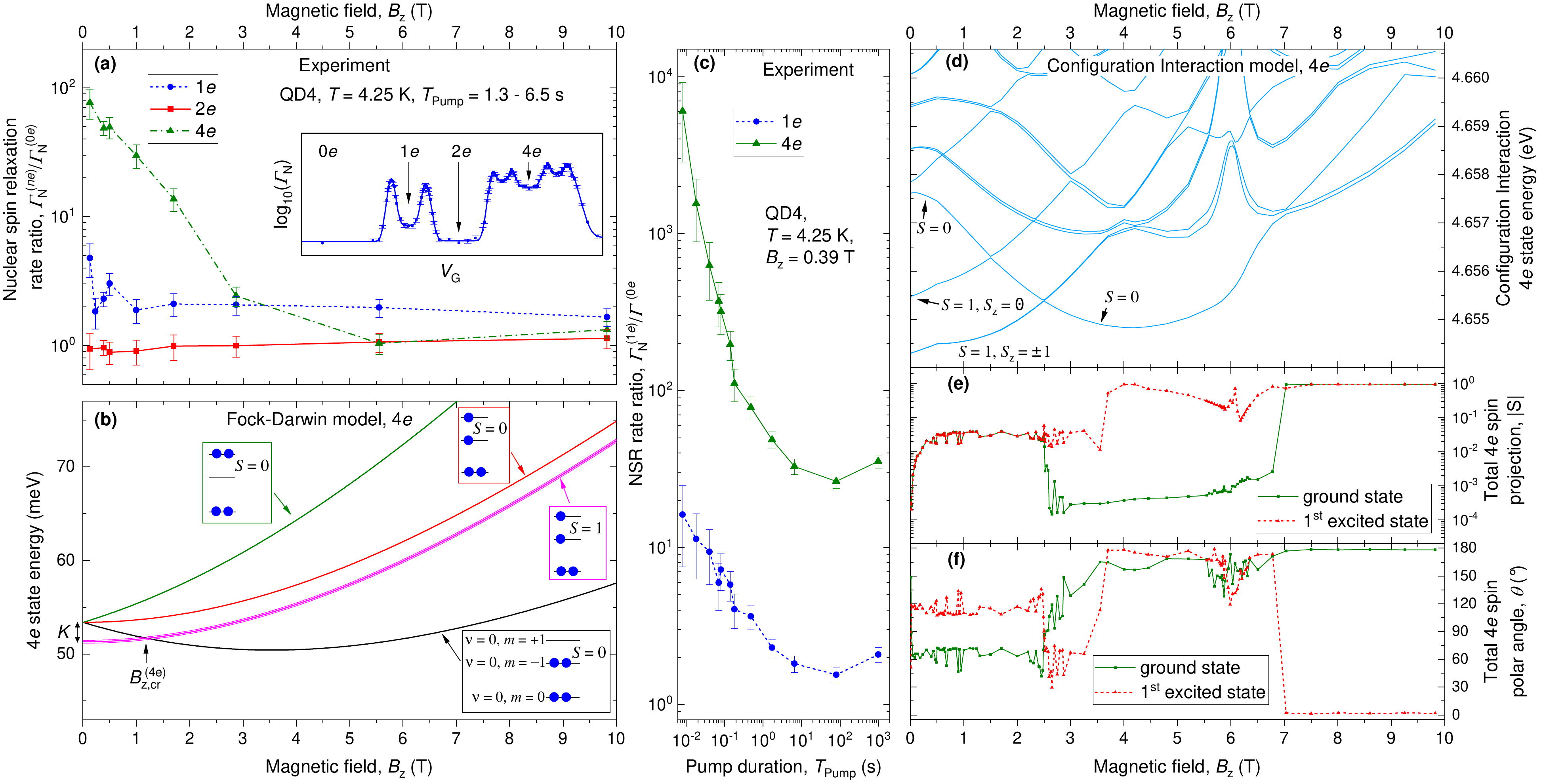}
\caption{\label{Fig:4e} {Phase transitions in a four-electron
quantum dot.} (a) Magnetic field dependence of the NSR rates
measured in a Coulomb blockade regime (see inset) where the QD is charged with $n$
electrons. For each magnetic field the $n$-electron NSR rate
$\varGamma_{\rm{N}}^{(ne)}$ is normalized by the rate
$\varGamma_{\rm{N}}^{(0e)}$ measured in an empty QD. (b) Energies of
the four lowest four-electron states as a function of magnetic
field $B_{\rm{z}}$, calculated from the Fock-Darwin model of
Eq.~\eqref{Eq:EnFD}, assuming exchange interaction energy $K\approx2$~meV.
Horizontal lines and balls depict the way the electrons fill up
the single-particle orbitals for each eigenstate branch. (c) NSR rate ratios $\varGamma_{\rm{N}}^{(ne)}/\varGamma_{\rm{N}}^{(0e)}$ as a function of optical pumping time $T_{\rm{Pump}}$ at constant $B_{\rm{z}}=0.39$~T. (d) Energies of the 4$e$ states as a function of magnetic field derived from the configuration interaction (CI) numerical model. (e) Calculated total spin projection of the ground and first excited state of the 4$e$ configuration. (f) Orientation of the spin projection for the ground and first excited state of the 4$e$ configuration, plotted as a polar angle between the spin projection and the $z$ axis. All error bars in this figure are 95\% confidence intervals.}
\end{figure*}

\newcommand{\FigNSRRatBz}{Fig.~\ref{Fig:4e}(a)}
\newcommand{\FigEnFD}{Fig.~\ref{Fig:4e}(b)}
\newcommand{\FigNSRTPump}{Fig.~\ref{Fig:4e}(c)}
\newcommand{\FigEnCI}{Fig.~\ref{Fig:4e}(d)}
\newcommand{\FigSTotCI}{Fig.~\ref{Fig:4e}(e)}
\newcommand{\FigSThetaCI}{Fig.~\ref{Fig:4e}(f)}

\newcommand{\FigNMRZeroFoure}{Fig.~\ref{Fig:NeNMR}(a)}
\newcommand{\FigNMROneThreee}{Fig.~\ref{Fig:NeNMR}(b)}

\newcommand{\FigTDiageSpin}{Fig.~\ref{Fig:3e}(a)}
\newcommand{\FigOneepump}{Fig.~\ref{Fig:3e}(b)}
\newcommand{\FigThreeepump}{Fig.~\ref{Fig:3e}(c)}
\newcommand{\FigOneeNopump}{Fig.~\ref{Fig:3e}(d)}
\newcommand{\FigThreeeNopump}{Fig.~\ref{Fig:3e}(e)}
\newcommand{\FigeDec}{Fig.~\ref{Fig:3e}(f)}
\newcommand{\FigeTeBz}{Fig.~\ref{Fig:3e}(g)}

\section{Introduction}

Classical semiconductor devices, both digital and analogue, operate with macroscopic voltages and currents, where the discrete nature of the charges can usually be ignored. In the opposite limit, which became accessible through advances in semiconductor technology, one seeks to operate on just a single electron. Semiconductor quantum dots (QDs) are ideally suited for this role as they can trap a controllable number of individual charges in a potential well defined through morphology or electric gating. QDs are a promising platform for implementation of scalable quantum logic gates (qubits) that use the lowest-energy quantum states of individual electrons \cite{Nakajima2019,Mortemousque2021,Zaporski2022}. However, any real semiconductor system possesses a spectrum of excited quantum states that can be populated through real or virtual processes, resulting in complex dynamics of the qubit states. In QDs the complete Hilbert space includes the excited single-particle states as well as the multi-electron configurations. Few-electron quantum states have been a subject of intense research \cite{Kouwenhoven2001,Reimann2002,Hanson2007} both with a practical purpose of building quantum information processing devices, and in pursuit of understanding the fundamental physics of the many-body quantum interactions and correlations.

A variety of techniques have been developed to this end. In gate-defined QDs, typically operating at milli-Kelvin temperatures, electron transport is a powerful tool of choice for probing both the ground \citep{Tarucha1996} and excited \citep{Kouwenhoven1997Science} states of the few-electron configurations. On the other hand, epitaxial QDs can operate at higher (few-Kelvin) temperatures and offer the benefit of excellent quantum optical properties. However, charge transport in such structures is rather an exception, as it requires complex device fabrication \citep{Ota2004,Jung2005,Amaha2008,Kanai2011}, often incompatible with optical operation. Consequently, a range of alternative techniques has been applied to epitaxial QDs. Photoluminescence (PL) \citep{Dalgarno2008,Mlinar2009,Holtkemper2018,Huber2019}, resonance fluorescence \citep{Vamivakas2009,Nawrath2021} and PL excitation \citep{Hawrylak2000,Ware2005} spectroscopies are versatile techniques, but are usually limited by the radiative broadening of the short-lived states where electrons and holes appear in pairs. Auger spectroscopy \citep{Lobl2020} circumvents many of these issues, but provides information only about single-particle states. The multi-electron states can be probed through capacitance-voltage spectroscopy \citep{Drexler1994,Miller1997,Wetzler2000,REUTER2008}, but the technique normally requires QD ensembles and therefore suffers from inhomogeneous broadening. Thus, understanding of the few-electron states in QDs remains limited.

Here we develop and implement a new approach for probing few-electron states in QDs. The method uses the spins of the crystal atom nuclei as a sensor. We measure spatial flow of spin momentum (spin current) which takes place without the actual flow of charge carriers. In this way we circumvent the issues arising from electric currents, while keeping the advantages of the transport technique. Historically, nuclear spins have been used to probe electronic states in a variety of applications, such as phase transitions in alkali fullerides \citep{Tycko1991,Tycko19931713} and high temperature superconductors \citep{Walstedt1990,Martindale1992}. In GaAs based structures, where all nuclear spins are non-zero, they have been used to probe ferromagnet \citep{Smet2002} and canted antiferromagnet \citep{Kumada2006} phases as well as Skyrmion excitations \citep{Barrett1995} in two-dimensional electron systems. Due to their small magnetic moments and weak interactions, the nuclei probe the electronic states in a nearly non-invasive manner, which is difficult to achieve with optical and charge transport techniques.

We conduct our experiments on GaAs/AlGaAs QDs grown by in-situ etching and infilling of nanoholes. This new generation of epitaxial structures is attracting attention due to their excellent properties as quantum light emitters \citep{Liu2019,Zhai2020,Tomm2021}, electron \citep{Zaporski2022} and nuclear \citep{Chekhovich2020} spin qubits. By measuring the nuclear spin relaxation rate as a function of the gate bias, we extract the ground state energies of different charge configurations with up to 7 electrons. Experiments reveal filling of the $s$, $p$ and $d$ shells in individual QDs of different sizes. The measurements are backed up by first-principles configuration interaction (CI) numerical modelling, which yields good agreement with experiments and offers insights such as prediction that small QDs can operate as spin qubits at elevated temperatures of up to $\approx 20$~K. Beyond the energy state spectroscopy, our  techniques reveal a wealth of nuanced information: this includes detection of the phase transition between spinless and spin-polarized four-electron ground states and observation of strong spin-orbit interaction in a five-electron configuration. Using nuclear spin sensing, we further demonstrate spin initialization, single-shot readout and long spin lifetimes exceeding $\approx 50$~ms for unpaired electrons in the $s$ and $p$ shells. Using nuclear magnetic resonance (NMR) spectroscopy, we find a factor of $\approx 2$ reduction in electron-nuclear spin interaction when going from the single-electron to three-electron configuration. These results suggest multi-electron states as a route for tailored spin coherence \citep{Leon2020} and fast electrical control of spin-orbit qubits \citep{Nowack2007,NadjPerge2010,Yoneda2014}. Furthermore, we find that a four-electron configuration gives rise to an unexpectedly efficient mechanism of nuclear spin diffusion, which is an important tool in hyperpolarization and signal enhancement in NMR spectroscopy \cite{Schmidt1992, Demco1995,Hall1997,vanderWel2006,Manolikas2008,Rossini2014,VigerGravel2018,vonWitte2025}.

\section{Results}

\subsection{Quantum dot device design and characterization}

We study samples grown by in-situ etching of nanoholes
\cite{Heyn2009,Atkinson2012} in an AlGaAs layer, which are then
infilled with GaAs to form the QDs. The structure
is processed into a $p-i-n$ diode. The width of the tunnel barrier
on the $n$-doped layer side ($\approx25~$nm) is chosen to be much
smaller than on the $p$-doped layer side ($\approx270~$nm) so that
an external gate bias $V_{\rm{G}}$ charges QDs deterministically
with individual electrons, while preventing the tunneling of the holes. The band edge profiles are sketched in
\FigBand. 

The sample is placed in a cryostat with a variable temperature $T\geq4.2$~K. A static
magnetic field $B_{\rm{z}}$ is applied in Faraday geometry along
the growth axis $z$. Confocal microscopy configuration is used to
excite and collect photoluminescence of individual QDs. PL is
analyzed on a double grating spectrometer with a charge-coupled
device (CCD) detector. Initial characterization of the QD charge
state control is examined via bias dependent PL. The PL spectra
shown in \FigBiasPL\; reveal a series of excitonic
features. The neutral exciton $X^0$, consisting of one electron and
one valence band hole, is identified from its fine structure
splitting. The negatively charged trion $X^{1-}$ (two electrons
and one hole) is identified from spectroscopy of the forbidden
transitions in an oblique magnetic field. The excitons
charged with multiple electrons up to $X^{5-}$ are identified by
counting the bright PL lines appearing when $V_{\rm{G}}$ is increased.

A typical QD contains $N\approx10^5$ atomic nuclei, each bearing a 3/2 spin moment (in units of $\hbar$), coupled to the
electron spins via the hyperfine (hf) magnetic interaction. The hf Hamiltonian is
\begin{equation}
\mathcal{\hat{H}}_{\rm{hf}} = \Sigma_{k}a_{k}\hat{\textbf{s}} \cdot \hat{\textbf{I}}_{k}, \label{Eq:Hhf}
\end{equation}
where $a_{k}$ describes the coupling between the vector of spin operators $\hat{\textbf{s}}$ of the resident electron and the $k$-th nuclear spin vector $\hat{\textbf{I}}_{k}$. The hf interaction manifests in two ways. Firstly, in addition to the bare Larmor frequency $\nu_{\rm{N}}$, each nucleus acquires a Knight shift $s_{\rm{z}} a_{k}/(2h)$, which is on the order of 100~kHz \cite{Dyte2023}. Secondly, the electron states with $s_{\rm{z}}=\pm1/2$ acquire the (Overhauser) hyperfine shifts $\pm E_{\rm{hf}}/2$, resulting from the net polarization of the QD nuclear spins. The average hyperfine shift can be written as $E_{\rm{hf}}=\Sigma_{k}a_{k}\langle\hat{I}_{\rm{z,k}} \rangle$, where $\langle...\rangle$ denotes the expectation value. 

\subsection{Nuclear spin relaxation measurements}

In order to use nuclear spin relaxation (NSR) measurements as a probe of the QD electronic state we implement an optical pump-probe experiment with a timing sequence shown in \FigTDiagNSR. A radio frequency (rf) pulse is first applied to erase any remnant nuclear spin polarization \citep{Barrett1995,MillingtonHotze2022} and then a reproducible nuclear spin state with polarization
degree of $\approx45\%$ is generated by a circularly polarized pump laser focused on a chosen individual QD \citep{Paget1982,Tycko1995,Hayashi2008,Nikolaenko2009,MillingtonHotze2023}.
A reverse bias is applied to empty the QD (0$e$ configuration) during optical pumping of the nuclear spin. Once optical pumping is switched off, the gate bias $V_{\rm{G}}$ is set to an arbitrary desired level for a ``dark time'' $T_{\rm{Dark}}$: this way evolution under an arbitrary QD charge state is studied for nominally identical initial nuclear spin polarizations. The cycle is completed by recording a PL spectrum under a short probe laser pulse. The changes in the splitting $\Delta E_{\rm{PL}}$ of a
negatively charged trion $X^-$, observed in the probe PL spectra (\FigPL), are used to derive the hyperfine (Overhauser) shift $E_{\rm{hf}}$. This hf shift measures the average nuclear spin polarization degree \citep{Urbaszek2013} that remains in the QD following NSR during $T_{\rm{Dark}}$. The hf shift $E_{\rm{hf}}$ is related to the total electron spin energy splitting $\Delta E_{\rm{e}}=\mu_{\rm{B}}g_{\rm{e}}B_{\rm{z}}+E_{\rm{hf}}$, where the electron $g$-factor is negative $g_{\rm{e}}\approx-0.1$ \cite{MillingtonHotze2022}. The polarization of the pump laser is chosen to produce a negative $E_{\rm{hf}}$, thus avoiding the electron-nuclear spin feedback at $\Delta E_{\rm{e}}\approx0$ where nonlinear electron-nuclear spin dynamics make PL probing difficult.

The measured dependencies of $E_{\rm{hf}}$ on $T_{\rm{Dark}}$, shown by the symbols in \FigNucDec, are fitted with stretched exponentials $E_{\rm{hf}}=E_{\rm{hf}}(0) 2^{-(T_{\rm{Dark}}/T_{1,\rm{N}})^\beta}$ to derive the characteristic NSR half-lifetimes $T_{1,\rm{N}}$. Here, $\beta$ is the stretching (compression) parameter for $\beta<1$ ($\beta>1$). Results shown for three different sample biases $V_{\rm{G}}$ during the dark time reveal pronounced variations in the NSR dynamics. The detailed dependencies of $T_{1,\rm{N}}$, and the corresponding NSR rate $\varGamma_{\rm{N}}=1/T_{1,\rm{N}}$, are then measured systematically as a function of the gate bias $V_{\rm{G}}$ applied during NSR in the dark. The resulting $\varGamma_{\rm{N}}(V_{\rm{G}})$ traces are shown in \FigNSRBias\; for five different individual QDs and reveal a series of peaks, indicating accelerated nuclear spin relaxation at the biases comparable to those where PL switches between the different charge states (\FigBiasPL). We therefore ascribe the peaks in $\varGamma_{\rm{N}}(V_{\rm{G}})$ to QD charging with individual electrons. These results demonstrate that NSR can be used as a spectroscopy tool to probe multi-electron states in QDs. Importantly, for the entire range of biases used in \FigNSRBias, the electric current through the QD semiconductor structure is negligible. Thus, NSR spectroscopy probes the few-electron states close to equilibrium.

\subsection{Few-electron quantum dot states}

In order to interpret the NSR spectroscopy data we outline the background theory, starting from a basic description of the QD charging process. For $n$ electrons in the dot the electrostatic energy is $n(n-1)E_{\rm{C}}/2-n e V_{\rm{G}}/\lambda$, where $E_{\rm{C}}=e^2/C$ is the charging energy per electron, $V_{\rm{G}}$ is the sample gate bias, and the ``lever arm'' $\lambda>1$ is approximately the total distance between the doped layers divided by the electron tunnel barrier thickness. This energy contribution is associated with the classical capacitance $C$ of the QD \citep{Averin1991,Beenakker1991,Kouwenhoven2001}. In order to account for the quantum effects we further include the confinement energies of all the single-particle electron states, whose energies $E({\nu,m})$ are indexed by the quantum numbers $\nu$ and $m$. We further include the energy $F_n$ arising from the quantum correlations between single particles in the $n$-electron system. The total ground state energy of the $n$-electron configuration is then:
\begin{equation}
U_n=n(n-1)E_{\rm{C}}/2-n e
V_{\rm{G}}/\lambda+\sum_{i=1}^{n}E({\nu_i,m_i})+F_n, \label{Eq:Un}
\end{equation}

The electrochemical potential of the QD is the difference of the
energies of the adjacent ($n$ and $n-1$) ground state charge configurations:
\begin{equation}
\mu_n=U_n-U_{n-1}=(n-1)E_{\rm{C}}-e
V_{\rm{G}}/\lambda+E_{n}+(F_n-F_{n-1}),\label{Eq:mun}
\end{equation} where $E_n$ is the energy of the topmost filled
single-particle state $({\nu_n,m_n})$. The potentials $\mu_n$ are tuned
with the bias, and at certain $V_{\rm{G}}$, where $\mu_n$ equals
the Fermi energy $E_{\rm{F}}$ of the electron Fermi reservoir, the
dot has equal probability to be occupied by $n-1$ or $n$
electrons. Under these conditions cotunneling is enhanced, so that
an electron can transition back and forth between the QD and the
Fermi reservoir of the $n$-type doped layer, carrying the nuclear spin polarization out of the QD \citep{Latta2011,Gillard2021}.
Thus a peak in the NSR rate $\varGamma_{\rm{N}}$ observed at a resonant bias $V_{\rm{G}}=V_{\rm{R}}^{(ne)}$ (see \FigNSRBias) corresponds to a cotunneling resonance
$\mu_n=E_{\rm{F}}$ for a certain $n$. No other peaks are observed
when reducing the gate bias down to $V_{\rm{G}}<-3$~V, hence we
attribute the first peak (the one observed at the lowest $V_{\rm{G}}$ for each QD)
to resonant injection of one electron into an empty QD. The
valleys between the peaks correspond to Coulomb blockade regimes
where a QD is occupied by a stable number $n$ of electrons ($e$), as
labeled for QD4 in \FigNSRBias. The data shown for five individual dots QD1$-$QD5 reveals a clear
correlation between the tunneling peaks appearing at lower biases $V_{\rm{R}}^{(1e)}$ and
a lower ground state PL energy ($E_{X^{0}}$): the reduction in both quantities is attributed to a
deeper confinement of the electrons in larger QDs.

Resonant electron cotunnelling requires that: (i) the electron
reservoir can provide electrons at energy $\mu_n$ to tunnel into
the QD, and (ii) the reservoir has unfilled states at the same energy
which can accept the electrons tunneling back from the QD. Thus
the rate of the nuclear spin transfer can be modelled as $\propto
f(1-f)$, where $f=1/(1+e^{(\mu_n-E_{\rm{F}})/k_{\rm{B}}T})$ is the
Fermi-Dirac distribution of the electronic energies in the Fermi reservoir, and $k_{\rm{B}}$ is the Boltzmann
constant. By modelling each peak in this way (lines in
\FigNSRBias, see details in Supplementary Section 2F) we
find a very good agreement with experiments. The energy broadening of each
peak is $k_{\rm{B}}T$, which corresponds to $w=\lambda k_{\rm{B}}T/e$ for the width of the peak measured as a function of the gate bias $V_{\rm{G}}$. For
QD5 the dependencies $\varGamma_{\rm{N}}(V_{\rm{G}})$ were
measured at several temperatures, as shown in the bottom panel of
\FigNSRBias. The dependence of the width of the first
peak on the temperature $T$ is shown by the squares in \FigwTemp\;
(left scale) -- it is well described by a linear fit (straight
line) and reveals the ``lever arm'' $\lambda\approx18.3\pm0.6$,
which we use to convert the applied gate bias $V_{\rm{G}}$ into resonant electrochemical potentials $\mu_n$. We also
find a small temperature-induced shift of the peaks, observed as
repulsion of the first two peaks (circles in
\FigwTemp). These shifts are due to the two-fold
spin degeneracy, which leads to a difference in tunnelling rates to
and from the QD, leading in turn to an offsets in the cotunneling peak
positions from the exact $\mu_n=E_{\rm{F}}$ condition \citep{Cockins2010,Beenakker1991}.

\subsection{Electron addition spectroscopy in epitaxial quantum dots}

In order to characterize the few-electron interactions and correlations in epitaxial QDs, we further define the addition energies
\begin{equation}
\Delta\mu_{n\leftrightarrow n+1}=\mu_{n+1}-\mu_n=E_{\rm{C}}+(E_{n+1}-E_{n})+(F_{n+1} +F_{n-1}-2F_{n}), \label{Eq:Dmu}
\end{equation}
which can be derived from the
separation of the adjacent cotunnelling peaks in
$\varGamma_{\rm{N}}(V_{\rm{G}})$ dependencies
\citep{Franceschetti2000,Kouwenhoven2001} and have no explicit dependence on $V_{\rm{G}}$. In the absence of
quantum effects (i.e. in the classical limit of a large QD) the addition energies
would all be equal to $E_{\rm{C}}$. The addition energies measured
at a small magnetic field $B_{\rm{z}}=0.5$~T are shown by the symbols in
\FigEnAdd\; and reveal a systematic dependence on the
number of electrons already present in the QD. In order to interpret the addition
energies, we use the two-dimensional parabolic-potential model of Fock-Darwin, which is applicable since the electron confinement in the sample plane ($xy$) is weak compared to the out-of-plane confinement in the growth ($z$) direction \citep{Reimann2002}. The single-particle electron eigenenergies in the presence of magnetic field $B_{\rm{z}}$ are: 
\begin{equation}
E({\nu,m})=\hbar(2\nu+\lvert
m\rvert+1)\sqrt{\omega_0^2+\omega_c^2/4}+\hbar\omega_c m/2,
\label{Eq:EnFD}
\end{equation} where $\nu=0,1,2,...$ is the radial quantum number, $m=0,\pm1,\pm2,...$ is the angular quantum number, $\hbar\omega_0$ is the confinement energy of the parabolic QD potential, $\omega_c=eB_{\rm{z}}/m_{\rm{e}}^*$ is the cyclotron angular frequency of the electron with an effective mass $m_{\rm{e}}^*$ (taken to be $m_{\rm{e}}^*\approx 0.067 m_{\rm{e}}$ in GaAs \cite{Vurgaftman2001}), and each state is twice degenerate due to the electron spin. The first electron occupies the lowest-energy state $(\nu=0,m=0)$. When the second electron is added, it occupies the same single-particle state but with the opposite spin, forming a two-electron spin singlet $S=0$. Thus for the first addition energy we have:
$\Delta\mu_{1\leftrightarrow 2}=E_{\rm{C}}+(F_{2}+F_{0}-2F_{1})$.
Correlations require at least two particles, so that $F_{0}=F_{1}=0$. Moreover, the two electrons occupying the same orbital $(\nu=0,m=0)$ are restricted by the Pauli principle to have opposite spin projections. Therefore, we can interpret the $F_{2}$ term as pure Coulomb interaction and define the effective charging energy
$E_{\rm{C},1\leftrightarrow2}=\Delta\mu_{1\leftrightarrow 2}=(E_{\rm{C}}+F_{2})$. In this definition, we would like to set $B_{\rm{z}}=0$, so that $E_{\rm{C},1\leftrightarrow2}$, as well as other parameters discussed below, describe the QD itself, rather than the externally applied fields. However, a NSR measurement in a QD requires a finite magnetic field of at least a few hundred mT. Luckily, the contribution of the finite cyclotron frequency $\vert\omega_c\vert>0$ in Eq.~\eqref{Eq:EnFD} is relatively small even at $B_{\rm{z}}=0.5$~T. This allows us to use the low-field data, such as shown in \FigNSRBias, to estimate the charging energies at zero magnetic field.
The results are plotted by the squares in \FigEnvsEnPL, revealing $E_{\rm{C},1\leftrightarrow2}\approx9$~meV, consistent across all studied QDs. 

Since the first two electrons in the $n=2$ configuration form a closed $s$-shell singlet, the third electron has to occupy the next $(\nu=0,m=\pm1)$ single-particle Fock-Darwin state, belonging to the higher-energy $p$-shell. As a result, the addition energy $\Delta\mu_{2\leftrightarrow 3}$ is considerably larger than $\Delta\mu_{1\leftrightarrow 2}$ (see \FigEnAdd). Neglecting the exchange interaction between the $s$-shell $(\nu=0,m=0)$ and $p$-shell $(\nu=0,m=\pm1)$ electrons (i.e. assuming $F_3\approx 3F_2$),
we use Eqns.~\eqref{Eq:Un}, \eqref{Eq:mun}, \eqref{Eq:Dmu} to estimate the
single-particle confinement energy as $\hbar\omega_0\approx
E(0,1)-E(0,\pm1)\approx\Delta\mu_{2\leftrightarrow
3}-\Delta\mu_{1\leftrightarrow 2}$. The results derived from the data at $B_{\rm{z}}=0.5$~T are
plotted in \FigEnvsEnPL\; by the circles and are within
$\hbar\omega_0\approx9-11$~meV, in reasonable agreement with
$\hbar\omega_0\approx14$~meV obtained from Auger optical
spectroscopy on a similar type of epitaxial GaAs
QDs \citep{Lobl2020}. 

Noting that the separation of the 4th and 5th resonant peaks is larger than the separation of the 3rd and 4th peaks, we follow Ref. \citep{Kouwenhoven2001} to estimate the exchange interaction energy within the $p$-shell as $\vert K\vert=(\Delta\mu_{4\leftrightarrow 5}-\Delta\mu_{3\leftrightarrow 4})/2=(\mu_{3}-2\mu_{4}+\mu_{5})/2$. The results are shown by the triangles in \FigEnvsEnPL, revealing exchange energies between $K\approx1-3$~meV for different individual QDs.

At $n=6$ the $p$-shell originating from the $(\nu=0,m=\pm1)$ orbitals is completely filled, so that the seventh electron has to occupy the single-particle orbital of the next $d$-shell (corresponding to $\nu=1,m=0$ and $\nu=0,m=\pm2$), resulting in an even larger addition energy. The onset of cotunneling of the seventh electron is observed only in QD1 (\FigNSRBias), limited by the maximum bias $V_{\rm{G}}$ that can be applied to the sample. The seventh peak is indeed characterized by a large addition energy $\Delta\mu_{6\leftrightarrow 7}\gtrsim23$~meV required to populate the $d$-shell. For this particular QD1, the NSR peaks exhibit additional structure, making model fitting impossible. The origin of the additional structure is not clear, but might be related to electric field fluctuations arising from the charge traps adjacent to QD1.

This analysis shows how NSR spectroscopy can be used to quantify the key properties of the few-electron states in epitaxial QDs, including the single-particle confinement energy $\hbar\omega_0$, the charging energy $E_{\rm{C},1\leftrightarrow2}$, and the exchange energy $K$.


\subsection{Few-electron states in external magnetic field: experiment and configuration interaction model}

We now extend NSR spectroscopy of multi-electron states to strong magnetic fields $B_{\rm{z}}$.
Application of external magnetic field modifies the
electron state energies, especially in the $p$-shell, as observed
from the $\varGamma_{\rm{N}}(V_{\rm{G}})$ dependencies in
\FigNSRBz. \FigEnChargeBz\; shows the magnetic field dependence of the resonant peak biases $V_R^{(ne)}$, which correspond to a situation where the electrochemical potential $\mu_n$ of the $n-1\leftrightarrow n$ transition matches the electron Fermi energy $E_{\rm{F}}$. We observe a smooth increase (decrease) of $V_R^{(ne)}$ for
$n=1,2$ ($n=3,4$) in the entire range of $B_{\rm{z}}\leq10$~T. By
contrast, for $n=5,6$ the $V_R^{(5e)}(B_{\rm{z}})$ and $V_R^{(6e)}(B_{\rm{z}})$ dependencies are
clearly non-monotonic. The cusps and kinks in magnetic field dependence of
the resonant tunneling energies are indications of non-trivial
electronic shell filling, including phase transitions
\cite{Muller1996,Tavernier2003,Nishi2007}, where the many-body ground
state abruptly changes its angular momentum and spatial profile.

In order to explain the complex evolution of the few-electron ground state energies with magnetic field, we seek a description of the many-body electron problem that goes beyond the simple Fock-Darwin model. To this end, we model QDs with two types of geometries: either a spherical cap (lens) shape, or a shape derived from the atomic force microscopy (AFM) experiments conducted on a sample where growth was interrupted before the nanoholes were infilled with GaAs \citep{MillingtonHotze2022}. The spectrum of single-particle electron states for the chosen three-dimensional QD model is computed from the 8-band $k\cdot p$ model. The multi-electron basis states are then constructed from these single-particle basis orbitals as Slater determinants of appropriate symmetry. The multi-electron states are calculated using the CI method \cite{Klenovsky2017,Klenovsky2025}, taking into account electron-electron interactions and the applied electric and magnetic fields. Further details of the model can be found in Supplementary Section 5.

The calculated electrochemical potentials for an AFM-based QD model are shown in \FigEnChargeBzCI. The overall energy scales are well reproduced, within a few meV. The model predicts smooth dependencies of the electrochemical potentials on the external magnetic field for small electron numbers ($n=1,2$) with cusps and overall non-monotonic dependence for higher $n$, in agreement with experiment. However, the exact energies and positions of the cusps and kinks do not match the experiment, indicating the limitations of the model. The quantitative discrepancy is more pronounced for higher electron numbers $n\geq 3$. This is expected, since the many-body correlation effects become more pronounced with an increasing number of particles, and therefore more sensitive to the unknown nuances of the QD morphology and the single-electron spectrum. On the other hand, it is worth noting that the overall good agreement between the CI model and experimental few-electron spectra is achieved without any fitting. We thus conclude that the CI method is accurate enough to predict the key properties of the multi-electron energy spectrum where the accuracy within a few meV is sufficient.

Using the CI method, we explore the dependence of the few-electron interaction parameters for a wide range of QD morphologies that are not yet available in experiment. The effective charging energy $E_{\rm{C},1\leftrightarrow2}$ introduced above is an important parameter for operating the QD as an electron spin qubit, since the Coulomb blockade regime is only possible when $E_{\rm{C},1\leftrightarrow2}\gg k_{\rm{B}}T$. We calculate $E_{\rm{C},1\leftrightarrow2}$ (solid symbols and left scale in \FigEnCharge) together with the ground state exciton transition energy $E_{X^{0}}$ (open symbols and right scale in \FigEnCharge) for the lens-shaped (spherical cap) QDs characterised by the radius $R$ of the cap base and the cap height $H$. We explore two options: a fixed $H=10$~nm with variable $R$ (left panel in \FigEnCharge), or a fixed  $R=40$~nm with variable $H$ (right panel in \FigEnCharge). At $R=40$~nm and $H=10$~nm, which corresponds approximately to the shape of the QD deduced from AFM, the calculated charging energy is $E_{\rm{C},1\leftrightarrow2}\approx 9$~meV, matching the experiment. When reducing $R$ at fixed $H$ (solid triangles in \FigEnCharge) the charging energy increases, reaching the peak value $\approx58$~meV for a small radius of $R=5$~nm, corresponding to a half-sphere QD shape. By contrast, the reduction of $H$ at fixed $R=10$~nm (solid stars) results in a pronounced blueshift of the exciton optical transition energy $E_{X^0}$, but has little effect on $E_{\rm{C},1\leftrightarrow2}$. These results confirm that the magnitude of the electron-electron Coulomb interaction is dominated by the lateral confinement within the GaAs layer, whereas the optical transition energy is defined mainly by the confinement in the out-of-plane (sample growth) direction. Therefore, individual variation of the QD radius and height can be used to tune independently the optical energy and the thermal stability of the QD spin qubits. In particular, a factor of 5 increase in the charging energy (from the present $\approx10$~meV to the predicted $\approx50$~meV) may give a corresponding five-fold improvement in the QD spin qubit operating temperature from the current $T\approx4.2$~K to $T\approx20$~K, which would benefit deployability in applications including quantum communications and computing.

\subsection{Phase transitions in the ground state of a four-electron configuration}

Phase transitions of a multi-electron ground state were reported previously in mesa-etched and gate-defined QD structures
\citep{Tarucha1996,Kouwenhoven2001} based on the observation of
cusps and kinks in magnetic field dependence of the resonant charge carrier transport. Although such non-monotonic dependence, including
cusps, is also observed in \FigEnChargeBz, we now
demonstrate how nuclear spins can be used to probe the electronic
state in a more direct way.

Rather than operating in the resonant cotunnelling regime, where the QD electronic configuration is out of equilibrium, we now set the gate bias $V_{\rm{G}}$ either below the first resonant cotunnelling peak in order to empty the QD ($n=0$), or in the middle of the plateau between the two adjacent resonant
peaks, which results in a stable number of electrons $n\geq1$ in the QD. This Coulomb blockade regime has a significant advantage of allowing us to probe the few-electron state near equilibrium. We label the
NSR rates measured in the $n$-electron Coulomb blockade as $\varGamma_{\rm{N}}^{(ne)}$. The
empty-QD rate $\varGamma_{\rm{N}}^{(0e)}\approx0.01$~s$^{-1}$  is dominated by nuclear spin diffusion \cite{MillingtonHotze2022}. As a result, $\varGamma_{\rm{N}}^{(0e)}$ depends weakly on
$B_{\rm{z}}$, but is strongly affected by the nuclear spin pumping conditions (e.g. the duration and the power of the optical pump). Therefore $\varGamma_{\rm{N}}^{(0e)}$  is used as a reference to plot the ratios
$\varGamma_{\rm{N}}^{(ne)}/\varGamma_{\rm{N}}^{(0e)}$ as a function of
$B_{\rm{z}}$ for different charge states $n=1,2,4$, as shown in
\FigNSRRatBz. The ratios $\varGamma_{\rm{N}}^{(ne)}/\varGamma_{\rm{N}}^{(0e)}$ are less sensitive to experimental parameters, but are sensitive to the spin state of the $n$-electron configuration, which is of interest. (The absolute rates $\varGamma_{\rm{N}}^{(ne)}$ are presented in Supplementary Section 4B). 

We start with the simple cases of one and two electrons. The spin of a single unpaired electron ($n=1$, circles in \FigNSRRatBz) accelerates NSR ($\varGamma_{\rm{N}}^{(1e)}/\varGamma_{\rm{N}}^{(0e)}>1$) in the entire available range of fields $B_{\rm{z}}\leq10$~T, which is explained by electron-mediated
nuclear-nuclear interactions \cite{Latta2011,MillingtonHotze2022}. When two electrons ($2e$)
occupy the lowest Fock-Darwin orbital $(\nu=0,m=0)$ they form a
spin singlet $S=0$ which does not couple to the nuclear spins, and
indeed we find $\varGamma_{\rm{N}}^{(2e)}/\varGamma_{\rm{N}}^{(0e)}\approx1$ for
the entire range $B_{\rm{z}}\leq10$~T.

We now focus on the four-electron ($n=4$) configuration. While there is only a weak non-monotonic
character in the $\mu_4(B_{\rm{z}})$ dependence, without any pronounced kinks, we find strong acceleration of NSR at low magnetic fields $B_{\rm{z}}\leq2$~T, with $\varGamma_{\rm{N}}^{(4e)}/\varGamma_{\rm{N}}^{(0e)}$ as large as $\approx70$. By contrast, at high magnetic field $B_{\rm{z}}\geq5$~T we observe $\varGamma_{\rm{N}}^{(4e)}/\varGamma_{\rm{N}}^{(0e)}\approx1$,
indicating that the four-electron system transitions into a state
akin to the two-electron spin singlet.

In order to explain this pronounced change in $\varGamma_{\rm{N}}^{(4e)}/\varGamma_{\rm{N}}^{(0e)}$ with magnetic field we recall that the first two electrons of a $4e$ configuration form a closed $s$-shell, while the remaining two electrons result in a half-filled $p$-shell. There are four single-particle states in the $p$-shell, arising from the two orbital states $(\nu=0,m=\pm1)$, each two-fold degenerate due to spin. Thus there are six possible states of a half-filled $p$-shell: There are two singlet states $S=0$ when both $p$-shell electrons have the same orbital quantum number $m$, and there are further four states spanned by a singlet $S=0$ and a triplet $S=1$ when the two $p$-shell electrons have unequal $m$ (\FigEnFD). At $B_{\rm{z}}=0$ the exchange interaction lifts the degeneracy, resulting in a triplet ground state, with the three singlets spilt off by the exchange energy $K$. When $B_{\rm{z}}$ is increased, the energies of the four-electron states evolve according to the quantum numbers $m$ of the underlying Fock-Darwin single-particle orbitals. As a result, at
some critical field $B_{\rm{z,cr}}^{(4e)}$ the $(\nu=0,m=-1)$ singlet becomes the ground state as shown in the eigenenergy dependencies of \FigEnFD, calculated according to
Eq.~\eqref{Eq:EnFD}. The nature of the ground state phase transition can be understood intuitively as follows. The low-field ground state is dominated by Coulomb repulsion, which forces the two $p$-shell electrons to occupy different orbitals ($m=-1$ and $m=+1$), resulting in a ground-state triplet \citep{Tavernier2003}. At a sufficiently high magnetic field $B_{\rm{z}}>B_{\rm{z,cr}}^{(4e)}$ the energy cost of promoting an electron from the $m=-1$ to the $m=+1$ single-particle orbital exceeds the exchange energy $K$. As a result, it is energetically favorable for the two $p$-shell electrons to occupy the same $m=-1$ orbital, resulting in a ground-state singlet.

The exchange splitting can be estimated from the difference in the fourth and third addition energies \citep{Tarucha1996,Kouwenhoven2001}. The typical values for the studied QDs, shown by the triangles in \FigEnCharge\;, are
$\lvert K\rvert\approx1-3$~meV, which corresponds to $B_{\rm{z,cr}}^{(4e)}\approx 0.6 - 1.7$~T. For QD4, the estimated exchange interaction is $\lvert K\rvert\approx2$~meV, which corresponds to $B_{\rm{z,cr}}^{(4e)}\approx 1.2$~T. This estimate of the critical field, derived from a simple Fock-Darwin model, matches the range of magnetic fields where the measured NSR rate ratio $\varGamma_{\rm{N}}^{(4e)}/\varGamma_{\rm{N}}^{(0e)}$ depends strongly on $B_{\rm{z}}$ in QD4. Just as in case of kinks and cusps in the charge carrier transport spectra, the $\varGamma_{\rm{N}}^{(4e)}(B_{\rm{z}})$ dependence gives a somewhat indirect evidence of the changes in the 4$e$ ground state. Luckily, as we now show, nuclear spins allow for direct probing of the spin character of the few-electron ground state via NMR spectroscopy. 

The NMR spectra are measured on individual QDs with the optical pump-rf-probe protocol, with the timing diagram shown in the inset of \FigNMRZeroFoure. Further details of the NMR spectroscopy techniques can be found in Supplementary Section 3. By choosing the gate bias during the rf pulse, we measure NMR spectra of the QD nuclei subject to hyperfine (Knight) field of the different electronic configurations. \FigNMRZeroFoure\; shows NMR spectra of the $^{75}$As spin-3/2 nuclei measured on a magnetic-dipole transition between the states with $I_{\rm{z}}=\pm1/2$ projections of the nuclear spins. At high magnetic field ($B_{\rm{z}}=5.3$~T, top part of the graph) we find similarly narrow NMR spectra for an empty QD (0$e$) and for the 4$e$ Coulomb blockade state. The NMR full width at half maximum of $\approx1$~kHz is limited by the second-order quadrupolar and the nuclear dipole-dipole spin interactions \citep{Chekhovich2020,Dyte2025}, indicating the absence of the Knight shifts and confirming the spin-0 phase of the four-electron ground state. 

\begin{figure*}
\includegraphics[width=0.95\linewidth]{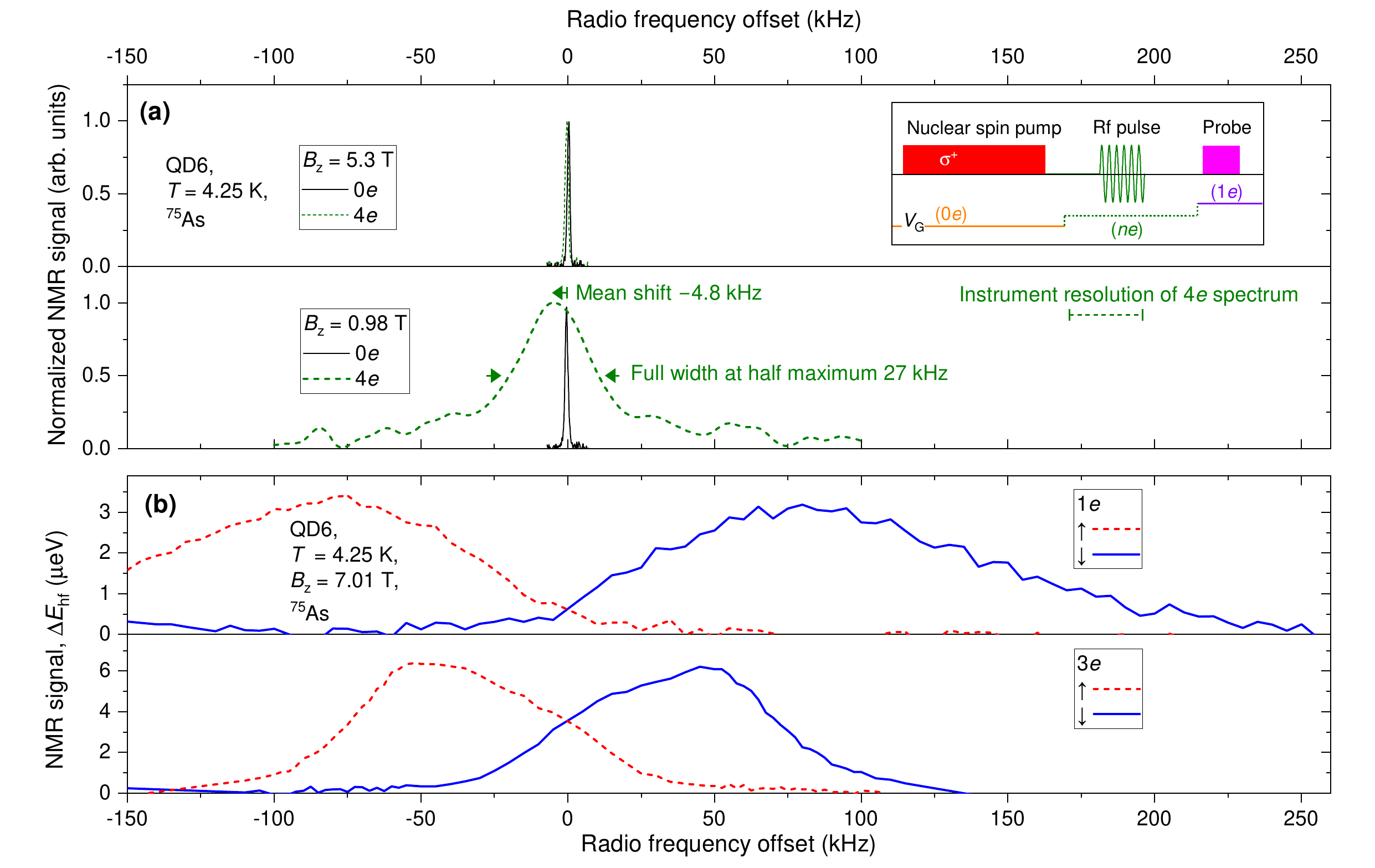}
\caption{\label{Fig:NeNMR} {Nuclear magnetic resonance spectra of individual quantum dots.} (a)  The inset shows the timing diagram of the NMR measurement cycle, including optical pumping of the nuclear spins, radiofrequency (rf) depolarization, optical probing, and the variation of the gate bias $V_{\rm{G}}$ during the cycle. The main graph shows NMR spectra of the $N\approx10^4$ $^{75}$As nuclei in a single QD measured at high magnetic field ($B_{\rm{z}}=5.3$~T, top part) and low field ($B_{\rm{z}}=0.98$~T, bottom part). Solid (dashed)
lines show spectra measured in an empty QD state 0$e$ (QD in a 4$e$ Coulomb blockade regime). The spectra are derived from Fourier transform free induction decay spectroscopy. The exception is the 4$e$ spectrum at $B_{\rm{z}}=0.98$~T, recorded by varying the frequency offset of the rf pulse, in which case the spectral resolution is limited by the pulse duration and equals $\approx25$~kHz. See Supplementary Section 3 for details on the NMR techniques. (b) NMR spectra measured in a 1$e$ Coulomb blockade regime (top part) and in a 3$e$ Coulomb blockade regime (bottom part). These NMR spectra are measured in a single-shot regime, without averaging over multiple pump-rf-probe cycles.  At finite rf offsets the two distinct values of $E_{\rm{hf}}$ are detected randomly, corresponding to spin-up (dashed lines) and spin-down (solid lines) states of the unpaired electron spin.}
\end{figure*}

At low magnetic field ($B_{\rm{z}}=0.98$~T, bottom part of \FigNMRZeroFoure) the 0$e$ NMR spectrum remains narrow, whereas the 4$e$ spectrum is significantly broadened, providing direct evidence of the hyperfine (Knight) field induced by the uncompensated electron spins. This low-field ground state is often referred to as polarized \cite{MacDonald1993,Muller1996,Kouwenhoven2001}. However, the Knight shifts observed in NMR are determined by both the ground state and the excited electronic states within the thermal energy $\propto k_{\rm{B}}T$. This explains why the low-field 4$e$ NMR spectrum stretches both into positive and negative frequency offsets, indicating that during the NMR rf pulse (which has a full width at half maximum of 40~$\mu$s for the 4$e$ NMR measurement) the QD intermittently goes through the states with both positive and negative $z$-projections of the electron spin. The Knight shift broadening, characterized by the full width at half maximum of $\approx27$~kHz, is accompanied by a $\approx-4.8$~kHz mean shift of the NMR peak. The shift is a combined effect of the unequal (thermal) population of the ground and excited electronic states, and possibly the unequal electron spin projections of these states. 

NMR spectroscopy gives the most direct confirmation that the 4$e$ ground state undergoes a phase transition from a polarized phase (also referred to as maximum density droplet phase \citep{Muller1996,Tavernier2003}) at low magnetic fields to a spin-0 phase (also refereed to as Wigner molecule in the literature) at high magnetic fields. The NMR data also agrees with the NSR data (\FigNSRRatBz), which yields $\varGamma_{\rm{N}}^{(4e)}/\varGamma_{\rm{N}}^{(0e)}>1$ in the polarized phase and $\varGamma_{\rm{N}}^{(4e)}/\varGamma_{\rm{N}}^{(0e)}\approx1$ in the spin-0 phase. According to the NSR data, the critical field $B_{\rm{z,cr}}^{(4e)}$ is between 3 and 5.5~T. At the same time, the data of \FigNSRRatBz\; shows that below the critical field the NSR rate $\varGamma_{\rm{N}}^{(4e)}$ is not constant. Instead, $\varGamma_{\rm{N}}^{(4e)}$ depends strongly on magnetic field $B_{\rm{z}}$, suggesting that the four-electron ground state evolves with $B_{\rm{z}}$ even within the same (polarized) phase. We next discuss this strong dependence with the aid of CI model calculations.

\subsection{Anomalously fast nuclear spin relaxation and diffusion in the four-electron configuration}

In order to explain fast NSR in the 4$e$ configuration we first discuss the mechanisms of NSR. Nuclear spin diffusion is a process where the localized optically pumped nuclear spin polarization in a QD is transferred into the surrounding barriers through spin-exchange flip-flops between the adjacent nuclear spins. Spin diffusion happens both in the dark and under optical pumping of nuclear spins. Under long pumping, spin diffusion polarizes the nuclei not only in the QD, but also in the surrounding material. As a result, subsequent spin diffusion in the dark (after optical pumping is switched off) is suppressed. Thus, the duration of optical pumping can be used as a tool to distinguish spin-diffusion and non-spin-diffusion NSR mechanisms. \FigNSRTPump\; shows NSR rate ratios $\varGamma_{\rm{N}}^{(ne)}/\varGamma_{\rm{N}}^{(0e)}$ measured as a function of optical pumping time $T_{\rm{Pump}}$ for 1$e$ ($n=1$) and 4$e$ ($n=4$) Coulomb blockade charge configurations. Spin diffusion in the 1$e$ configuration (circles in \FigNSRTPump) has been studied previously \cite{MillingtonHotze2022}, so we discuss this effect only briefly. Acceleration of spin diffusion in the 1$e$ configuration occurs because the electron spin can mediate spin exchange between distant nuclear spins, as opposed to direct dipole-dipole interaction that couples only adjacent nuclei. According to \FigNSRTPump\;, in the limit of long pumping the electron has only a small effect on the NSR, characterized by $\varGamma_{\rm{N}}^{(1e)}/\varGamma_{\rm{N}}^{(0e)}\lesssim2$. For short pumping, this increases to $\varGamma_{\rm{N}}^{(1e)}/\varGamma_{\rm{N}}^{(0e)}\gtrsim15$. Thus the effect of the spin of a single electron on NSR is almost entirely through acceleration of nuclear spin diffusion.  

The results for the spin-polarized 4$e$ configuration (measured at $B_{\rm{z}}=0.39$~T, which is below the critical field) are quite different and more complicated. For the longest available pumping time ($T_{\rm{Pump}}\approx1000$~s in \FigNSRTPump) NSR in the 4$e$ configuration is a factor of $\varGamma_{\rm{N}}^{(4e)}/\varGamma_{\rm{N}}^{(0e)}\approx30$ faster than in an empty QD (0$e$). When the optical pumping is shortened ($T_{\rm{Pump}}\approx0.008$~s is the shortest available in experiment), NSR becomes even faster, characterized by $\varGamma_{\rm{N}}^{(4e)}/\varGamma_{\rm{N}}^{(0e)}\gtrsim6000$. These results show that there are multiple NSR mechanisms induced by the 4$e$ configuration: both the non-diffusion NSR (dominant at long $T_{\rm{Pump}}$) as well as acceleration of spin diffusion (responsible for fast NSR at short $T_{\rm{Pump}}$). In order to unravel these mechanisms we calculate the spectrum of the 4$e$ eigenstates using the CI model.

The CI spectrum of the 4$e$ eigenenergies as a function of the external magnetic field $B_{\rm{z}}$ is shown in \FigEnCI. For each eigenstate with wavefunction $\psi$ we calculate the magnitude of the total electron spin projection $\vert S \vert=\sqrt{\langle \psi \vert \hat{S}_{\rm{x}} \vert \psi \rangle^2+\langle \psi \vert \hat{S}_{\rm{y}} \vert \psi \rangle^2+\langle \psi \vert \hat{S}_{\rm{z}} \vert \psi \rangle^2}$, shown in \FigSTotCI. We further define the polar angle of the total spin quantization axis $\theta = \arctan{\left(\sqrt{\langle \psi \vert \hat{S}_{\rm{x}} \vert \psi \rangle^2+\langle \psi \vert \hat{S}_{\rm{y}} \vert \psi \rangle^2}\big/\langle \psi \vert \hat{S}_{\rm{z}} \vert \psi \rangle\right)}$, shown in \FigSThetaCI. Further details can be found in Supplementary Section 5C. The lowest energy manifold predicted by the CI is a doublet of states originating from the triplet ($S=1$) states with $z$-projections $S_{\rm{z}}=\pm1$. Exchange interaction and electron-electron correlations in the anisotropic quantum dot potential shift the $S_{\rm{z}}=0$ projection state to higher energies and further mix the $S_{\rm{z}}=\pm1$ states. The resulting eigenstates are characterized by a reduced but non-zero spin projections $\vert S \vert \approx 3\times10^{-2}$. Moreover the spins of these two ground states are quantized along the axis that is nearly orthogonal to the external magnetic field ($\theta \approx 70^\circ$ and $\theta \approx 110^\circ$). Such 4$e$ electronic states are very different both from the 2$e$ configuration, where the spin of the singlet state is very small $\vert S \vert <10^{-4}$, and the 3$e$ and 5$e$ configuration where the spin predicted by CI is close to $\vert S \vert \approx 1/2$ and is quantized approximately along the external magnetic field $z$ axis (see details in Supplementary Section 5C). We ascribe the anomalously fast NSR to this peculiar spin character of the 4$e$ ground state doublet. Next, we discuss how the structure of the 4$e$ spectrum affects the non-diffusion and spin diffusion mechanisms.

We start with the non-diffusion mechanism. The smaller spin projections $\vert S \vert$ of the 4$e$ polarized states predicted in CI are corroborated by the NMR spectra, which show considerably smaller hyperfine shifts (narrower NMR spectrum) experienced by the nuclei in the 4$e$ configuration, when compared to the 1$e$ and 3$e$ configurations (Fig.~\ref{Fig:NeNMR}). Although $\vert S \vert$ is small, the deviation of the 4$e$ electron spin quantization axis from the external magnetic field is what leads to the non-diffusion contribution towards the anomalously fast NSR. In order to understand the NSR mechanism, we first note that the nuclear spins are quantized essentially along the external magnetic field (due to the small electric quadrupolar effects in lattice-matched GaAs/AlGaAs structures). If the electron spin is also quantized along the external field, as is the case with good accuracy for 1$e$, 3$e$, and 5$e$ ground states, then the energy mismatch between the electron and nuclear spin Zeeman splittings precludes electron-nuclear spin flip-flops. In this ``collinear'' case, the strong magnetic field effectively truncates the hyperfine Hamiltonian (Eq.~\eqref{Eq:Hhf}) leaving only the $\propto \hat{s}_{\rm{z}}\hat{I}_{{\rm{z}},k}$ terms, under which the electron can change the nuclear spin splitting (via Knight shift) but cannot flip the nuclear spin from one $\hat{I}_{{\rm{z}}}$ eigenstate to the other. The situation is radically different if the electron spin is quantized along a tilted axis, as is the case for the 4$e$ ground state according to the CI model calculations. In this ``noncollinear'' case the electron spin can cause nuclear spin precession. If the electron spin undergoes spin jumps, it generates an effective stochastic transverse magnetic field that acts on the nuclei, leading to NSR. Indeed, from NMR measurements (see Section~\ref{subsec:SpinQubits}) the electron spin lifetime is shorter than the $40~\mu$s rf pulse, suggesting that the jumps in the 4$e$ configuration occur on a timescale of tens of $\mu$s, or shorter. Such telegraph-noise-like \cite{Dyte2023} random magnetic field contains a sufficiently strong spectral component that is resonant with the nuclei and provides coupling of the nuclear spins to the lattice phonons (via the electron spin) resulting in NSR without nuclear spin diffusion. The measured non-diffusion NSR rate driven by the 4$e$ configuration is on the order of  $\varGamma_{\rm{N}}^{(4e)}\approx0.12~{\rm{s}}^{-1}$ (at $B_{\rm{z}}=0.39$~T, $T_{\rm{Pump}}\approx1000$~s, see Supplementary Fig.~5(c)), which is much faster than the NSR rate $\varGamma_{\rm{N}}^{(0e)}\approx0.0033~{\rm{s}}^{-1}$ in an empty (0$e$) QD under the same conditions. The measured $\varGamma_{\rm{N}}$ agrees with results of numerical modelling on a simple system of one nucleus coupled to the electron spin. This simple model also confirms that the relaxation mechanism requires both the non-collinearity (i.e. the tilt of the electron quantization axis) and the electron spin jumps. We note a similarity with noncollinear hyperfine interaction in Stranski-Krastanov QDs \citep{Huang2010,Latta2011}, except that the roles of the nuclei and electrons are reversed there: the large strain-induced quadrupolar effects tilt the effective quantization for the nuclei in Stranski-Krastanov QDs \cite{Bulutay2012}, while the electron is quantized approximately along the magnetic field.


We now turn to the spin-diffusion component of the anomalously fast NSR induced by the polarized 4$e$ configuration. For magnetic fields exceeding tens of mT, the direct electron-nuclear spin flip-flops are suppressed by the energy mismatch. However, the electron spin can mediate a flip-flop of two nuclear spin. For two nuclei $j$ and $k$ with hyperfine constants $a_j$, $a_k$ the effective flip-flop coupling can be derived as a second order perturbation \citep{Coish2008,vonWitte2025} and scales as $\propto a_j a_k/\Delta E_{\rm{e}}$. This interaction plays an important role in nuclear spin diffusion \citep{Klauser2008,Reilly2010,Gong2011,MillingtonHotze2022}, since it allows spin transfer between the distant nuclei, whose direct dipole-dipole coupling is otherwise negligible. Electron-mediated spin diffusion can be derived for various configurations, including a two-electron two-nucleus system as demonstrated recently \citep{vonWitte2025}. However, such derivations are complex, meaning that rigorous treatment of spin-diffusion in a 4$e$ configuration is outside the scope of this work. Instead, we gain insights from numerical modelling of a simple system of two nuclei quantized along the $z$ axis and coupled via hyperfine interaction (Eq.~\eqref{Eq:Hhf}) to a common electron whose quantization axis is tilted away from the $z$ axis. Unlike with direct (non-diffusion) NSR discussed above, we find that the tilt of the electron quantization axis has no significant impact on spin-diffusion NSR. By contrast, the short electron spin lifetime of the 4$e$ configuration plays an important role in allowing flip-flops between two nuclei with slightly different effective Zeeman energies (non-resonant nuclei). The fluctuating hyperfine (Knight) field of a rapidly flipping electron spin broadens the nuclear spin energy levels. Such broadening bridges the energy gap of the nuclei that are otherwise made non-resonant by the inhomogeneous strain-induced quadrupolar shifts and chemical shifts. Furthermore, the small spin projection $\vert S \vert$ of the 4$e$ configuration reduces the electron spin splitting $\Delta E_{\rm{e}}$ in the energy denominator, while the large spatial extent of the 4$e$ wavefunction provides electron mediated coupling for a larger number of nuclear spins. This interpretation agrees with our experimental results, which reveal very pronounced acceleration of nuclear spin diffusion in the 4$e$ configuration (at least a factor of $\approx 300$ faster than spin diffusion driven by a single electron in 1$e$ configuration, see \FigNSRTPump). Interestingly, the measured absolute NSR rates in the 4$e$ diffusion regime ( $\varGamma_{\rm{N}}^{(4e)}\gtrsim 270~{\rm{s}}^{-1}$ at $B_{\rm{z}}=0.39$~T, $T_{\rm{Pump}}\approx0.008$~s, see Supplementary Fig.~5(c)) are comparable to the rates at zero magnetic field $\varGamma_{\rm{N}}\approx10^3~{\rm{s}}^{-1}$, where dipole-dipole interactions do not conserve spin and NSR is very fast. 

In summary, the CI model provides an accurate qualitative description of subtle many-body phenomena such as reduced spin projection in the 4$e$ triplet state, and the resulting anomalously fast NSR and nuclear spin diffusion. Spin diffusion is an important tool in NMR signal enhancement and structural analysis \cite{Schmidt1992, Demco1995,Hall1997,vanderWel2006,Manolikas2008,Rossini2014,VigerGravel2018}. Experimental results on the 4$e$ triplet give an example of how spin diffusion can be engineered through control of the few-electron configuration. On the other hand, the CI numerical model is inevitably limited by a variety of factors, including the accuracy of the material parameters and approximate knowledge of QD morphology. This leads to quantitative discrepancies, such as prediction of the second phase transition of the 4$e$ ground state at $B_{\rm{z}}\approx 7$~T, which is not observed in experiments up to $B_{\rm{z}}\approx 10$~T. Further work would be needed examine in more detail the mechanisms of NSR and spin diffusion in presence of few-electron configurations.


\subsection{Spin orbit interaction in the five-electron configuration}\label{subsec:SOI}

Experimental data in \FigNSRBz\; exhibits an additional peak in the NSR rate in the middle of
the 5$e$ Coulomb blockade valley (at $V_{\rm{G}}\approx1.3$~V) but only at the strongest magnetic field $B_{\rm{z}}=9.82$~T available in this work. Such extra NSR rate peak has been predicted previously to arise from the
spin-orbit interaction (SOI), but to our knowledge
has not been observed experimentally. In the previous theoretical work, conducted in the context of the gate-defined QDs \citep{LyandaGeller2002,LyandaGeller2003}, the appearance of an extra peak was classified as a weak SOI regime. In the present study on epitaxial QDs, this regime turns out to be the strongest achievable SOI because no other additional peaks are observed in the middle of the Coulomb blockade plateaus. Observation of the peak in the 5$e$ plateau and its absence for other configurations with an unpaired electron spin (1$e$ and 3$e$) agrees with the
general expectation that the spin orbit interactions are enhanced
by the many-body effects. In the same manner, the expected
increase of the spin orbit effects with magnetic field
\citep{Baruffa2010,Camenzind2018} explains why the extra NSR
peak is seen only at the highest $B_{\rm{z}}$ and already becomes
indistinct when the field is reduced to $B_{\rm{z}}=8$~T
(see additional data Supplementary Section 4A). Experimentally, the additional peak is observed only at one value of the external field, hence our attribution of this peak to SOI is tentative. Further experimental studies would be needed and would benefit from stronger magnetic fields, lower temperatures, and higher electron addition energies, which would facilitate observation of the peculiar peak-in-the-plateau structure.

\begin{figure}
\includegraphics[width=0.5\linewidth]{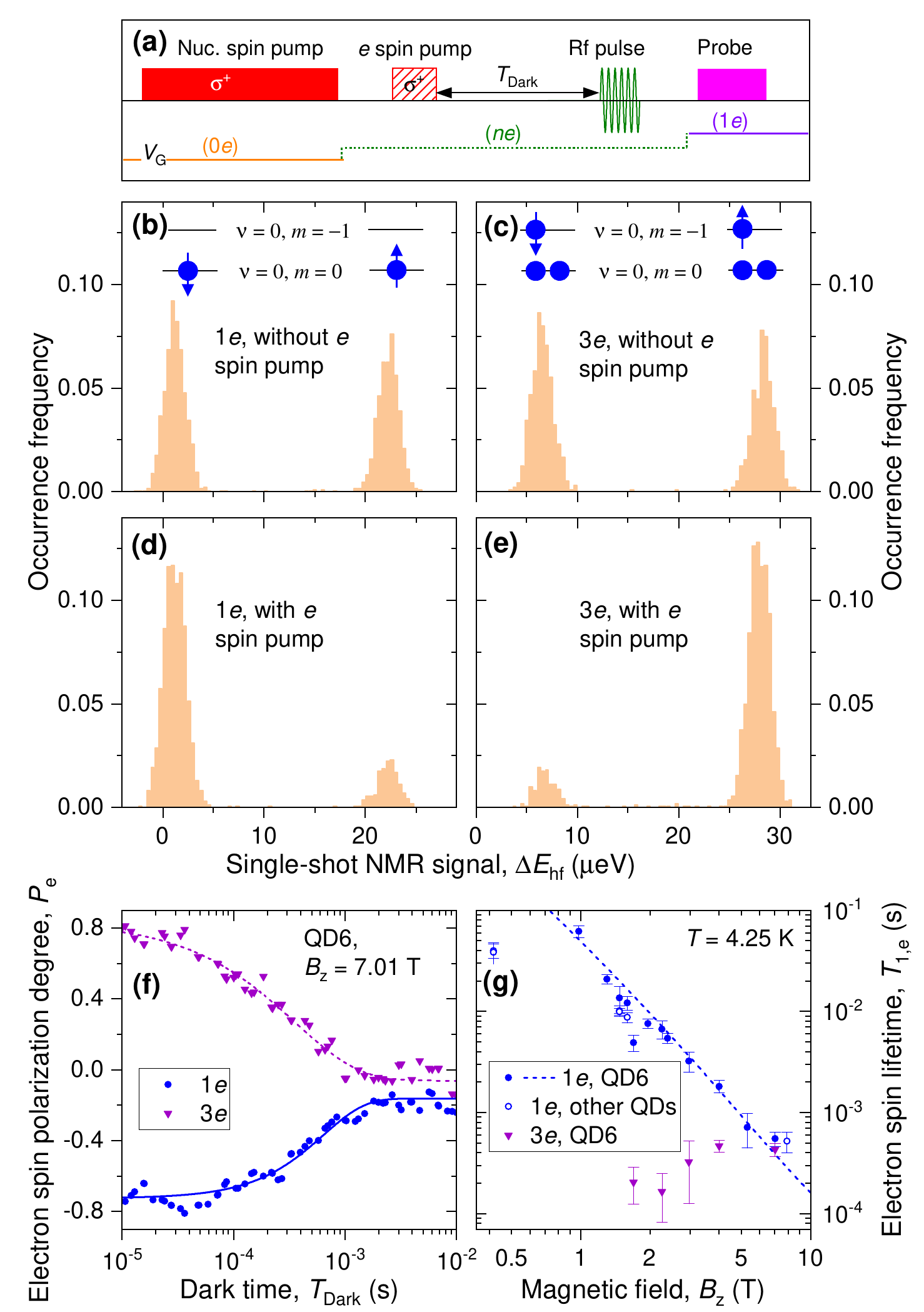}
\caption{\label{Fig:3e} {{Quantum dot few-electron spin qubits.} (a) Timing diagram of the experiment to probe electron and nuclear spins. The cycle starts with a quasiresonant circularly polarized laser pulse to polarize the nuclear spins in an empty (0$e$) QD. After the bias is changed to load QD with $n$ electrons, an optional second pulse of the same laser can be applied to polarize the unpaired electron spin. Following the dark evolution delay $T_{\rm{Dark}}$, a detuned rf $\pi$ rotation pulse is applied to convert electron spin into a nuclear spin hyperfine shift $E_{\rm{hf}}$, which is then read out by measuring the $X^{1-}$ PL spectrum under a nonresonant optical probe \cite{Gillard2022,Dyte2023}. (b) Histogram of the hyperfine shift variation values $\Delta E_{\rm{hf}}$ measured at an optimal rf offset in 1$e$ charge configuration without electron spin pumping. Two distinct histogram modes correspond to spin-up ($\uparrow$) and spin-down ($\downarrow$) states of an unpaired $s$-shell electron (orbital state $(\nu=0,m=0)$), depicted with balls and arrows. (c) Same as (b) but measured in a 3$e$ charge configuration with the two histogram modes corresponding to spin-up and spin-down states of an unpaired $p$-shell electron (orbital state $(\nu=0,m=-1)$). (d) Same as (b) but including the electron spin pumping pulse. The delay time is $T_{\rm{Dark}}\approx10~\mu$s. The balance of peaks reveals predominant population of the spin-down state. (e) same as (c) but with electron spin pumping. Here, the wavelength is tuned to populate preferentially the spin-up state of the unpaired $p$-shell electron. (f) Decay of the $s$-shell (circles) and $p$-shell (triangles) electron spin polarization measured at $B_{\rm{z}}=7.01$~T by varying $T_{\rm{Dark}}$. Lines show compressed exponential fitting yielding the characteristic $1/e$ spin lifetimes $T_{1,{\rm{e}}}$. (g) Magnetic field dependence of $T_{1,{\rm{e}}}$. Error bars are 95\% confidence intervals.}}
\end{figure}

\subsection{Few-electron spin qubits and electron spin lifetimes}\label{subsec:SpinQubits}

The few-electron configurations, observed in this work via nuclear spin relaxation spectroscopy, offer a variety of electronic states that can be used as spin qubits. We demonstrate this by performing initialization and readout of the resident electron spin in different shells. The timing diagram of the relevant experiment is shown in \FigTDiageSpin. In addition to optical nuclear spin pumping, a second optical pulse (typically 100~$\mu$s long) is applied optionally in order to polarize the electron spin after QD is loaded with $n$ electrons. Adapting the techniques developed recently on InGaAs QDs \citep{Gillard2022}, we use nuclear spins for single-shot readout of the electron spin \citep{Dyte2023}. An electron spin up $\uparrow$ (spin down $\downarrow$) shifts the NMR to lower (higher) frequencies via hyperfine (Knight) shift. Thus, an rf $\pi$ pulse detuned to lower frequencies rotates (does not rotate) the nuclear spins if the electron is in the $\uparrow$ ($\downarrow$) state. Following the detuned rf pulse, which performs quantum non-demolition measurement of the electron spin $s_{\rm{z}}$ projection, the nuclear hyperfine shift $E_{\rm{hf}}$ is probed via photoluminescence spectroscopy. Importantly, the measurement is conducted in a single-shot mode, without averaging over multiple pump-rf-probe cycles. A large variation $\Delta E_{\rm{hf}}$ of the hyperfine shift $E_{\rm{hf}}$ (a small variation of $E_{\rm{hf}}$) produced by the detuned rf pulse heralds the detection of a $\uparrow$ ($\downarrow$) electron spin state. In the experiment, a large number of pump-rf-probe cycles is measured. The resulting individual single-shot NMR signals are then plotted as histograms in Figs.~\ref{Fig:3e}(b), \ref{Fig:3e}(c), confirming that $\Delta E_{\rm{hf}}$ indeed takes on two distinct values, observed as a bimodal statistical distribution. This method of electron spin readout has been applied recently to the $s$-shell electron spin qubit in the 1$e$ configuration \cite{Dyte2023}. Here we apply this technique to the 3$e$ configuration by choosing the gate bias $V_{\rm{G}}$ during the rf pulse to be at the center of the 3$e$ Coulomb blockade plateau. The results are shown in Figs.~\ref{Fig:3e}(c), \ref{Fig:3e}(e). The two distinct modes in the histogram of the single-shot NMR signals are attributed to the spin up ($\uparrow$) and spin down ($\downarrow$) states of the unpaired $p$-shell electron, demonstrating that high-fidelity single-shot readout can be performed on the qubit states of the few-electron configurations. 

When the electron spin pumping pulse is omitted, the $\uparrow$ and $\downarrow$ readouts have nearly the same probabilities, as demonstrated in the histograms of Figs.~\ref{Fig:3e}(b), \ref{Fig:3e}(c) measured with rf detunings optimized for the spin readout in the 1$e$ and 3$e$ configurations, respectively. By activating the electron spin optical pumping pulse and tuning its wavelength we are able to prepare a non-equilibrium electronic state either with a positive or a negative spin polarization. This is shown in Figs.~\ref{Fig:3e}(d), (e) (for 1$e$ and 3$e$ configurations, respectively), where one finds a pronounced difference in the probabilities to detect the two possible $\Delta E_{\rm{hf}}$ values. For simplicity we use the same photon energy (typically tuned $\approx10~$meV above the ground state trion recombination energy) both for the nuclear and electron spin pumping. Such quasiresonant optical pumping limits the electron spin polarization degrees to $\vert P_{\rm{e}} \vert \lesssim 0.8$, but is sufficient for a proof-of-principle demonstration that $s$ and $p$ shell electrons can be initialized optically. Further improvement in electron spin preparation fidelity can be achieved using the well-known resonant optical spin pumping techniques \citep{Atature2006}.

Next we apply the single-shot electron spin readout protocol to perform NMR spectroscopy and probe the electron-nuclear interactions. To this end we measure the histograms of the single shot hyperfine shifts $\Delta E_{\rm{hf}}$, such as shown in Figs.~\ref{Fig:3e}(b) - \ref{Fig:3e}(e), but at variable frequency detunings of the rf pulse. In these experiments the electron spin pumping pulse is omitted, so that the histograms include the contributions from both electron spin projection states $\uparrow$ and $\downarrow$. For each rf frequency offset we find the centres of gravity of the two histogram modes and plot them as shown in \FigNMROneThreee. The traces obtained from the two histogram modes correspond to the two branches of the NMR spectra, measured selectively for the $\uparrow$ (dashed lines) and $\downarrow$ (solid lines) states of the unpaired electron spin. The method is applicable both to the 1$e$ (top part of \FigNMROneThreee) and 3$e$ (bottom part of \FigNMROneThreee) charge configurations. The NMR spectra of \FigNMROneThreee\; show that the average Knight shifts produced by an unpaired $p$-shell electron of the 3$e$ configuration are a factor of $\approx2$ smaller than those of the $s$-shell electron in the 1$e$ configuration. This observation is consistent with the expected larger spatial extent of the envelope wavefunction of the more delocalized $p$-shell electron. The smaller Knight shift indicates that the $p$-shell electron envelope wavefunction is spread over a $\approx2$ times larger number $N$ of the nuclei of the crystal lattice. An even more pronounced increase in $N$ would be expected for higher electron numbers $n$. Since the electron spin qubit dephasing time scales as $N^{-1/2}$ the use of the higher-shell electrons as spin qubits \citep{Leon2020} can be a route to an improved qubit coherence.

Employing the spin preparation and readout techniques we vary the delay $T_{\rm{Dark}}$ between the electron spin optical pump and the rf spin detection pulse to measure the decay of the electron spin polarization. The results are averaged over many pump-delay-rf-probe cycles and are shown by the symbols in \FigeDec\; revealing a gradual decay of the optically-pumped electron spin polarization. Using exponential fitting (lines) we derive the electron spin lifetimes $T_{1,{\rm{e}}}$, which are plotted in \FigeTeBz\; as a function of magnetic field $B_{\rm{z}}$. Electron spin relaxation is known to be dominated by the phonon-related mechanism, and in the high-temperature regime $\Delta E_{\rm{e}}\gg k_{\rm{B}}T$ it leads to the following electron spin lifetime scaling \citep{Khaetskii2000,Kroutvar2004,Camenzind2018}:
\begin{equation}
T_{1,{\rm{e}}}\propto (k_{\rm{B}}T)^{-1} B_{\rm{z}}^{-2} \Delta E_{\rm{e}}^{-2} \label{Eq:T1eLaw}
\end{equation}
For InGaAs QDs, Zeeman splitting dominates over the nuclear hyperfine (Overhauser) field $\vert\mu_{\rm{B}}g_{\rm{e}}B_{\rm{z}}\vert\gg \vert E_{\rm{hf}}\vert$ in a wide range of magnetic fields. As a result, the total electron spin splitting is dominated by the Zeeman effect $\Delta E_{\rm{e}}\propto B_{\rm{z}}$. This leads to the well-known $T_{1,{\rm{e}}}\propto B_{\rm{z}}^{-4}$ (high temperature regime \citep{Gillard2021}) or $T_{1,{\rm{e}}}\propto B_{\rm{z}}^{-5}$ dependence (low temperature regime \citep{Kroutvar2004,Lu2010}). By contrast, in the GaAs epitaxial QDs studied here the electron $g$-factor is small \citep{Ulhaq2016,MillingtonHotze2022} ($g_{\rm{e}}\approx-0.04$ in QD6). With the typical $E_{\rm{hf}}\approx-50~\mu$eV in our electron spin relaxation experiments, the hyperfine shift dominates over the Zeeman effect ($\vert E_{\rm{hf}}\vert >\vert\mu_{\rm{B}}g_{\rm{e}}B_{\rm{z}}\vert$), resulting in an unusual field dependence close to $T_{1,{\rm{e}}}\propto B_{\rm{z}}^{-2}$. Using a factor of proportionality as the only fitting parameter, we plot Eq.~\eqref{Eq:T1eLaw} by the dashed line in \FigeTeBz\; to find good agreement with the experiment on the $s$-shell electron spin (1$e$). At low magnetic fields $B_{\rm{z}}\lesssim1$~T, the measured lifetimes saturate at $T_{1,{\rm{e}}}\approx50$~ms, indicating that the field-independent cotunneling-related mechanism \cite{Gillard2021} of electron spin relaxation becomes dominant. The rate of the cotunneling-related electron spin relaxation is controlled by the thickness of the tunnel barrier between the QD and the electron Fermi reservoir, and is therefore specific to the studied semiconductor device. By contrast the phonon-related electron spin relaxation rate has a fairly universal dependence on electron spin energy splitting, temperature, and a rather weak dependence on the size of the QD.  Therefore, the $T_{1,{\rm{e}}}$ relaxation times shown in  \FigeTeBz\; for the 1$e$ configuration above $B_{\rm{z}}\gtrsim1$~T can be used as a practical calibration curve: the spin lifetime $T_{1,{\rm{e}}}$ is an important parameter since it sets an upper limit for the electron spin coherence time $T_{2,{\rm{e}}}$. Our results show that for the $s$-shell electrons in GaAs/AlGaAs QDs the lifetimes range from $T_{1,{\rm{e}}}\approx50$~ms to $\approx0.5$~ms when magnetic field is changed between $B_{\rm{z}}\gtrsim1$~T and 8~T. Even longer $T_{1,{\rm{e}}}$ can be achieved in principle at low magnetic fields by increasing the thickness of the tunnel barrier, although this comes at the expense of a more prominent Auger recombination, with some unwanted consequences for optical manipulation of the electron spin qubit \citep{Gillard2021}. 

The spin lifetimes of the unpaired $p$-shell electron in the 3$e$ configuration (triangles in \FigeTeBz) show a rather different trend with a moderate increase in $T_{1,{\rm{e}}}$ with the applied magnetic field. The relaxation of the $p$-shell electron is likely dominated by the field-independent cotunnelling coupling to the electron Fermi reservoir in the entire range of the studied magnetic fields $B_{\rm{z}}<8$~T. This explanation is consistent with the smaller width of the 3$e$ Coulomb blockade valley (\FigNSRBz), which becomes even narrower at low magnetic fields, enhancing both the NSR and the electron spin decay. We expect that the $p$-shell electron spin lifetimes can be increased significantly by effectively widening the 3$e$ Coulomb blockade plateau. This in turn can be achieved  by a moderate reduction of the base temperature (from $T=4.25$~K to $T\approx2$~K) or by reducing the lateral size of the QD in order to increase the electron charging energy (see \FigEnCharge). With these improvements, a more in-depth investigation of the few-electron spin lifetimes should be possible, followed by investigation of spin coherence in the few-electron qubits of epitaxial QDs.

We also conduct the single-shot NMR experiments in the spin-polarized (low-field) phase of the 4$e$ configuration. NMR is measured at $B_{\rm{z}}=0.98$~T with $T_{\rm{rf}}=80~\mu$s rf pulses (full width at half maximum is  $40~\mu$s) but, unlike for 1$e$ and 3$e$ configurations, no bimodal distribution can be observed in the histogram of the single-shot NMR signals. Such lack of bimodality can be easily explained by assuming that the electron spin lifetime of the 4$e$ polarized state is shorter than $40~\mu$s. In that case the electron spin polarization goes through the opposite states during each rf pulse, so that the NMR measures the time-averaged Knight shifts rather than the eigenstate values. While the rf pulse can be made much shorter than $40~\mu$s, this leads to spectral broadening of the rf pulse in excess of the Knight shift splitting, which is small due to the small spin projections of the 4$e$ polarized states. As a result the opposite 4$e$ spin states cannot be resolved in NMR even with short rf pulses. However, NMR spectroscopy is sufficient to conclude that the spin lifetimes of the polarized 4$e$ phase states are very short, at least 3 orders of magnitude shorter than in the 1$e$ spin qubit. The short $T_{1,{\rm{e}}}$ agrees with the CI calculations, which indicate strong mixing of the spin and orbital degrees of freedom as well as tilting of the quantization axis of the 4$e$ ground states with respect to the external magnetic field. Further investigation of the fast 4$e$ electron spin dynamics can be attempted using optical techniques for spin pumping and detection \cite{Atature2006,Lu2010,Gillard2021}, and would shed more light on the nature of the 4$e$ states and the physics of the ground state phase transitions.

\section{Discussion and outlook}

The results presented in this work demonstrate the rich variety of the few-electron states that can be constructed in optically active epitaxial GaAs QDs. Such $n$-electron states may be used to build qubits with favorable properties. The large spin-orbit effects observed with the increasing number of the electrons in a QD offer a route to spin-orbit qubits \citep{Nowack2007,NadjPerge2010} where coherent control can be implemented with oscillating electric fields, to achieve scalability and high speeds \citep{Yoneda2014} that are not attainable with oscillating magnetic fields
\citep{Koppens2006}. Such an electric dipole spin resonance (EDSR) control, previously achieved only in the milli-Kelvin regime \citep{Nowack2007,NadjPerge2010,Shafiei2013}, could thus be adapted to epitaxial QDs where higher operating temperatures (few-Kelvin) and efficient spin-optical interfaces offer advantages for implementation of quantum devices.

Bias-dependent nuclear spin relaxation (NSR) spectroscopy demonstrated here is a versatile probe of both the energy spectrum and the spin properties of the few-electron states. The technique does not rely on electronic transport, which is often unavailable in the optically active semiconductor nanostructures, and can thus be extended to a wider range of devices. The absence of electric currents is itself an advantage, as it permits probing of the few-electron quantum status near equilibrium. The NSR spectroscopy technique is versatile and can be applied to a wide range of QD structures, without requiring any specialised design of the semiconductor device.   

Although the hyperfine coupling of the valence band carriers is small compared to the conduction band electrons, we expect the technique to be applicable to spectroscopy of the few-hole quantum dot states, where even more rich physics can be expected, due to the interplay of Coulomb, exchange, and heavy hole - light hole coupling contributions. Future work will investigate the effect of structural parameters (such as size, shape and chemical composition of the QD) on the multi-electronic energy spectrum and the qubit performance characteristics, such as spin lifetimes, spin coherence and the possibility of the electric spin control.

\begin{acknowledgments}
\textit{Acknowledgements}: P.M-H. and H.E.D. were supported through EPSRC doctoral training grants. G.G., P.M-H., H.E.D., and E.A.C. were supported by an EPSRC grant EP/V048333/1. E.A.C. was supported by a Royal Society University Research
Fellowship, the Leverhulme Trust grant RPG-2023-141, and the QuantERA award MEEDGARD. A.R. acknowledges support of the Austrian Science Fund
(FWF) via the Research Group FG5, I 4320, I 4380, I 3762, the
European Union's Horizon 2020 research and innovation program
under Grant Agreements No. 899814 (Qurope) and No. 871130
(Ascent+), the Linz Institute of Technology (LIT), and the LIT
Secure and Correct Systems Lab, supported by the State of Upper
Austria. P.K. acknowledges funding from the European Innovation Council Pathfinder program under grant agreement No 101185617 (QCEED), projects 20IND05 QADeT, 20FUN05 SEQUME which received funding from the EMPIR programme co-financed by the Participating States and from the European Union’s Horizon 2020 research and innovation programme. P.K. was also supported by the project Quantum materials for applications in sustainable technologies, CZ.02.01.01/00/22\_008/0004572, and partly funded by Institutional Subsidy for Long-Term Conceptual Development of a Research Organization granted to the Czech Metrology Institute by the Ministry of Industry and Trade of the Czech Republic. \textit{Author contributions}: S.M., S.F.C.S and A.R.
developed, grew and processed the quantum dot samples. P.M-H. and
E.A.C. conducted multi-electron spectroscopy experiments. P.K. conducted Configuration Interaction modelling. H.E.D. and G.G. conducted electron spin lifetime and NMR experiments. E.A.C. drafted the manuscript with input from all authors. E.A.C. coordinated the project.
\end{acknowledgments}


\begin{thebibliography}{90}%
\makeatletter
\providecommand \@ifxundefined [1]{%
 \@ifx{#1\undefined}
}%
\providecommand \@ifnum [1]{%
 \ifnum #1\expandafter \@firstoftwo
 \else \expandafter \@secondoftwo
 \fi
}%
\providecommand \@ifx [1]{%
 \ifx #1\expandafter \@firstoftwo
 \else \expandafter \@secondoftwo
 \fi
}%
\providecommand \natexlab [1]{#1}%
\providecommand \enquote  [1]{``#1''}%
\providecommand \bibnamefont  [1]{#1}%
\providecommand \bibfnamefont [1]{#1}%
\providecommand \citenamefont [1]{#1}%
\providecommand \href@noop [0]{\@secondoftwo}%
\providecommand \href [0]{\begingroup \@sanitize@url \@href}%
\providecommand \@href[1]{\@@startlink{#1}\@@href}%
\providecommand \@@href[1]{\endgroup#1\@@endlink}%
\providecommand \@sanitize@url [0]{\catcode `\\12\catcode `\$12\catcode
  `\&12\catcode `\#12\catcode `\^12\catcode `\_12\catcode `\%12\relax}%
\providecommand \@@startlink[1]{}%
\providecommand \@@endlink[0]{}%
\providecommand \url  [0]{\begingroup\@sanitize@url \@url }%
\providecommand \@url [1]{\endgroup\@href {#1}{\urlprefix }}%
\providecommand \urlprefix  [0]{URL }%
\providecommand \Eprint [0]{\href }%
\providecommand \doibase [0]{https://doi.org/}%
\providecommand \selectlanguage [0]{\@gobble}%
\providecommand \bibinfo  [0]{\@secondoftwo}%
\providecommand \bibfield  [0]{\@secondoftwo}%
\providecommand \translation [1]{[#1]}%
\providecommand \BibitemOpen [0]{}%
\providecommand \bibitemStop [0]{}%
\providecommand \bibitemNoStop [0]{.\EOS\space}%
\providecommand \EOS [0]{\spacefactor3000\relax}%
\providecommand \BibitemShut  [1]{\csname bibitem#1\endcsname}%
\let\auto@bib@innerbib\@empty
\bibitem [{\citenamefont {Nakajima}\ \emph {et~al.}(2019)\citenamefont
  {Nakajima}, \citenamefont {Noiri}, \citenamefont {Yoneda}, \citenamefont
  {Delbecq}, \citenamefont {Stano}, \citenamefont {Otsuka}, \citenamefont
  {Takeda}, \citenamefont {Amaha}, \citenamefont {Allison}, \citenamefont
  {Kawasaki}, \citenamefont {Ludwig}, \citenamefont {Wieck}, \citenamefont
  {Loss},\ and\ \citenamefont {Tarucha}}]{Nakajima2019}%
  \BibitemOpen
  \bibfield  {author} {\bibinfo {author} {\bibfnamefont {T.}~\bibnamefont
  {Nakajima}}, \bibinfo {author} {\bibfnamefont {A.}~\bibnamefont {Noiri}},
  \bibinfo {author} {\bibfnamefont {J.}~\bibnamefont {Yoneda}}, \bibinfo
  {author} {\bibfnamefont {M.~R.}\ \bibnamefont {Delbecq}}, \bibinfo {author}
  {\bibfnamefont {P.}~\bibnamefont {Stano}}, \bibinfo {author} {\bibfnamefont
  {T.}~\bibnamefont {Otsuka}}, \bibinfo {author} {\bibfnamefont
  {K.}~\bibnamefont {Takeda}}, \bibinfo {author} {\bibfnamefont
  {S.}~\bibnamefont {Amaha}}, \bibinfo {author} {\bibfnamefont
  {G.}~\bibnamefont {Allison}}, \bibinfo {author} {\bibfnamefont
  {K.}~\bibnamefont {Kawasaki}}, \bibinfo {author} {\bibfnamefont
  {A.}~\bibnamefont {Ludwig}}, \bibinfo {author} {\bibfnamefont {A.~D.}\
  \bibnamefont {Wieck}}, \bibinfo {author} {\bibfnamefont {D.}~\bibnamefont
  {Loss}},\ and\ \bibinfo {author} {\bibfnamefont {S.}~\bibnamefont
  {Tarucha}},\ }\bibfield  {title} {\bibinfo {title} {Quantum non-demolition
  measurement of an electron spin qubit},\ }\href
  {https://doi.org/10.1038/s41565-019-0426-x} {\bibfield  {journal} {\bibinfo
  {journal} {Nat. Nanotechnol.}\ }\textbf {\bibinfo {volume} {14}},\ \bibinfo
  {pages} {555} (\bibinfo {year} {2019})}\BibitemShut {NoStop}%
\bibitem [{\citenamefont {Mortemousque}\ \emph {et~al.}(2021)\citenamefont
  {Mortemousque}, \citenamefont {Chanrion}, \citenamefont {Jadot},
  \citenamefont {Flentje}, \citenamefont {Ludwig}, \citenamefont {Wieck},
  \citenamefont {Urdampilleta}, \citenamefont {B{\"a}uerle},\ and\
  \citenamefont {Meunier}}]{Mortemousque2021}%
  \BibitemOpen
  \bibfield  {author} {\bibinfo {author} {\bibfnamefont {P.-A.}\ \bibnamefont
  {Mortemousque}}, \bibinfo {author} {\bibfnamefont {E.}~\bibnamefont
  {Chanrion}}, \bibinfo {author} {\bibfnamefont {B.}~\bibnamefont {Jadot}},
  \bibinfo {author} {\bibfnamefont {H.}~\bibnamefont {Flentje}}, \bibinfo
  {author} {\bibfnamefont {A.}~\bibnamefont {Ludwig}}, \bibinfo {author}
  {\bibfnamefont {A.~D.}\ \bibnamefont {Wieck}}, \bibinfo {author}
  {\bibfnamefont {M.}~\bibnamefont {Urdampilleta}}, \bibinfo {author}
  {\bibfnamefont {C.}~\bibnamefont {B{\"a}uerle}},\ and\ \bibinfo {author}
  {\bibfnamefont {T.}~\bibnamefont {Meunier}},\ }\bibfield  {title} {\bibinfo
  {title} {Coherent control of individual electron spins in a two-dimensional
  quantum dot array},\ }\href {https://doi.org/10.1038/s41565-020-00816-w}
  {\bibfield  {journal} {\bibinfo  {journal} {Nat. Nanotechnol.}\ }\textbf
  {\bibinfo {volume} {16}},\ \bibinfo {pages} {296} (\bibinfo {year}
  {2021})}\BibitemShut {NoStop}%
\bibitem [{\citenamefont {Zaporski}\ \emph {et~al.}(2023)\citenamefont
  {Zaporski}, \citenamefont {Shofer}, \citenamefont {Bodey}, \citenamefont
  {Manna}, \citenamefont {Gillard}, \citenamefont {Appel}, \citenamefont
  {Schimpf}, \citenamefont {Covre~da Silva}, \citenamefont {Jarman},
  \citenamefont {Delamare},\ and\ \citenamefont {et~al.}}]{Zaporski2022}%
  \BibitemOpen
  \bibfield  {author} {\bibinfo {author} {\bibfnamefont {L.}~\bibnamefont
  {Zaporski}}, \bibinfo {author} {\bibfnamefont {N.}~\bibnamefont {Shofer}},
  \bibinfo {author} {\bibfnamefont {J.~H.}\ \bibnamefont {Bodey}}, \bibinfo
  {author} {\bibfnamefont {S.}~\bibnamefont {Manna}}, \bibinfo {author}
  {\bibfnamefont {G.}~\bibnamefont {Gillard}}, \bibinfo {author} {\bibfnamefont
  {M.~H.}\ \bibnamefont {Appel}}, \bibinfo {author} {\bibfnamefont
  {C.}~\bibnamefont {Schimpf}}, \bibinfo {author} {\bibfnamefont {S.~F.}\
  \bibnamefont {Covre~da Silva}}, \bibinfo {author} {\bibfnamefont
  {J.}~\bibnamefont {Jarman}}, \bibinfo {author} {\bibfnamefont
  {G.}~\bibnamefont {Delamare}},\ and\ \bibinfo {author} {\bibnamefont
  {et~al.}},\ }\bibfield  {title} {\bibinfo {title} {Ideal refocusing of an
  optically active spin qubit under strong hyperfine interactions},\ }\href
  {https://doi.org/10.1038/s41565-022-01282-2} {\bibfield  {journal} {\bibinfo
  {journal} {Nat. Nanotechnol.}\ }\textbf {\bibinfo {volume} {18}},\ \bibinfo
  {pages} {257} (\bibinfo {year} {2023})}\BibitemShut {NoStop}%
\bibitem [{\citenamefont {Kouwenhoven}\ \emph {et~al.}(2001)\citenamefont
  {Kouwenhoven}, \citenamefont {Austing},\ and\ \citenamefont
  {Tarucha}}]{Kouwenhoven2001}%
  \BibitemOpen
  \bibfield  {author} {\bibinfo {author} {\bibfnamefont {L.~P.}\ \bibnamefont
  {Kouwenhoven}}, \bibinfo {author} {\bibfnamefont {D.~G.}\ \bibnamefont
  {Austing}},\ and\ \bibinfo {author} {\bibfnamefont {S.}~\bibnamefont
  {Tarucha}},\ }\bibfield  {title} {\bibinfo {title} {Few-electron quantum
  dots},\ }\href {https://doi.org/10.1088/0034-4885/64/6/201} {\bibfield
  {journal} {\bibinfo  {journal} {Rep. Prog. Phys.}\ }\textbf {\bibinfo
  {volume} {64}},\ \bibinfo {pages} {701} (\bibinfo {year} {2001})}\BibitemShut
  {NoStop}%
\bibitem [{\citenamefont {Reimann}\ and\ \citenamefont
  {Manninen}(2002)}]{Reimann2002}%
  \BibitemOpen
  \bibfield  {author} {\bibinfo {author} {\bibfnamefont {S.~M.}\ \bibnamefont
  {Reimann}}\ and\ \bibinfo {author} {\bibfnamefont {M.}~\bibnamefont
  {Manninen}},\ }\bibfield  {title} {\bibinfo {title} {Electronic structure of
  quantum dots},\ }\href {https://doi.org/10.1103/RevModPhys.74.1283}
  {\bibfield  {journal} {\bibinfo  {journal} {Rev. Mod. Phys.}\ }\textbf
  {\bibinfo {volume} {74}},\ \bibinfo {pages} {1283} (\bibinfo {year}
  {2002})}\BibitemShut {NoStop}%
\bibitem [{\citenamefont {Hanson}\ \emph {et~al.}(2007)\citenamefont {Hanson},
  \citenamefont {Kouwenhoven}, \citenamefont {Petta}, \citenamefont {Tarucha},\
  and\ \citenamefont {Vandersypen}}]{Hanson2007}%
  \BibitemOpen
  \bibfield  {author} {\bibinfo {author} {\bibfnamefont {R.}~\bibnamefont
  {Hanson}}, \bibinfo {author} {\bibfnamefont {L.~P.}\ \bibnamefont
  {Kouwenhoven}}, \bibinfo {author} {\bibfnamefont {J.~R.}\ \bibnamefont
  {Petta}}, \bibinfo {author} {\bibfnamefont {S.}~\bibnamefont {Tarucha}},\
  and\ \bibinfo {author} {\bibfnamefont {L.~M.~K.}\ \bibnamefont
  {Vandersypen}},\ }\bibfield  {title} {\bibinfo {title} {Spins in few-electron
  quantum dots},\ }\href {https://doi.org/10.1103/RevModPhys.79.1217}
  {\bibfield  {journal} {\bibinfo  {journal} {Rev. Mod. Phys.}\ }\textbf
  {\bibinfo {volume} {79}},\ \bibinfo {pages} {1217} (\bibinfo {year}
  {2007})}\BibitemShut {NoStop}%
\bibitem [{\citenamefont {Tarucha}\ \emph {et~al.}(1996)\citenamefont
  {Tarucha}, \citenamefont {Austing}, \citenamefont {Honda}, \citenamefont
  {van~der Hage},\ and\ \citenamefont {Kouwenhoven}}]{Tarucha1996}%
  \BibitemOpen
  \bibfield  {author} {\bibinfo {author} {\bibfnamefont {S.}~\bibnamefont
  {Tarucha}}, \bibinfo {author} {\bibfnamefont {D.~G.}\ \bibnamefont
  {Austing}}, \bibinfo {author} {\bibfnamefont {T.}~\bibnamefont {Honda}},
  \bibinfo {author} {\bibfnamefont {R.~J.}\ \bibnamefont {van~der Hage}},\ and\
  \bibinfo {author} {\bibfnamefont {L.~P.}\ \bibnamefont {Kouwenhoven}},\
  }\bibfield  {title} {\bibinfo {title} {Shell filling and spin effects in a
  few electron quantum dot},\ }\href
  {https://doi.org/10.1103/PhysRevLett.77.3613} {\bibfield  {journal} {\bibinfo
   {journal} {Phys. Rev. Lett.}\ }\textbf {\bibinfo {volume} {77}},\ \bibinfo
  {pages} {3613} (\bibinfo {year} {1996})}\BibitemShut {NoStop}%
\bibitem [{\citenamefont {Kouwenhoven}\ \emph {et~al.}(1997)\citenamefont
  {Kouwenhoven}, \citenamefont {Oosterkamp}, \citenamefont {Danoesastro},
  \citenamefont {Eto}, \citenamefont {Austing}, \citenamefont {Honda},\ and\
  \citenamefont {Tarucha}}]{Kouwenhoven1997Science}%
  \BibitemOpen
  \bibfield  {author} {\bibinfo {author} {\bibfnamefont {L.~P.}\ \bibnamefont
  {Kouwenhoven}}, \bibinfo {author} {\bibfnamefont {T.~H.}\ \bibnamefont
  {Oosterkamp}}, \bibinfo {author} {\bibfnamefont {M.~W.~S.}\ \bibnamefont
  {Danoesastro}}, \bibinfo {author} {\bibfnamefont {M.}~\bibnamefont {Eto}},
  \bibinfo {author} {\bibfnamefont {D.~G.}\ \bibnamefont {Austing}}, \bibinfo
  {author} {\bibfnamefont {T.}~\bibnamefont {Honda}},\ and\ \bibinfo {author}
  {\bibfnamefont {S.}~\bibnamefont {Tarucha}},\ }\bibfield  {title} {\bibinfo
  {title} {Excitation spectra of circular, few-electron quantum dots},\ }\href
  {https://doi.org/10.1126/science.278.5344.1788} {\bibfield  {journal}
  {\bibinfo  {journal} {Science}\ }\textbf {\bibinfo {volume} {278}},\ \bibinfo
  {pages} {1788} (\bibinfo {year} {1997})}\BibitemShut {NoStop}%
\bibitem [{\citenamefont {Ota}\ \emph {et~al.}(2004)\citenamefont {Ota},
  \citenamefont {Ono}, \citenamefont {Stopa}, \citenamefont {Hatano},
  \citenamefont {Tarucha}, \citenamefont {Song}, \citenamefont {Nakata},
  \citenamefont {Miyazawa}, \citenamefont {Ohshima},\ and\ \citenamefont
  {Yokoyama}}]{Ota2004}%
  \BibitemOpen
  \bibfield  {author} {\bibinfo {author} {\bibfnamefont {T.}~\bibnamefont
  {Ota}}, \bibinfo {author} {\bibfnamefont {K.}~\bibnamefont {Ono}}, \bibinfo
  {author} {\bibfnamefont {M.}~\bibnamefont {Stopa}}, \bibinfo {author}
  {\bibfnamefont {T.}~\bibnamefont {Hatano}}, \bibinfo {author} {\bibfnamefont
  {S.}~\bibnamefont {Tarucha}}, \bibinfo {author} {\bibfnamefont {H.~Z.}\
  \bibnamefont {Song}}, \bibinfo {author} {\bibfnamefont {Y.}~\bibnamefont
  {Nakata}}, \bibinfo {author} {\bibfnamefont {T.}~\bibnamefont {Miyazawa}},
  \bibinfo {author} {\bibfnamefont {T.}~\bibnamefont {Ohshima}},\ and\ \bibinfo
  {author} {\bibfnamefont {N.}~\bibnamefont {Yokoyama}},\ }\bibfield  {title}
  {\bibinfo {title} {Single-dot spectroscopy via elastic single-electron
  tunneling through a pair of coupled quantum dots},\ }\href
  {https://doi.org/10.1103/PhysRevLett.93.066801} {\bibfield  {journal}
  {\bibinfo  {journal} {Phys. Rev. Lett.}\ }\textbf {\bibinfo {volume} {93}},\
  \bibinfo {pages} {066801} (\bibinfo {year} {2004})}\BibitemShut {NoStop}%
\bibitem [{\citenamefont {Jung}\ \emph {et~al.}(2005)\citenamefont {Jung},
  \citenamefont {Machida}, \citenamefont {Hirakawa}, \citenamefont {Komiyama},
  \citenamefont {Nakaoka}, \citenamefont {Ishida},\ and\ \citenamefont
  {Arakawa}}]{Jung2005}%
  \BibitemOpen
  \bibfield  {author} {\bibinfo {author} {\bibfnamefont {M.}~\bibnamefont
  {Jung}}, \bibinfo {author} {\bibfnamefont {T.}~\bibnamefont {Machida}},
  \bibinfo {author} {\bibfnamefont {K.}~\bibnamefont {Hirakawa}}, \bibinfo
  {author} {\bibfnamefont {S.}~\bibnamefont {Komiyama}}, \bibinfo {author}
  {\bibfnamefont {T.}~\bibnamefont {Nakaoka}}, \bibinfo {author} {\bibfnamefont
  {S.}~\bibnamefont {Ishida}},\ and\ \bibinfo {author} {\bibfnamefont
  {Y.}~\bibnamefont {Arakawa}},\ }\bibfield  {title} {\bibinfo {title} {Shell
  structures in self-assembled {{InAs}} quantum dots probed by lateral electron
  tunneling structures},\ }\href {https://doi.org/10.1063/1.2131177} {\bibfield
   {journal} {\bibinfo  {journal} {Appl. Phys. Lett.}\ }\textbf {\bibinfo
  {volume} {87}},\ \bibinfo {pages} {203109} (\bibinfo {year}
  {2005})}\BibitemShut {NoStop}%
\bibitem [{\citenamefont {Amaha}\ \emph {et~al.}(2008)\citenamefont {Amaha},
  \citenamefont {Hatano}, \citenamefont {Teraoka}, \citenamefont {Shibatomi},
  \citenamefont {Tarucha}, \citenamefont {Nakata}, \citenamefont {Miyazawa},
  \citenamefont {Oshima}, \citenamefont {Usuki},\ and\ \citenamefont
  {Yokoyama}}]{Amaha2008}%
  \BibitemOpen
  \bibfield  {author} {\bibinfo {author} {\bibfnamefont {S.}~\bibnamefont
  {Amaha}}, \bibinfo {author} {\bibfnamefont {T.}~\bibnamefont {Hatano}},
  \bibinfo {author} {\bibfnamefont {S.}~\bibnamefont {Teraoka}}, \bibinfo
  {author} {\bibfnamefont {A.}~\bibnamefont {Shibatomi}}, \bibinfo {author}
  {\bibfnamefont {S.}~\bibnamefont {Tarucha}}, \bibinfo {author} {\bibfnamefont
  {Y.}~\bibnamefont {Nakata}}, \bibinfo {author} {\bibfnamefont
  {T.}~\bibnamefont {Miyazawa}}, \bibinfo {author} {\bibfnamefont
  {T.}~\bibnamefont {Oshima}}, \bibinfo {author} {\bibfnamefont
  {T.}~\bibnamefont {Usuki}},\ and\ \bibinfo {author} {\bibfnamefont
  {N.}~\bibnamefont {Yokoyama}},\ }\bibfield  {title} {\bibinfo {title}
  {Laterally coupled self-assembled inas quantum dots embedded in resonant
  tunnel diode with multigate electrodes},\ }\href
  {https://doi.org/10.1063/1.2920205} {\bibfield  {journal} {\bibinfo
  {journal} {Appl. Phys. Lett.}\ }\textbf {\bibinfo {volume} {92}},\ \bibinfo
  {pages} {202109} (\bibinfo {year} {2008})}\BibitemShut {NoStop}%
\bibitem [{\citenamefont {Kanai}\ \emph {et~al.}(2011)\citenamefont {Kanai},
  \citenamefont {Deacon}, \citenamefont {Takahashi}, \citenamefont {Oiwa},
  \citenamefont {Yoshida}, \citenamefont {Shibata}, \citenamefont {Hirakawa},
  \citenamefont {Tokura},\ and\ \citenamefont {Tarucha}}]{Kanai2011}%
  \BibitemOpen
  \bibfield  {author} {\bibinfo {author} {\bibfnamefont {Y.}~\bibnamefont
  {Kanai}}, \bibinfo {author} {\bibfnamefont {R.~S.}\ \bibnamefont {Deacon}},
  \bibinfo {author} {\bibfnamefont {S.}~\bibnamefont {Takahashi}}, \bibinfo
  {author} {\bibfnamefont {A.}~\bibnamefont {Oiwa}}, \bibinfo {author}
  {\bibfnamefont {K.}~\bibnamefont {Yoshida}}, \bibinfo {author} {\bibfnamefont
  {K.}~\bibnamefont {Shibata}}, \bibinfo {author} {\bibfnamefont
  {K.}~\bibnamefont {Hirakawa}}, \bibinfo {author} {\bibfnamefont
  {Y.}~\bibnamefont {Tokura}},\ and\ \bibinfo {author} {\bibfnamefont
  {S.}~\bibnamefont {Tarucha}},\ }\bibfield  {title} {\bibinfo {title}
  {Electrically tuned spin--orbit interaction in an {{InAs}} self-assembled
  quantum dot},\ }\href {https://doi.org/10.1038/nnano.2011.103} {\bibfield
  {journal} {\bibinfo  {journal} {Nat. Nanotechnol.}\ }\textbf {\bibinfo
  {volume} {6}},\ \bibinfo {pages} {511} (\bibinfo {year} {2011})}\BibitemShut
  {NoStop}%
\bibitem [{\citenamefont {Dalgarno}\ \emph {et~al.}(2008)\citenamefont
  {Dalgarno}, \citenamefont {Smith}, \citenamefont {McFarlane}, \citenamefont
  {Gerardot}, \citenamefont {Karrai}, \citenamefont {Badolato}, \citenamefont
  {Petroff},\ and\ \citenamefont {Warburton}}]{Dalgarno2008}%
  \BibitemOpen
  \bibfield  {author} {\bibinfo {author} {\bibfnamefont {P.~A.}\ \bibnamefont
  {Dalgarno}}, \bibinfo {author} {\bibfnamefont {J.~M.}\ \bibnamefont {Smith}},
  \bibinfo {author} {\bibfnamefont {J.}~\bibnamefont {McFarlane}}, \bibinfo
  {author} {\bibfnamefont {B.~D.}\ \bibnamefont {Gerardot}}, \bibinfo {author}
  {\bibfnamefont {K.}~\bibnamefont {Karrai}}, \bibinfo {author} {\bibfnamefont
  {A.}~\bibnamefont {Badolato}}, \bibinfo {author} {\bibfnamefont {P.~M.}\
  \bibnamefont {Petroff}},\ and\ \bibinfo {author} {\bibfnamefont {R.~J.}\
  \bibnamefont {Warburton}},\ }\bibfield  {title} {\bibinfo {title} {Coulomb
  interactions in single charged self-assembled quantum dots: Radiative
  lifetime and recombination energy},\ }\href
  {https://doi.org/10.1103/PhysRevB.77.245311} {\bibfield  {journal} {\bibinfo
  {journal} {Phys. Rev. B}\ }\textbf {\bibinfo {volume} {77}},\ \bibinfo
  {pages} {245311} (\bibinfo {year} {2008})}\BibitemShut {NoStop}%
\bibitem [{\citenamefont {Mlinar}\ \emph {et~al.}(2009)\citenamefont {Mlinar},
  \citenamefont {Bozkurt}, \citenamefont {Ulloa}, \citenamefont {Ediger},
  \citenamefont {Bester}, \citenamefont {Badolato}, \citenamefont {Koenraad},
  \citenamefont {Warburton},\ and\ \citenamefont {Zunger}}]{Mlinar2009}%
  \BibitemOpen
  \bibfield  {author} {\bibinfo {author} {\bibfnamefont {V.}~\bibnamefont
  {Mlinar}}, \bibinfo {author} {\bibfnamefont {M.}~\bibnamefont {Bozkurt}},
  \bibinfo {author} {\bibfnamefont {J.~M.}\ \bibnamefont {Ulloa}}, \bibinfo
  {author} {\bibfnamefont {M.}~\bibnamefont {Ediger}}, \bibinfo {author}
  {\bibfnamefont {G.}~\bibnamefont {Bester}}, \bibinfo {author} {\bibfnamefont
  {A.}~\bibnamefont {Badolato}}, \bibinfo {author} {\bibfnamefont {P.~M.}\
  \bibnamefont {Koenraad}}, \bibinfo {author} {\bibfnamefont {R.~J.}\
  \bibnamefont {Warburton}},\ and\ \bibinfo {author} {\bibfnamefont
  {A.}~\bibnamefont {Zunger}},\ }\bibfield  {title} {\bibinfo {title}
  {Structure of quantum dots as seen by excitonic spectroscopy versus
  structural characterization: Using theory to close the loop},\ }\href
  {https://doi.org/10.1103/PhysRevB.80.165425} {\bibfield  {journal} {\bibinfo
  {journal} {Phys. Rev. B}\ }\textbf {\bibinfo {volume} {80}},\ \bibinfo
  {pages} {165425} (\bibinfo {year} {2009})}\BibitemShut {NoStop}%
\bibitem [{\citenamefont {Holtkemper}\ \emph {et~al.}(2018)\citenamefont
  {Holtkemper}, \citenamefont {Reiter},\ and\ \citenamefont
  {Kuhn}}]{Holtkemper2018}%
  \BibitemOpen
  \bibfield  {author} {\bibinfo {author} {\bibfnamefont {M.}~\bibnamefont
  {Holtkemper}}, \bibinfo {author} {\bibfnamefont {D.~E.}\ \bibnamefont
  {Reiter}},\ and\ \bibinfo {author} {\bibfnamefont {T.}~\bibnamefont {Kuhn}},\
  }\bibfield  {title} {\bibinfo {title} {Influence of the quantum dot geometry
  on $p$-shell transitions in differently charged quantum dots},\ }\href
  {https://doi.org/10.1103/PhysRevB.97.075308} {\bibfield  {journal} {\bibinfo
  {journal} {Phys. Rev. B}\ }\textbf {\bibinfo {volume} {97}},\ \bibinfo
  {pages} {075308} (\bibinfo {year} {2018})}\BibitemShut {NoStop}%
\bibitem [{\citenamefont {Huber}\ \emph {et~al.}(2019)\citenamefont {Huber},
  \citenamefont {Lehner}, \citenamefont {Csontosov\'a}, \citenamefont {Reindl},
  \citenamefont {Schuler}, \citenamefont {Covre~da Silva}, \citenamefont
  {Klenovsk\'y},\ and\ \citenamefont {Rastelli}}]{Huber2019}%
  \BibitemOpen
  \bibfield  {author} {\bibinfo {author} {\bibfnamefont {D.}~\bibnamefont
  {Huber}}, \bibinfo {author} {\bibfnamefont {B.~U.}\ \bibnamefont {Lehner}},
  \bibinfo {author} {\bibfnamefont {D.}~\bibnamefont {Csontosov\'a}}, \bibinfo
  {author} {\bibfnamefont {M.}~\bibnamefont {Reindl}}, \bibinfo {author}
  {\bibfnamefont {S.}~\bibnamefont {Schuler}}, \bibinfo {author} {\bibfnamefont
  {S.~F.}\ \bibnamefont {Covre~da Silva}}, \bibinfo {author} {\bibfnamefont
  {P.}~\bibnamefont {Klenovsk\'y}},\ and\ \bibinfo {author} {\bibfnamefont
  {A.}~\bibnamefont {Rastelli}},\ }\bibfield  {title} {\bibinfo {title}
  {Single-particle-picture breakdown in laterally weakly confining {{GaAs}}
  quantum dots},\ }\href {https://doi.org/10.1103/PhysRevB.100.235425}
  {\bibfield  {journal} {\bibinfo  {journal} {Phys. Rev. B}\ }\textbf {\bibinfo
  {volume} {100}},\ \bibinfo {pages} {235425} (\bibinfo {year}
  {2019})}\BibitemShut {NoStop}%
\bibitem [{\citenamefont {Nick~Vamivakas}\ \emph {et~al.}(2009)\citenamefont
  {Nick~Vamivakas}, \citenamefont {Zhao}, \citenamefont {Lu},\ and\
  \citenamefont {Atat{\"u}re}}]{Vamivakas2009}%
  \BibitemOpen
  \bibfield  {author} {\bibinfo {author} {\bibfnamefont {A.}~\bibnamefont
  {Nick~Vamivakas}}, \bibinfo {author} {\bibfnamefont {Y.}~\bibnamefont
  {Zhao}}, \bibinfo {author} {\bibfnamefont {C.-Y.}\ \bibnamefont {Lu}},\ and\
  \bibinfo {author} {\bibfnamefont {M.}~\bibnamefont {Atat{\"u}re}},\
  }\bibfield  {title} {\bibinfo {title} {Spin-resolved quantum-dot resonance
  fluorescence},\ }\href {https://doi.org/10.1038/nphys1182} {\bibfield
  {journal} {\bibinfo  {journal} {Nat. Phys.}\ }\textbf {\bibinfo {volume}
  {5}},\ \bibinfo {pages} {198} (\bibinfo {year} {2009})}\BibitemShut {NoStop}%
\bibitem [{\citenamefont {Nawrath}\ \emph {et~al.}(2021)\citenamefont
  {Nawrath}, \citenamefont {Vural}, \citenamefont {Fischer}, \citenamefont
  {Schaber}, \citenamefont {Portalupi}, \citenamefont {Jetter},\ and\
  \citenamefont {Michler}}]{Nawrath2021}%
  \BibitemOpen
  \bibfield  {author} {\bibinfo {author} {\bibfnamefont {C.}~\bibnamefont
  {Nawrath}}, \bibinfo {author} {\bibfnamefont {H.}~\bibnamefont {Vural}},
  \bibinfo {author} {\bibfnamefont {J.}~\bibnamefont {Fischer}}, \bibinfo
  {author} {\bibfnamefont {R.}~\bibnamefont {Schaber}}, \bibinfo {author}
  {\bibfnamefont {S.~L.}\ \bibnamefont {Portalupi}}, \bibinfo {author}
  {\bibfnamefont {M.}~\bibnamefont {Jetter}},\ and\ \bibinfo {author}
  {\bibfnamefont {P.}~\bibnamefont {Michler}},\ }\bibfield  {title} {\bibinfo
  {title} {Resonance fluorescence of single {{In(Ga)As}} quantum dots emitting
  in the telecom {{C-band}}},\ }\href {https://doi.org/10.1063/5.0048695}
  {\bibfield  {journal} {\bibinfo  {journal} {Appl. Phys. Lett.}\ }\textbf
  {\bibinfo {volume} {118}},\ \bibinfo {pages} {244002} (\bibinfo {year}
  {2021})}\BibitemShut {NoStop}%
\bibitem [{\citenamefont {Hawrylak}\ \emph {et~al.}(2000)\citenamefont
  {Hawrylak}, \citenamefont {Narvaez}, \citenamefont {Bayer},\ and\
  \citenamefont {Forchel}}]{Hawrylak2000}%
  \BibitemOpen
  \bibfield  {author} {\bibinfo {author} {\bibfnamefont {P.}~\bibnamefont
  {Hawrylak}}, \bibinfo {author} {\bibfnamefont {G.~A.}\ \bibnamefont
  {Narvaez}}, \bibinfo {author} {\bibfnamefont {M.}~\bibnamefont {Bayer}},\
  and\ \bibinfo {author} {\bibfnamefont {A.}~\bibnamefont {Forchel}},\
  }\bibfield  {title} {\bibinfo {title} {Excitonic absorption in a quantum
  dot},\ }\href {https://doi.org/10.1103/PhysRevLett.85.389} {\bibfield
  {journal} {\bibinfo  {journal} {Phys. Rev. Lett.}\ }\textbf {\bibinfo
  {volume} {85}},\ \bibinfo {pages} {389} (\bibinfo {year} {2000})}\BibitemShut
  {NoStop}%
\bibitem [{\citenamefont {Ware}\ \emph {et~al.}(2005)\citenamefont {Ware},
  \citenamefont {Stinaff}, \citenamefont {Gammon}, \citenamefont {Doty},
  \citenamefont {Bracker}, \citenamefont {Gershoni}, \citenamefont {Korenev},
  \citenamefont {B\ifmmode~\u{a}\else \u{a}\fi{}descu}, \citenamefont
  {Lyanda-Geller},\ and\ \citenamefont {Reinecke}}]{Ware2005}%
  \BibitemOpen
  \bibfield  {author} {\bibinfo {author} {\bibfnamefont {M.~E.}\ \bibnamefont
  {Ware}}, \bibinfo {author} {\bibfnamefont {E.~A.}\ \bibnamefont {Stinaff}},
  \bibinfo {author} {\bibfnamefont {D.}~\bibnamefont {Gammon}}, \bibinfo
  {author} {\bibfnamefont {M.~F.}\ \bibnamefont {Doty}}, \bibinfo {author}
  {\bibfnamefont {A.~S.}\ \bibnamefont {Bracker}}, \bibinfo {author}
  {\bibfnamefont {D.}~\bibnamefont {Gershoni}}, \bibinfo {author}
  {\bibfnamefont {V.~L.}\ \bibnamefont {Korenev}}, \bibinfo {author}
  {\bibfnamefont {i.~m. c.~C.}\ \bibnamefont {B\ifmmode~\u{a}\else
  \u{a}\fi{}descu}}, \bibinfo {author} {\bibfnamefont {Y.}~\bibnamefont
  {Lyanda-Geller}},\ and\ \bibinfo {author} {\bibfnamefont {T.~L.}\
  \bibnamefont {Reinecke}},\ }\bibfield  {title} {\bibinfo {title} {Polarized
  fine structure in the photoluminescence excitation spectrum of a negatively
  charged quantum dot},\ }\href {https://doi.org/10.1103/PhysRevLett.95.177403}
  {\bibfield  {journal} {\bibinfo  {journal} {Phys. Rev. Lett.}\ }\textbf
  {\bibinfo {volume} {95}},\ \bibinfo {pages} {177403} (\bibinfo {year}
  {2005})}\BibitemShut {NoStop}%
\bibitem [{\citenamefont {L{\"o}bl}\ \emph {et~al.}(2020)\citenamefont
  {L{\"o}bl}, \citenamefont {Spinnler}, \citenamefont {Javadi}, \citenamefont
  {Zhai}, \citenamefont {Nguyen}, \citenamefont {Ritzmann}, \citenamefont
  {Midolo}, \citenamefont {Lodahl}, \citenamefont {Wieck}, \citenamefont
  {Ludwig},\ and\ \citenamefont {Warburton}}]{Lobl2020}%
  \BibitemOpen
  \bibfield  {author} {\bibinfo {author} {\bibfnamefont {M.~C.}\ \bibnamefont
  {L{\"o}bl}}, \bibinfo {author} {\bibfnamefont {C.}~\bibnamefont {Spinnler}},
  \bibinfo {author} {\bibfnamefont {A.}~\bibnamefont {Javadi}}, \bibinfo
  {author} {\bibfnamefont {L.}~\bibnamefont {Zhai}}, \bibinfo {author}
  {\bibfnamefont {G.~N.}\ \bibnamefont {Nguyen}}, \bibinfo {author}
  {\bibfnamefont {J.}~\bibnamefont {Ritzmann}}, \bibinfo {author}
  {\bibfnamefont {L.}~\bibnamefont {Midolo}}, \bibinfo {author} {\bibfnamefont
  {P.}~\bibnamefont {Lodahl}}, \bibinfo {author} {\bibfnamefont {A.~D.}\
  \bibnamefont {Wieck}}, \bibinfo {author} {\bibfnamefont {A.}~\bibnamefont
  {Ludwig}},\ and\ \bibinfo {author} {\bibfnamefont {R.~J.}\ \bibnamefont
  {Warburton}},\ }\bibfield  {title} {\bibinfo {title} {Radiative {{Auger}}
  process in the single-photon limit},\ }\href
  {https://doi.org/10.1038/s41565-020-0697-2} {\bibfield  {journal} {\bibinfo
  {journal} {Nat. Nanotechnol.}\ }\textbf {\bibinfo {volume} {15}},\ \bibinfo
  {pages} {558} (\bibinfo {year} {2020})}\BibitemShut {NoStop}%
\bibitem [{\citenamefont {Drexler}\ \emph {et~al.}(1994)\citenamefont
  {Drexler}, \citenamefont {Leonard}, \citenamefont {Hansen}, \citenamefont
  {Kotthaus},\ and\ \citenamefont {Petroff}}]{Drexler1994}%
  \BibitemOpen
  \bibfield  {author} {\bibinfo {author} {\bibfnamefont {H.}~\bibnamefont
  {Drexler}}, \bibinfo {author} {\bibfnamefont {D.}~\bibnamefont {Leonard}},
  \bibinfo {author} {\bibfnamefont {W.}~\bibnamefont {Hansen}}, \bibinfo
  {author} {\bibfnamefont {J.~P.}\ \bibnamefont {Kotthaus}},\ and\ \bibinfo
  {author} {\bibfnamefont {P.~M.}\ \bibnamefont {Petroff}},\ }\bibfield
  {title} {\bibinfo {title} {Spectroscopy of quantum levels in charge-tunable
  {{InGaAs}} quantum dots},\ }\href
  {https://doi.org/10.1103/PhysRevLett.73.2252} {\bibfield  {journal} {\bibinfo
   {journal} {Phys. Rev. Lett.}\ }\textbf {\bibinfo {volume} {73}},\ \bibinfo
  {pages} {2252} (\bibinfo {year} {1994})}\BibitemShut {NoStop}%
\bibitem [{\citenamefont {Miller}\ \emph {et~al.}(1997)\citenamefont {Miller},
  \citenamefont {Hansen}, \citenamefont {Manus}, \citenamefont {Luyken},
  \citenamefont {Lorke}, \citenamefont {Kotthaus}, \citenamefont {Huant},
  \citenamefont {Medeiros-Ribeiro},\ and\ \citenamefont
  {Petroff}}]{Miller1997}%
  \BibitemOpen
  \bibfield  {author} {\bibinfo {author} {\bibfnamefont {B.~T.}\ \bibnamefont
  {Miller}}, \bibinfo {author} {\bibfnamefont {W.}~\bibnamefont {Hansen}},
  \bibinfo {author} {\bibfnamefont {S.}~\bibnamefont {Manus}}, \bibinfo
  {author} {\bibfnamefont {R.~J.}\ \bibnamefont {Luyken}}, \bibinfo {author}
  {\bibfnamefont {A.}~\bibnamefont {Lorke}}, \bibinfo {author} {\bibfnamefont
  {J.~P.}\ \bibnamefont {Kotthaus}}, \bibinfo {author} {\bibfnamefont
  {S.}~\bibnamefont {Huant}}, \bibinfo {author} {\bibfnamefont
  {G.}~\bibnamefont {Medeiros-Ribeiro}},\ and\ \bibinfo {author} {\bibfnamefont
  {P.~M.}\ \bibnamefont {Petroff}},\ }\bibfield  {title} {\bibinfo {title}
  {Few-electron ground states of charge-tunable self-assembled quantum dots},\
  }\href {https://doi.org/10.1103/PhysRevB.56.6764} {\bibfield  {journal}
  {\bibinfo  {journal} {Phys. Rev. B}\ }\textbf {\bibinfo {volume} {56}},\
  \bibinfo {pages} {6764} (\bibinfo {year} {1997})}\BibitemShut {NoStop}%
\bibitem [{\citenamefont {Wetzler}\ \emph {et~al.}(2000)\citenamefont
  {Wetzler}, \citenamefont {Wacker}, \citenamefont {Sch\"{o}ll}, \citenamefont
  {Kapteyn}, \citenamefont {Heitz},\ and\ \citenamefont
  {Bimberg}}]{Wetzler2000}%
  \BibitemOpen
  \bibfield  {author} {\bibinfo {author} {\bibfnamefont {R.}~\bibnamefont
  {Wetzler}}, \bibinfo {author} {\bibfnamefont {A.}~\bibnamefont {Wacker}},
  \bibinfo {author} {\bibfnamefont {E.}~\bibnamefont {Sch\"{o}ll}}, \bibinfo
  {author} {\bibfnamefont {C.~M.~A.}\ \bibnamefont {Kapteyn}}, \bibinfo
  {author} {\bibfnamefont {R.}~\bibnamefont {Heitz}},\ and\ \bibinfo {author}
  {\bibfnamefont {D.}~\bibnamefont {Bimberg}},\ }\bibfield  {title} {\bibinfo
  {title} {Capacitance-voltage characteristics of {{InAs/GaAs}} quantum dots
  embedded in a pn structure},\ }\href {https://doi.org/10.1063/1.1290137}
  {\bibfield  {journal} {\bibinfo  {journal} {Appl. Phys. Lett.}\ }\textbf
  {\bibinfo {volume} {77}},\ \bibinfo {pages} {1671} (\bibinfo {year}
  {2000})}\BibitemShut {NoStop}%
\bibitem [{\citenamefont {Reuter}\ \emph {et~al.}(2008)\citenamefont {Reuter},
  \citenamefont {Roescu}, \citenamefont {Mehta}, \citenamefont {Richter},\ and\
  \citenamefont {Wieck}}]{REUTER2008}%
  \BibitemOpen
  \bibfield  {author} {\bibinfo {author} {\bibfnamefont {D.}~\bibnamefont
  {Reuter}}, \bibinfo {author} {\bibfnamefont {R.}~\bibnamefont {Roescu}},
  \bibinfo {author} {\bibfnamefont {M.}~\bibnamefont {Mehta}}, \bibinfo
  {author} {\bibfnamefont {M.}~\bibnamefont {Richter}},\ and\ \bibinfo {author}
  {\bibfnamefont {A.}~\bibnamefont {Wieck}},\ }\bibfield  {title} {\bibinfo
  {title} {Capacitance-voltage spectroscopy of post-growth annealed {{InAs}}
  quantum dots},\ }\href
  {https://doi.org/https://doi.org/10.1016/j.physe.2007.09.061} {\bibfield
  {journal} {\bibinfo  {journal} {Physica E: Low-dimens. Syst. Nanostruct.}\
  }\textbf {\bibinfo {volume} {40}},\ \bibinfo {pages} {1961} (\bibinfo {year}
  {2008})}\BibitemShut {NoStop}%
\bibitem [{\citenamefont {Tycko}\ \emph {et~al.}(1991)\citenamefont {Tycko},
  \citenamefont {Dabbagh}, \citenamefont {Rosseinsky}, \citenamefont {Murphy},
  \citenamefont {Fleming}, \citenamefont {Ramirez},\ and\ \citenamefont
  {Tully}}]{Tycko1991}%
  \BibitemOpen
  \bibfield  {author} {\bibinfo {author} {\bibfnamefont {R.}~\bibnamefont
  {Tycko}}, \bibinfo {author} {\bibfnamefont {G.}~\bibnamefont {Dabbagh}},
  \bibinfo {author} {\bibfnamefont {M.~J.}\ \bibnamefont {Rosseinsky}},
  \bibinfo {author} {\bibfnamefont {D.~W.}\ \bibnamefont {Murphy}}, \bibinfo
  {author} {\bibfnamefont {R.~M.}\ \bibnamefont {Fleming}}, \bibinfo {author}
  {\bibfnamefont {A.~P.}\ \bibnamefont {Ramirez}},\ and\ \bibinfo {author}
  {\bibfnamefont {J.~C.}\ \bibnamefont {Tully}},\ }\bibfield  {title} {\bibinfo
  {title} {{$^{13}$C NMR Spectroscopy of K$_x$C$_{60}$: Phase Separation,
  Molecular Dynamics, and Metallic Properties}},\ }\href
  {https://doi.org/10.1126/science.253.5022.884} {\bibfield  {journal}
  {\bibinfo  {journal} {Science}\ }\textbf {\bibinfo {volume} {253}},\ \bibinfo
  {pages} {884} (\bibinfo {year} {1991})}\BibitemShut {NoStop}%
\bibitem [{\citenamefont {Tycko}(1993)}]{Tycko19931713}%
  \BibitemOpen
  \bibfield  {author} {\bibinfo {author} {\bibfnamefont {R.}~\bibnamefont
  {Tycko}},\ }\bibfield  {title} {\bibinfo {title} {Electronic properties and
  phase transitions of alkali fullerides: Investigations by nuclear magnetic
  resonance spectroscopy},\ }\href
  {https://doi.org/https://doi.org/10.1016/0022-3697(93)90286-Z} {\bibfield
  {journal} {\bibinfo  {journal} {J. Phys. Chem. Solids.}\ }\textbf {\bibinfo
  {volume} {54}},\ \bibinfo {pages} {1713} (\bibinfo {year}
  {1993})}\BibitemShut {NoStop}%
\bibitem [{\citenamefont {Walstedt}\ and\ \citenamefont
  {Warren}(1990)}]{Walstedt1990}%
  \BibitemOpen
  \bibfield  {author} {\bibinfo {author} {\bibfnamefont {R.~E.}\ \bibnamefont
  {Walstedt}}\ and\ \bibinfo {author} {\bibfnamefont {W.~W.}\ \bibnamefont
  {Warren}},\ }\bibfield  {title} {\bibinfo {title} {Nuclear resonance
  properties of {YBa{$_2$}Cu{$_3$}O{$_{6+x}$}} superconductors},\ }\href
  {https://doi.org/10.1126/science.248.4959.1082} {\bibfield  {journal}
  {\bibinfo  {journal} {Science}\ }\textbf {\bibinfo {volume} {248}},\ \bibinfo
  {pages} {1082} (\bibinfo {year} {1990})}\BibitemShut {NoStop}%
\bibitem [{\citenamefont {Martindale}\ \emph {et~al.}(1992)\citenamefont
  {Martindale}, \citenamefont {Barrett}, \citenamefont {Klug}, \citenamefont
  {O'Hara}, \citenamefont {DeSoto}, \citenamefont {Slichter}, \citenamefont
  {Friedmann},\ and\ \citenamefont {Ginsberg}}]{Martindale1992}%
  \BibitemOpen
  \bibfield  {author} {\bibinfo {author} {\bibfnamefont {J.~A.}\ \bibnamefont
  {Martindale}}, \bibinfo {author} {\bibfnamefont {S.~E.}\ \bibnamefont
  {Barrett}}, \bibinfo {author} {\bibfnamefont {C.~A.}\ \bibnamefont {Klug}},
  \bibinfo {author} {\bibfnamefont {K.~E.}\ \bibnamefont {O'Hara}}, \bibinfo
  {author} {\bibfnamefont {S.~M.}\ \bibnamefont {DeSoto}}, \bibinfo {author}
  {\bibfnamefont {C.~P.}\ \bibnamefont {Slichter}}, \bibinfo {author}
  {\bibfnamefont {T.~A.}\ \bibnamefont {Friedmann}},\ and\ \bibinfo {author}
  {\bibfnamefont {D.~M.}\ \bibnamefont {Ginsberg}},\ }\bibfield  {title}
  {\bibinfo {title} {{Anisotropy and magnetic field dependence of the planar
  copper NMR spin-lattice relaxation rate in the superconducting state of
  ${\mathrm{YBa}}_{2}$${\mathrm{Cu}}_{3}$${\mathrm{O}}_{7}$}},\ }\href
  {https://doi.org/10.1103/PhysRevLett.68.702} {\bibfield  {journal} {\bibinfo
  {journal} {Phys. Rev. Lett.}\ }\textbf {\bibinfo {volume} {68}},\ \bibinfo
  {pages} {702} (\bibinfo {year} {1992})}\BibitemShut {NoStop}%
\bibitem [{\citenamefont {Smet}\ \emph {et~al.}(2002)\citenamefont {Smet},
  \citenamefont {Deutschmann}, \citenamefont {Ertl}, \citenamefont
  {Wegscheider}, \citenamefont {Abstreiter},\ and\ \citenamefont {von
  Klitzing}}]{Smet2002}%
  \BibitemOpen
  \bibfield  {author} {\bibinfo {author} {\bibfnamefont {J.~H.}\ \bibnamefont
  {Smet}}, \bibinfo {author} {\bibfnamefont {R.~A.}\ \bibnamefont
  {Deutschmann}}, \bibinfo {author} {\bibfnamefont {F.}~\bibnamefont {Ertl}},
  \bibinfo {author} {\bibfnamefont {W.}~\bibnamefont {Wegscheider}}, \bibinfo
  {author} {\bibfnamefont {G.}~\bibnamefont {Abstreiter}},\ and\ \bibinfo
  {author} {\bibfnamefont {K.}~\bibnamefont {von Klitzing}},\ }\bibfield
  {title} {\bibinfo {title} {Gate-voltage control of spin interactions between
  electrons and nuclei in a semiconductor},\ }\href
  {https://doi.org/10.1038/415281a} {\bibfield  {journal} {\bibinfo  {journal}
  {Nature}\ }\textbf {\bibinfo {volume} {415}},\ \bibinfo {pages} {281}
  (\bibinfo {year} {2002})}\BibitemShut {NoStop}%
\bibitem [{\citenamefont {Kumada}\ \emph {et~al.}(2006)\citenamefont {Kumada},
  \citenamefont {Muraki},\ and\ \citenamefont {Hirayama}}]{Kumada2006}%
  \BibitemOpen
  \bibfield  {author} {\bibinfo {author} {\bibfnamefont {N.}~\bibnamefont
  {Kumada}}, \bibinfo {author} {\bibfnamefont {K.}~\bibnamefont {Muraki}},\
  and\ \bibinfo {author} {\bibfnamefont {Y.}~\bibnamefont {Hirayama}},\
  }\bibfield  {title} {\bibinfo {title} {Low-frequency spin dynamics in a
  canted antiferromagnet},\ }\href {https://doi.org/10.1126/science.1127094}
  {\bibfield  {journal} {\bibinfo  {journal} {Science}\ }\textbf {\bibinfo
  {volume} {313}},\ \bibinfo {pages} {329} (\bibinfo {year}
  {2006})}\BibitemShut {NoStop}%
\bibitem [{\citenamefont {Barrett}\ \emph {et~al.}(1995)\citenamefont
  {Barrett}, \citenamefont {Dabbagh}, \citenamefont {Pfeiffer}, \citenamefont
  {West},\ and\ \citenamefont {Tycko}}]{Barrett1995}%
  \BibitemOpen
  \bibfield  {author} {\bibinfo {author} {\bibfnamefont {S.~E.}\ \bibnamefont
  {Barrett}}, \bibinfo {author} {\bibfnamefont {G.}~\bibnamefont {Dabbagh}},
  \bibinfo {author} {\bibfnamefont {L.~N.}\ \bibnamefont {Pfeiffer}}, \bibinfo
  {author} {\bibfnamefont {K.~W.}\ \bibnamefont {West}},\ and\ \bibinfo
  {author} {\bibfnamefont {R.}~\bibnamefont {Tycko}},\ }\bibfield  {title}
  {\bibinfo {title} {Optically pumped {{NMR}} evidence for finite-size
  skyrmions in {{GaAs}} quantum wells near {{Landau}} level filling
  $\mathit{\ensuremath{\nu}}\phantom{\rule{0ex}{0ex}}=\phantom{\rule{0ex}{0ex}}1$},\
  }\href {https://doi.org/10.1103/PhysRevLett.74.5112} {\bibfield  {journal}
  {\bibinfo  {journal} {Phys. Rev. Lett.}\ }\textbf {\bibinfo {volume} {74}},\
  \bibinfo {pages} {5112} (\bibinfo {year} {1995})}\BibitemShut {NoStop}%
\bibitem [{\citenamefont {Liu}\ \emph {et~al.}(2019)\citenamefont {Liu},
  \citenamefont {Su}, \citenamefont {Wei}, \citenamefont {Yao}, \citenamefont
  {Silva}, \citenamefont {Yu}, \citenamefont {Iles-Smith}, \citenamefont
  {Srinivasan}, \citenamefont {Rastelli}, \citenamefont {Li},\ and\
  \citenamefont {Wang}}]{Liu2019}%
  \BibitemOpen
  \bibfield  {author} {\bibinfo {author} {\bibfnamefont {J.}~\bibnamefont
  {Liu}}, \bibinfo {author} {\bibfnamefont {R.}~\bibnamefont {Su}}, \bibinfo
  {author} {\bibfnamefont {Y.}~\bibnamefont {Wei}}, \bibinfo {author}
  {\bibfnamefont {B.}~\bibnamefont {Yao}}, \bibinfo {author} {\bibfnamefont
  {S.~F. C.~d.}\ \bibnamefont {Silva}}, \bibinfo {author} {\bibfnamefont
  {Y.}~\bibnamefont {Yu}}, \bibinfo {author} {\bibfnamefont {J.}~\bibnamefont
  {Iles-Smith}}, \bibinfo {author} {\bibfnamefont {K.}~\bibnamefont
  {Srinivasan}}, \bibinfo {author} {\bibfnamefont {A.}~\bibnamefont
  {Rastelli}}, \bibinfo {author} {\bibfnamefont {J.}~\bibnamefont {Li}},\ and\
  \bibinfo {author} {\bibfnamefont {X.}~\bibnamefont {Wang}},\ }\bibfield
  {title} {\bibinfo {title} {A solid-state source of strongly entangled photon
  pairs with high brightness and indistinguishability},\ }\href
  {https://doi.org/10.1038/s41565-019-0435-9} {\bibfield  {journal} {\bibinfo
  {journal} {Nat. Nanotechnol.}\ }\textbf {\bibinfo {volume} {14}},\ \bibinfo
  {pages} {586} (\bibinfo {year} {2019})}\BibitemShut {NoStop}%
\bibitem [{\citenamefont {Zhai}\ \emph {et~al.}(2020)\citenamefont {Zhai},
  \citenamefont {L{\"o}bl}, \citenamefont {Nguyen}, \citenamefont {Ritzmann},
  \citenamefont {Javadi}, \citenamefont {Spinnler}, \citenamefont {Wieck},
  \citenamefont {Ludwig},\ and\ \citenamefont {Warburton}}]{Zhai2020}%
  \BibitemOpen
  \bibfield  {author} {\bibinfo {author} {\bibfnamefont {L.}~\bibnamefont
  {Zhai}}, \bibinfo {author} {\bibfnamefont {M.~C.}\ \bibnamefont {L{\"o}bl}},
  \bibinfo {author} {\bibfnamefont {G.~N.}\ \bibnamefont {Nguyen}}, \bibinfo
  {author} {\bibfnamefont {J.}~\bibnamefont {Ritzmann}}, \bibinfo {author}
  {\bibfnamefont {A.}~\bibnamefont {Javadi}}, \bibinfo {author} {\bibfnamefont
  {C.}~\bibnamefont {Spinnler}}, \bibinfo {author} {\bibfnamefont {A.~D.}\
  \bibnamefont {Wieck}}, \bibinfo {author} {\bibfnamefont {A.}~\bibnamefont
  {Ludwig}},\ and\ \bibinfo {author} {\bibfnamefont {R.~J.}\ \bibnamefont
  {Warburton}},\ }\bibfield  {title} {\bibinfo {title} {{Low-noise GaAs quantum
  dots for quantum photonics}},\ }\href
  {https://doi.org/10.1038/s41467-020-18625-z} {\bibfield  {journal} {\bibinfo
  {journal} {Nat. Commun.}\ }\textbf {\bibinfo {volume} {11}},\ \bibinfo
  {pages} {4745} (\bibinfo {year} {2020})}\BibitemShut {NoStop}%
\bibitem [{\citenamefont {Tomm}\ \emph {et~al.}(2021)\citenamefont {Tomm},
  \citenamefont {Javadi}, \citenamefont {Antoniadis}, \citenamefont {Najer},
  \citenamefont {L{\"o}bl}, \citenamefont {Korsch}, \citenamefont {Schott},
  \citenamefont {Valentin}, \citenamefont {Wieck}, \citenamefont {Ludwig},\
  and\ \citenamefont {Warburton}}]{Tomm2021}%
  \BibitemOpen
  \bibfield  {author} {\bibinfo {author} {\bibfnamefont {N.}~\bibnamefont
  {Tomm}}, \bibinfo {author} {\bibfnamefont {A.}~\bibnamefont {Javadi}},
  \bibinfo {author} {\bibfnamefont {N.~O.}\ \bibnamefont {Antoniadis}},
  \bibinfo {author} {\bibfnamefont {D.}~\bibnamefont {Najer}}, \bibinfo
  {author} {\bibfnamefont {M.~C.}\ \bibnamefont {L{\"o}bl}}, \bibinfo {author}
  {\bibfnamefont {A.~R.}\ \bibnamefont {Korsch}}, \bibinfo {author}
  {\bibfnamefont {R.}~\bibnamefont {Schott}}, \bibinfo {author} {\bibfnamefont
  {S.~R.}\ \bibnamefont {Valentin}}, \bibinfo {author} {\bibfnamefont {A.~D.}\
  \bibnamefont {Wieck}}, \bibinfo {author} {\bibfnamefont {A.}~\bibnamefont
  {Ludwig}},\ and\ \bibinfo {author} {\bibfnamefont {R.~J.}\ \bibnamefont
  {Warburton}},\ }\bibfield  {title} {\bibinfo {title} {A bright and fast
  source of coherent single photons},\ }\href
  {https://doi.org/10.1038/s41565-020-00831-x} {\bibfield  {journal} {\bibinfo
  {journal} {Nat. Nanotechnol.}\ }\textbf {\bibinfo {volume} {16}},\ \bibinfo
  {pages} {399} (\bibinfo {year} {2021})}\BibitemShut {NoStop}%
\bibitem [{\citenamefont {Chekhovich}\ \emph {et~al.}(2020)\citenamefont
  {Chekhovich}, \citenamefont {da~Silva},\ and\ \citenamefont
  {Rastelli}}]{Chekhovich2020}%
  \BibitemOpen
  \bibfield  {author} {\bibinfo {author} {\bibfnamefont {E.~A.}\ \bibnamefont
  {Chekhovich}}, \bibinfo {author} {\bibfnamefont {S.~F.~C.}\ \bibnamefont
  {da~Silva}},\ and\ \bibinfo {author} {\bibfnamefont {A.}~\bibnamefont
  {Rastelli}},\ }\bibfield  {title} {\bibinfo {title} {Nuclear spin quantum
  register in an optically active semiconductor quantum dot},\ }\href
  {https://doi.org/10.1038/s41565-020-0769-3} {\bibfield  {journal} {\bibinfo
  {journal} {Nat. Nanotechnol.}\ }\textbf {\bibinfo {volume} {15}},\ \bibinfo
  {pages} {999} (\bibinfo {year} {2020})}\BibitemShut {NoStop}%
\bibitem [{\citenamefont {Leon}\ \emph {et~al.}(2020)\citenamefont {Leon},
  \citenamefont {Yang}, \citenamefont {Hwang}, \citenamefont {Lemyre},
  \citenamefont {Tanttu}, \citenamefont {Huang}, \citenamefont {Chan},
  \citenamefont {Tan}, \citenamefont {Hudson}, \citenamefont {Itoh},
  \citenamefont {Morello}, \citenamefont {Laucht}, \citenamefont
  {Pioro-Ladri{\`e}re}, \citenamefont {Saraiva},\ and\ \citenamefont
  {Dzurak}}]{Leon2020}%
  \BibitemOpen
  \bibfield  {author} {\bibinfo {author} {\bibfnamefont {R.~C.~C.}\
  \bibnamefont {Leon}}, \bibinfo {author} {\bibfnamefont {C.~H.}\ \bibnamefont
  {Yang}}, \bibinfo {author} {\bibfnamefont {J.~C.~C.}\ \bibnamefont {Hwang}},
  \bibinfo {author} {\bibfnamefont {J.~C.}\ \bibnamefont {Lemyre}}, \bibinfo
  {author} {\bibfnamefont {T.}~\bibnamefont {Tanttu}}, \bibinfo {author}
  {\bibfnamefont {W.}~\bibnamefont {Huang}}, \bibinfo {author} {\bibfnamefont
  {K.~W.}\ \bibnamefont {Chan}}, \bibinfo {author} {\bibfnamefont {K.~Y.}\
  \bibnamefont {Tan}}, \bibinfo {author} {\bibfnamefont {F.~E.}\ \bibnamefont
  {Hudson}}, \bibinfo {author} {\bibfnamefont {K.~M.}\ \bibnamefont {Itoh}},
  \bibinfo {author} {\bibfnamefont {A.}~\bibnamefont {Morello}}, \bibinfo
  {author} {\bibfnamefont {A.}~\bibnamefont {Laucht}}, \bibinfo {author}
  {\bibfnamefont {M.}~\bibnamefont {Pioro-Ladri{\`e}re}}, \bibinfo {author}
  {\bibfnamefont {A.}~\bibnamefont {Saraiva}},\ and\ \bibinfo {author}
  {\bibfnamefont {A.~S.}\ \bibnamefont {Dzurak}},\ }\bibfield  {title}
  {\bibinfo {title} {Coherent spin control of s-, p-, d- and f-electrons in a
  silicon quantum dot},\ }\href {https://doi.org/10.1038/s41467-019-14053-w}
  {\bibfield  {journal} {\bibinfo  {journal} {Nat. Commun.}\ }\textbf {\bibinfo
  {volume} {11}},\ \bibinfo {pages} {797} (\bibinfo {year} {2020})}\BibitemShut
  {NoStop}%
\bibitem [{\citenamefont {Nowack}\ \emph {et~al.}(2007)\citenamefont {Nowack},
  \citenamefont {Koppens}, \citenamefont {Nazarov},\ and\ \citenamefont
  {Vandersypen}}]{Nowack2007}%
  \BibitemOpen
  \bibfield  {author} {\bibinfo {author} {\bibfnamefont {K.~C.}\ \bibnamefont
  {Nowack}}, \bibinfo {author} {\bibfnamefont {F.~H.~L.}\ \bibnamefont
  {Koppens}}, \bibinfo {author} {\bibfnamefont {Y.~V.}\ \bibnamefont
  {Nazarov}},\ and\ \bibinfo {author} {\bibfnamefont {L.~M.~K.}\ \bibnamefont
  {Vandersypen}},\ }\bibfield  {title} {\bibinfo {title} {Coherent control of a
  single electron spin with electric fields},\ }\href
  {https://doi.org/10.1126/science.1148092} {\bibfield  {journal} {\bibinfo
  {journal} {Science}\ }\textbf {\bibinfo {volume} {318}},\ \bibinfo {pages}
  {1430} (\bibinfo {year} {2007})}\BibitemShut {NoStop}%
\bibitem [{\citenamefont {Nadj-Perge}\ \emph {et~al.}(2010)\citenamefont
  {Nadj-Perge}, \citenamefont {Frolov}, \citenamefont {Bakkers},\ and\
  \citenamefont {Kouwenhoven}}]{NadjPerge2010}%
  \BibitemOpen
  \bibfield  {author} {\bibinfo {author} {\bibfnamefont {S.}~\bibnamefont
  {Nadj-Perge}}, \bibinfo {author} {\bibfnamefont {S.~M.}\ \bibnamefont
  {Frolov}}, \bibinfo {author} {\bibfnamefont {E.~P. A.~M.}\ \bibnamefont
  {Bakkers}},\ and\ \bibinfo {author} {\bibfnamefont {L.~P.}\ \bibnamefont
  {Kouwenhoven}},\ }\bibfield  {title} {\bibinfo {title} {Spin--orbit qubit in
  a semiconductor nanowire},\ }\href {https://doi.org/10.1038/nature09682}
  {\bibfield  {journal} {\bibinfo  {journal} {Nature}\ }\textbf {\bibinfo
  {volume} {468}},\ \bibinfo {pages} {1084} (\bibinfo {year}
  {2010})}\BibitemShut {NoStop}%
\bibitem [{\citenamefont {Yoneda}\ \emph {et~al.}(2014)\citenamefont {Yoneda},
  \citenamefont {Otsuka}, \citenamefont {Nakajima}, \citenamefont {Takakura},
  \citenamefont {Obata}, \citenamefont {Pioro-Ladri\`ere}, \citenamefont {Lu},
  \citenamefont {Palmstr\o{}m}, \citenamefont {Gossard},\ and\ \citenamefont
  {Tarucha}}]{Yoneda2014}%
  \BibitemOpen
  \bibfield  {author} {\bibinfo {author} {\bibfnamefont {J.}~\bibnamefont
  {Yoneda}}, \bibinfo {author} {\bibfnamefont {T.}~\bibnamefont {Otsuka}},
  \bibinfo {author} {\bibfnamefont {T.}~\bibnamefont {Nakajima}}, \bibinfo
  {author} {\bibfnamefont {T.}~\bibnamefont {Takakura}}, \bibinfo {author}
  {\bibfnamefont {T.}~\bibnamefont {Obata}}, \bibinfo {author} {\bibfnamefont
  {M.}~\bibnamefont {Pioro-Ladri\`ere}}, \bibinfo {author} {\bibfnamefont
  {H.}~\bibnamefont {Lu}}, \bibinfo {author} {\bibfnamefont {C.~J.}\
  \bibnamefont {Palmstr\o{}m}}, \bibinfo {author} {\bibfnamefont {A.~C.}\
  \bibnamefont {Gossard}},\ and\ \bibinfo {author} {\bibfnamefont
  {S.}~\bibnamefont {Tarucha}},\ }\bibfield  {title} {\bibinfo {title} {Fast
  electrical control of single electron spins in quantum dots with vanishing
  influence from nuclear spins},\ }\href
  {https://doi.org/10.1103/PhysRevLett.113.267601} {\bibfield  {journal}
  {\bibinfo  {journal} {Phys. Rev. Lett.}\ }\textbf {\bibinfo {volume} {113}},\
  \bibinfo {pages} {267601} (\bibinfo {year} {2014})}\BibitemShut {NoStop}%
\bibitem [{\citenamefont {Schmidt-Rohr}\ \emph {et~al.}(1992)\citenamefont
  {Schmidt-Rohr}, \citenamefont {Clauss},\ and\ \citenamefont
  {Spiess}}]{Schmidt1992}%
  \BibitemOpen
  \bibfield  {author} {\bibinfo {author} {\bibfnamefont {K.}~\bibnamefont
  {Schmidt-Rohr}}, \bibinfo {author} {\bibfnamefont {J.}~\bibnamefont
  {Clauss}},\ and\ \bibinfo {author} {\bibfnamefont {H.}~\bibnamefont
  {Spiess}},\ }\bibfield  {title} {\bibinfo {title} {Correlation of structure,
  mobility, and morphological information in heterogeneous polymer materials by
  two-dimensional wideline-separation {{NMR}} spectroscopy},\ }\href@noop {}
  {\bibfield  {journal} {\bibinfo  {journal} {Macromolecules}\ }\textbf
  {\bibinfo {volume} {25}},\ \bibinfo {pages} {3273} (\bibinfo {year}
  {1992})}\BibitemShut {NoStop}%
\bibitem [{\citenamefont {Demco}\ \emph {et~al.}(1995)\citenamefont {Demco},
  \citenamefont {Johansson},\ and\ \citenamefont {Tegenfeldt}}]{Demco1995}%
  \BibitemOpen
  \bibfield  {author} {\bibinfo {author} {\bibfnamefont {D.~E.}\ \bibnamefont
  {Demco}}, \bibinfo {author} {\bibfnamefont {A.}~\bibnamefont {Johansson}},\
  and\ \bibinfo {author} {\bibfnamefont {J.}~\bibnamefont {Tegenfeldt}},\
  }\bibfield  {title} {\bibinfo {title} {Proton spin diffusion for spatial
  heterogeneity and morphology investigations of polymers},\ }\href
  {https://doi.org/https://doi.org/10.1016/0926-2040(94)00036-C} {\bibfield
  {journal} {\bibinfo  {journal} {Solid State Nucl. Magn. Reson.}\ }\textbf
  {\bibinfo {volume} {4}},\ \bibinfo {pages} {13} (\bibinfo {year}
  {1995})}\BibitemShut {NoStop}%
\bibitem [{\citenamefont {Hall}\ \emph {et~al.}(1997)\citenamefont {Hall},
  \citenamefont {Maus}, \citenamefont {Gerfen}, \citenamefont {Inati},
  \citenamefont {Becerra}, \citenamefont {Dahlquist},\ and\ \citenamefont
  {Griffin}}]{Hall1997}%
  \BibitemOpen
  \bibfield  {author} {\bibinfo {author} {\bibfnamefont {D.~A.}\ \bibnamefont
  {Hall}}, \bibinfo {author} {\bibfnamefont {D.~C.}\ \bibnamefont {Maus}},
  \bibinfo {author} {\bibfnamefont {G.~J.}\ \bibnamefont {Gerfen}}, \bibinfo
  {author} {\bibfnamefont {S.~J.}\ \bibnamefont {Inati}}, \bibinfo {author}
  {\bibfnamefont {L.~R.}\ \bibnamefont {Becerra}}, \bibinfo {author}
  {\bibfnamefont {F.~W.}\ \bibnamefont {Dahlquist}},\ and\ \bibinfo {author}
  {\bibfnamefont {R.~G.}\ \bibnamefont {Griffin}},\ }\bibfield  {title}
  {\bibinfo {title} {Polarization-enhanced {{NMR}} spectroscopy of biomolecules
  in frozen solution},\ }\href {https://doi.org/10.1126/science.276.5314.930}
  {\bibfield  {journal} {\bibinfo  {journal} {Science}\ }\textbf {\bibinfo
  {volume} {276}},\ \bibinfo {pages} {930} (\bibinfo {year}
  {1997})}\BibitemShut {NoStop}%
\bibitem [{\citenamefont {van~der Wel}\ \emph {et~al.}(2006)\citenamefont
  {van~der Wel}, \citenamefont {Hu}, \citenamefont {Lewandowski},\ and\
  \citenamefont {Griffin}}]{vanderWel2006}%
  \BibitemOpen
  \bibfield  {author} {\bibinfo {author} {\bibfnamefont {P.~C.~A.}\
  \bibnamefont {van~der Wel}}, \bibinfo {author} {\bibfnamefont {K.-N.}\
  \bibnamefont {Hu}}, \bibinfo {author} {\bibfnamefont {J.}~\bibnamefont
  {Lewandowski}},\ and\ \bibinfo {author} {\bibfnamefont {R.~G.}\ \bibnamefont
  {Griffin}},\ }\bibfield  {title} {\bibinfo {title} {Dynamic nuclear
  polarization of amyloidogenic peptide nanocrystals: {{GNNQQNY}}, a core
  segment of the yeast prion protein {{Sup35p}}},\ }\href
  {https://doi.org/10.1021/ja0626685} {\bibfield  {journal} {\bibinfo
  {journal} {J. Am. Chem. Soc.}\ }\textbf {\bibinfo {volume} {128}},\ \bibinfo
  {pages} {10840} (\bibinfo {year} {2006})}\BibitemShut {NoStop}%
\bibitem [{\citenamefont {Manolikas}\ \emph {et~al.}(2008)\citenamefont
  {Manolikas}, \citenamefont {Herrmann},\ and\ \citenamefont
  {Meier}}]{Manolikas2008}%
  \BibitemOpen
  \bibfield  {author} {\bibinfo {author} {\bibfnamefont {T.}~\bibnamefont
  {Manolikas}}, \bibinfo {author} {\bibfnamefont {T.}~\bibnamefont
  {Herrmann}},\ and\ \bibinfo {author} {\bibfnamefont {B.~H.}\ \bibnamefont
  {Meier}},\ }\bibfield  {title} {\bibinfo {title} {Protein structure
  determination from {{13C}} spin-diffusion solid-state {{NMR}} spectroscopy},\
  }\href {https://doi.org/10.1021/ja078039s} {\bibfield  {journal} {\bibinfo
  {journal} {J. Am. Chem. Soc.}\ }\textbf {\bibinfo {volume} {130}},\ \bibinfo
  {pages} {3959} (\bibinfo {year} {2008})}\BibitemShut {NoStop}%
\bibitem [{\citenamefont {Rossini}\ \emph {et~al.}(2014)\citenamefont
  {Rossini}, \citenamefont {Widdifield}, \citenamefont {Zagdoun}, \citenamefont
  {Lelli}, \citenamefont {Schwarzwalder}, \citenamefont {Coperet},
  \citenamefont {Lesage},\ and\ \citenamefont {Emsley}}]{Rossini2014}%
  \BibitemOpen
  \bibfield  {author} {\bibinfo {author} {\bibfnamefont {A.~J.}\ \bibnamefont
  {Rossini}}, \bibinfo {author} {\bibfnamefont {C.~M.}\ \bibnamefont
  {Widdifield}}, \bibinfo {author} {\bibfnamefont {A.}~\bibnamefont {Zagdoun}},
  \bibinfo {author} {\bibfnamefont {M.}~\bibnamefont {Lelli}}, \bibinfo
  {author} {\bibfnamefont {M.}~\bibnamefont {Schwarzwalder}}, \bibinfo {author}
  {\bibfnamefont {C.}~\bibnamefont {Coperet}}, \bibinfo {author} {\bibfnamefont
  {A.}~\bibnamefont {Lesage}},\ and\ \bibinfo {author} {\bibfnamefont
  {L.}~\bibnamefont {Emsley}},\ }\bibfield  {title} {\bibinfo {title} {Dynamic
  nuclear polarization enhanced {{NMR}} spectroscopy for pharmaceutical
  formulations},\ }\href {https://doi.org/10.1021/ja4092038} {\bibfield
  {journal} {\bibinfo  {journal} {J. Am. Chem. Soc.}\ }\textbf {\bibinfo
  {volume} {136}},\ \bibinfo {pages} {2324} (\bibinfo {year}
  {2014})}\BibitemShut {NoStop}%
\bibitem [{\citenamefont {Viger-Gravel}\ \emph {et~al.}(2018)\citenamefont
  {Viger-Gravel}, \citenamefont {Schantz}, \citenamefont {Pinon}, \citenamefont
  {Rossini}, \citenamefont {Schantz},\ and\ \citenamefont
  {Emsley}}]{VigerGravel2018}%
  \BibitemOpen
  \bibfield  {author} {\bibinfo {author} {\bibfnamefont {J.}~\bibnamefont
  {Viger-Gravel}}, \bibinfo {author} {\bibfnamefont {A.}~\bibnamefont
  {Schantz}}, \bibinfo {author} {\bibfnamefont {A.~C.}\ \bibnamefont {Pinon}},
  \bibinfo {author} {\bibfnamefont {A.~J.}\ \bibnamefont {Rossini}}, \bibinfo
  {author} {\bibfnamefont {S.}~\bibnamefont {Schantz}},\ and\ \bibinfo {author}
  {\bibfnamefont {L.}~\bibnamefont {Emsley}},\ }\bibfield  {title} {\bibinfo
  {title} {Structure of lipid nanoparticles containing {{siRNA}} or {{mRNA}} by
  dynamic nuclear polarization-enhanced {{NMR}} spectroscopy},\ }\href
  {https://doi.org/10.1021/acs.jpcb.7b10795} {\bibfield  {journal} {\bibinfo
  {journal} {J. Phys. Chem. B}\ }\textbf {\bibinfo {volume} {122}},\ \bibinfo
  {pages} {2073} (\bibinfo {year} {2018})}\BibitemShut {NoStop}%
\bibitem [{\citenamefont {von Witte}\ \emph {et~al.}(2025)\citenamefont {von
  Witte}, \citenamefont {Kozerke},\ and\ \citenamefont {Ernst}}]{vonWitte2025}%
  \BibitemOpen
  \bibfield  {author} {\bibinfo {author} {\bibfnamefont {G.}~\bibnamefont {von
  Witte}}, \bibinfo {author} {\bibfnamefont {S.}~\bibnamefont {Kozerke}},\ and\
  \bibinfo {author} {\bibfnamefont {M.}~\bibnamefont {Ernst}},\ }\bibfield
  {title} {\bibinfo {title} {Two-electron two-nucleus effective hamiltonian and
  the spin diffusion barrier},\ }\href {https://doi.org/10.1126/sciadv.adr7168}
  {\bibfield  {journal} {\bibinfo  {journal} {Sci. Adv.}\ }\textbf {\bibinfo
  {volume} {11}},\ \bibinfo {pages} {eadr7168} (\bibinfo {year}
  {2025})}\BibitemShut {NoStop}%
\bibitem [{\citenamefont {Heyn}\ \emph {et~al.}(2009)\citenamefont {Heyn},
  \citenamefont {Stemmann}, \citenamefont {Koppen}, \citenamefont {Strelow},
  \citenamefont {Kipp}, \citenamefont {Grave}, \citenamefont {Mendach},\ and\
  \citenamefont {Hansen}}]{Heyn2009}%
  \BibitemOpen
  \bibfield  {author} {\bibinfo {author} {\bibfnamefont {C.}~\bibnamefont
  {Heyn}}, \bibinfo {author} {\bibfnamefont {A.}~\bibnamefont {Stemmann}},
  \bibinfo {author} {\bibfnamefont {T.}~\bibnamefont {Koppen}}, \bibinfo
  {author} {\bibfnamefont {C.}~\bibnamefont {Strelow}}, \bibinfo {author}
  {\bibfnamefont {T.}~\bibnamefont {Kipp}}, \bibinfo {author} {\bibfnamefont
  {M.}~\bibnamefont {Grave}}, \bibinfo {author} {\bibfnamefont
  {S.}~\bibnamefont {Mendach}},\ and\ \bibinfo {author} {\bibfnamefont
  {W.}~\bibnamefont {Hansen}},\ }\bibfield  {title} {\bibinfo {title} {Highly
  uniform and strain-free {{GaAs}} quantum dots fabricated by filling of
  self-assembled nanoholes},\ }\href {https://doi.org/10.1063/1.3133338}
  {\bibfield  {journal} {\bibinfo  {journal} {Appl. Phys. Lett.}\ }\textbf
  {\bibinfo {volume} {94}},\ \bibinfo {pages} {183113} (\bibinfo {year}
  {2009})}\BibitemShut {NoStop}%
\bibitem [{\citenamefont {Atkinson}\ \emph {et~al.}(2012)\citenamefont
  {Atkinson}, \citenamefont {Zallo},\ and\ \citenamefont
  {Schmidt}}]{Atkinson2012}%
  \BibitemOpen
  \bibfield  {author} {\bibinfo {author} {\bibfnamefont {P.}~\bibnamefont
  {Atkinson}}, \bibinfo {author} {\bibfnamefont {E.}~\bibnamefont {Zallo}},\
  and\ \bibinfo {author} {\bibfnamefont {O.~G.}\ \bibnamefont {Schmidt}},\
  }\bibfield  {title} {\bibinfo {title} {Independent wavelength and density
  control of uniform {GaAs/AlGaAs} quantum dots grown by infilling
  self-assembled nanoholes},\ }\href
  {https://doi.org/ttp://dx.doi.org/10.1063/1.4748183} {\bibfield  {journal}
  {\bibinfo  {journal} {J. Appl. Phys.}\ }\textbf {\bibinfo {volume} {112}},\
  \bibinfo {pages} {054303} (\bibinfo {year} {2012})}\BibitemShut {NoStop}%
\bibitem [{\citenamefont {Dyte}\ \emph {et~al.}(2024)\citenamefont {Dyte},
  \citenamefont {Gillard}, \citenamefont {Manna}, \citenamefont {Covre~da
  Silva}, \citenamefont {Rastelli},\ and\ \citenamefont
  {Chekhovich}}]{Dyte2023}%
  \BibitemOpen
  \bibfield  {author} {\bibinfo {author} {\bibfnamefont {H.~E.}\ \bibnamefont
  {Dyte}}, \bibinfo {author} {\bibfnamefont {G.}~\bibnamefont {Gillard}},
  \bibinfo {author} {\bibfnamefont {S.}~\bibnamefont {Manna}}, \bibinfo
  {author} {\bibfnamefont {S.~F.}\ \bibnamefont {Covre~da Silva}}, \bibinfo
  {author} {\bibfnamefont {A.}~\bibnamefont {Rastelli}},\ and\ \bibinfo
  {author} {\bibfnamefont {E.~A.}\ \bibnamefont {Chekhovich}},\ }\bibfield
  {title} {\bibinfo {title} {Is wave function collapse necessary?
  {{Explaining}} quantum nondemolition measurement of a spin qubit within
  linear evolution},\ }\href {https://doi.org/10.1103/PhysRevLett.132.160804}
  {\bibfield  {journal} {\bibinfo  {journal} {Phys. Rev. Lett.}\ }\textbf
  {\bibinfo {volume} {132}},\ \bibinfo {pages} {160804} (\bibinfo {year}
  {2024})}\BibitemShut {NoStop}%
\bibitem [{\citenamefont {Millington-Hotze}\ \emph {et~al.}(2023)\citenamefont
  {Millington-Hotze}, \citenamefont {Manna}, \citenamefont {Covre~da Silva},
  \citenamefont {Rastelli},\ and\ \citenamefont
  {Chekhovich}}]{MillingtonHotze2022}%
  \BibitemOpen
  \bibfield  {author} {\bibinfo {author} {\bibfnamefont {P.}~\bibnamefont
  {Millington-Hotze}}, \bibinfo {author} {\bibfnamefont {S.}~\bibnamefont
  {Manna}}, \bibinfo {author} {\bibfnamefont {S.~F.}\ \bibnamefont {Covre~da
  Silva}}, \bibinfo {author} {\bibfnamefont {A.}~\bibnamefont {Rastelli}},\
  and\ \bibinfo {author} {\bibfnamefont {E.~A.}\ \bibnamefont {Chekhovich}},\
  }\bibfield  {title} {\bibinfo {title} {Nuclear spin diffusion in the central
  spin system of a {{GaAs/AlGaAs}} quantum dot},\ }\href
  {https://doi.org/10.1038/s41467-023-38349-0} {\bibfield  {journal} {\bibinfo
  {journal} {Nat. Commun.}\ }\textbf {\bibinfo {volume} {14}},\ \bibinfo
  {pages} {2677} (\bibinfo {year} {2023})}\BibitemShut {NoStop}%
\bibitem [{\citenamefont {Paget}(1982)}]{Paget1982}%
  \BibitemOpen
  \bibfield  {author} {\bibinfo {author} {\bibfnamefont {D.}~\bibnamefont
  {Paget}},\ }\bibfield  {title} {\bibinfo {title} {Optical detection of
  {{NMR}} in high-purity {{GaAs}}: Direct study of the relaxation of nuclei
  close to shallow donors},\ }\href {https://doi.org/10.1103/PhysRevB.25.4444}
  {\bibfield  {journal} {\bibinfo  {journal} {Phys. Rev. B}\ }\textbf {\bibinfo
  {volume} {25}},\ \bibinfo {pages} {4444} (\bibinfo {year}
  {1982})}\BibitemShut {NoStop}%
\bibitem [{\citenamefont {Tycko}\ \emph {et~al.}(1995)\citenamefont {Tycko},
  \citenamefont {Barrett}, \citenamefont {Dabbagh}, \citenamefont {Pfeiffer},\
  and\ \citenamefont {West}}]{Tycko1995}%
  \BibitemOpen
  \bibfield  {author} {\bibinfo {author} {\bibfnamefont {R.}~\bibnamefont
  {Tycko}}, \bibinfo {author} {\bibfnamefont {S.}~\bibnamefont {Barrett}},
  \bibinfo {author} {\bibfnamefont {G.}~\bibnamefont {Dabbagh}}, \bibinfo
  {author} {\bibfnamefont {L.}~\bibnamefont {Pfeiffer}},\ and\ \bibinfo
  {author} {\bibfnamefont {K.}~\bibnamefont {West}},\ }\bibfield  {title}
  {\bibinfo {title} {Electronic states in gallium arsenide quantum wells probed
  by optically pumped {{NMR}}},\ }\href
  {https://doi.org/10.1126/science.7539550} {\bibfield  {journal} {\bibinfo
  {journal} {Science}\ }\textbf {\bibinfo {volume} {268}},\ \bibinfo {pages}
  {1460} (\bibinfo {year} {1995})}\BibitemShut {NoStop}%
\bibitem [{\citenamefont {Hayashi}\ \emph {et~al.}(2008)\citenamefont
  {Hayashi}, \citenamefont {Itoh},\ and\ \citenamefont
  {Vlasenko}}]{Hayashi2008}%
  \BibitemOpen
  \bibfield  {author} {\bibinfo {author} {\bibfnamefont {H.}~\bibnamefont
  {Hayashi}}, \bibinfo {author} {\bibfnamefont {K.~M.}\ \bibnamefont {Itoh}},\
  and\ \bibinfo {author} {\bibfnamefont {L.~S.}\ \bibnamefont {Vlasenko}},\
  }\bibfield  {title} {\bibinfo {title} {Nuclear magnetic resonance linewidth
  and spin diffusion in $^{29}\text{S}\text{i}$ isotopically controlled
  silicon},\ }\href {https://doi.org/10.1103/PhysRevB.78.153201} {\bibfield
  {journal} {\bibinfo  {journal} {Phys. Rev. B}\ }\textbf {\bibinfo {volume}
  {78}},\ \bibinfo {pages} {153201} (\bibinfo {year} {2008})}\BibitemShut
  {NoStop}%
\bibitem [{\citenamefont {Nikolaenko}\ \emph {et~al.}(2009)\citenamefont
  {Nikolaenko}, \citenamefont {Chekhovich}, \citenamefont {Makhonin},
  \citenamefont {Drouzas}, \citenamefont {Van'kov}, \citenamefont
  {Skiba-Szymanska}, \citenamefont {Skolnick}, \citenamefont {Senellart},
  \citenamefont {Martrou}, \citenamefont {Lema\^{\i}tre},\ and\ \citenamefont
  {Tartakovskii}}]{Nikolaenko2009}%
  \BibitemOpen
  \bibfield  {author} {\bibinfo {author} {\bibfnamefont {A.~E.}\ \bibnamefont
  {Nikolaenko}}, \bibinfo {author} {\bibfnamefont {E.~A.}\ \bibnamefont
  {Chekhovich}}, \bibinfo {author} {\bibfnamefont {M.~N.}\ \bibnamefont
  {Makhonin}}, \bibinfo {author} {\bibfnamefont {I.~W.}\ \bibnamefont
  {Drouzas}}, \bibinfo {author} {\bibfnamefont {A.~B.}\ \bibnamefont
  {Van'kov}}, \bibinfo {author} {\bibfnamefont {J.}~\bibnamefont
  {Skiba-Szymanska}}, \bibinfo {author} {\bibfnamefont {M.~S.}\ \bibnamefont
  {Skolnick}}, \bibinfo {author} {\bibfnamefont {P.}~\bibnamefont {Senellart}},
  \bibinfo {author} {\bibfnamefont {D.}~\bibnamefont {Martrou}}, \bibinfo
  {author} {\bibfnamefont {A.}~\bibnamefont {Lema\^{\i}tre}},\ and\ \bibinfo
  {author} {\bibfnamefont {A.~I.}\ \bibnamefont {Tartakovskii}},\ }\bibfield
  {title} {\bibinfo {title} {Suppression of nuclear spin diffusion at a
  {$\text{GaAs}/{\text{Al}}_{x}{\text{Ga}}_{1\ensuremath{-}x}\text{As}$}
  interface measured with a single quantum-dot nanoprobe},\ }\href
  {https://doi.org/10.1103/PhysRevB.79.081303} {\bibfield  {journal} {\bibinfo
  {journal} {Phys. Rev. B}\ }\textbf {\bibinfo {volume} {79}},\ \bibinfo
  {pages} {081303} (\bibinfo {year} {2009})}\BibitemShut {NoStop}%
\bibitem [{\citenamefont {Millington-Hotze}\ \emph {et~al.}(2024)\citenamefont
  {Millington-Hotze}, \citenamefont {Dyte}, \citenamefont {Manna},
  \citenamefont {Covre~da Silva}, \citenamefont {Rastelli},\ and\ \citenamefont
  {Chekhovich}}]{MillingtonHotze2023}%
  \BibitemOpen
  \bibfield  {author} {\bibinfo {author} {\bibfnamefont {P.}~\bibnamefont
  {Millington-Hotze}}, \bibinfo {author} {\bibfnamefont {H.~E.}\ \bibnamefont
  {Dyte}}, \bibinfo {author} {\bibfnamefont {S.}~\bibnamefont {Manna}},
  \bibinfo {author} {\bibfnamefont {S.~F.}\ \bibnamefont {Covre~da Silva}},
  \bibinfo {author} {\bibfnamefont {A.}~\bibnamefont {Rastelli}},\ and\
  \bibinfo {author} {\bibfnamefont {E.~A.}\ \bibnamefont {Chekhovich}},\
  }\bibfield  {title} {\bibinfo {title} {Approaching a fully-polarized state of
  nuclear spins in a solid},\ }\href
  {https://doi.org/10.1038/s41467-024-45364-2} {\bibfield  {journal} {\bibinfo
  {journal} {Nat. Commun.}\ }\textbf {\bibinfo {volume} {15}},\ \bibinfo
  {pages} {985} (\bibinfo {year} {2024})}\BibitemShut {NoStop}%
\bibitem [{\citenamefont {Urbaszek}\ \emph {et~al.}(2013)\citenamefont
  {Urbaszek}, \citenamefont {Marie}, \citenamefont {Amand}, \citenamefont
  {Krebs}, \citenamefont {Voisin}, \citenamefont {Maletinsky}, \citenamefont
  {H\"ogele},\ and\ \citenamefont {Imamoglu}}]{Urbaszek2013}%
  \BibitemOpen
  \bibfield  {author} {\bibinfo {author} {\bibfnamefont {B.}~\bibnamefont
  {Urbaszek}}, \bibinfo {author} {\bibfnamefont {X.}~\bibnamefont {Marie}},
  \bibinfo {author} {\bibfnamefont {T.}~\bibnamefont {Amand}}, \bibinfo
  {author} {\bibfnamefont {O.}~\bibnamefont {Krebs}}, \bibinfo {author}
  {\bibfnamefont {P.}~\bibnamefont {Voisin}}, \bibinfo {author} {\bibfnamefont
  {P.}~\bibnamefont {Maletinsky}}, \bibinfo {author} {\bibfnamefont
  {A.}~\bibnamefont {H\"ogele}},\ and\ \bibinfo {author} {\bibfnamefont
  {A.}~\bibnamefont {Imamoglu}},\ }\bibfield  {title} {\bibinfo {title}
  {Nuclear spin physics in quantum dots: An optical investigation},\ }\href
  {https://doi.org/10.1103/RevModPhys.85.79} {\bibfield  {journal} {\bibinfo
  {journal} {Rev. Mod. Phys.}\ }\textbf {\bibinfo {volume} {85}},\ \bibinfo
  {pages} {79} (\bibinfo {year} {2013})}\BibitemShut {NoStop}%
\bibitem [{\citenamefont {Averin}\ \emph {et~al.}(1991)\citenamefont {Averin},
  \citenamefont {Korotkov},\ and\ \citenamefont {Likharev}}]{Averin1991}%
  \BibitemOpen
  \bibfield  {author} {\bibinfo {author} {\bibfnamefont {D.~V.}\ \bibnamefont
  {Averin}}, \bibinfo {author} {\bibfnamefont {A.~N.}\ \bibnamefont
  {Korotkov}},\ and\ \bibinfo {author} {\bibfnamefont {K.~K.}\ \bibnamefont
  {Likharev}},\ }\bibfield  {title} {\bibinfo {title} {Theory of
  single-electron charging of quantum wells and dots},\ }\href
  {https://doi.org/10.1103/PhysRevB.44.6199} {\bibfield  {journal} {\bibinfo
  {journal} {Phys. Rev. B}\ }\textbf {\bibinfo {volume} {44}},\ \bibinfo
  {pages} {6199} (\bibinfo {year} {1991})}\BibitemShut {NoStop}%
\bibitem [{\citenamefont {Beenakker}(1991)}]{Beenakker1991}%
  \BibitemOpen
  \bibfield  {author} {\bibinfo {author} {\bibfnamefont {C.~W.~J.}\
  \bibnamefont {Beenakker}},\ }\bibfield  {title} {\bibinfo {title} {Theory of
  coulomb-blockade oscillations in the conductance of a quantum dot},\ }\href
  {https://doi.org/10.1103/PhysRevB.44.1646} {\bibfield  {journal} {\bibinfo
  {journal} {Phys. Rev. B}\ }\textbf {\bibinfo {volume} {44}},\ \bibinfo
  {pages} {1646} (\bibinfo {year} {1991})}\BibitemShut {NoStop}%
\bibitem [{\citenamefont {Latta}\ \emph {et~al.}(2011)\citenamefont {Latta},
  \citenamefont {Srivastava},\ and\ \citenamefont {Imamo\ifmmode~\breve{g}\else
  \u{g}\fi{}lu}}]{Latta2011}%
  \BibitemOpen
  \bibfield  {author} {\bibinfo {author} {\bibfnamefont {C.}~\bibnamefont
  {Latta}}, \bibinfo {author} {\bibfnamefont {A.}~\bibnamefont {Srivastava}},\
  and\ \bibinfo {author} {\bibfnamefont {A.}~\bibnamefont
  {Imamo\ifmmode~\breve{g}\else \u{g}\fi{}lu}},\ }\bibfield  {title} {\bibinfo
  {title} {{Hyperfine Interaction-Dominated Dynamics of Nuclear Spins in
  Self-Assembled InGaAs Quantum Dots}},\ }\href
  {https://doi.org/10.1103/PhysRevLett.107.167401} {\bibfield  {journal}
  {\bibinfo  {journal} {Phys. Rev. Lett.}\ }\textbf {\bibinfo {volume} {107}},\
  \bibinfo {pages} {167401} (\bibinfo {year} {2011})}\BibitemShut {NoStop}%
\bibitem [{\citenamefont {Gillard}\ \emph {et~al.}(2021)\citenamefont
  {Gillard}, \citenamefont {Griffiths}, \citenamefont {Ragunathan},
  \citenamefont {Ulhaq}, \citenamefont {McEwan}, \citenamefont {Clarke},\ and\
  \citenamefont {Chekhovich}}]{Gillard2021}%
  \BibitemOpen
  \bibfield  {author} {\bibinfo {author} {\bibfnamefont {G.}~\bibnamefont
  {Gillard}}, \bibinfo {author} {\bibfnamefont {I.~M.}\ \bibnamefont
  {Griffiths}}, \bibinfo {author} {\bibfnamefont {G.}~\bibnamefont
  {Ragunathan}}, \bibinfo {author} {\bibfnamefont {A.}~\bibnamefont {Ulhaq}},
  \bibinfo {author} {\bibfnamefont {C.}~\bibnamefont {McEwan}}, \bibinfo
  {author} {\bibfnamefont {E.}~\bibnamefont {Clarke}},\ and\ \bibinfo {author}
  {\bibfnamefont {E.~A.}\ \bibnamefont {Chekhovich}},\ }\bibfield  {title}
  {\bibinfo {title} {Fundamental limits of electron and nuclear spin qubit
  lifetimes in an isolated self-assembled quantum dot},\ }\href
  {https://doi.org/10.1038/s41534-021-00378-2} {\bibfield  {journal} {\bibinfo
  {journal} {npj Quantum Inf.}\ }\textbf {\bibinfo {volume} {7}},\ \bibinfo
  {pages} {43} (\bibinfo {year} {2021})}\BibitemShut {NoStop}%
\bibitem [{\citenamefont {Cockins}\ \emph {et~al.}(2010)\citenamefont
  {Cockins}, \citenamefont {Miyahara}, \citenamefont {Bennett}, \citenamefont
  {Clerk}, \citenamefont {Studenikin}, \citenamefont {Poole}, \citenamefont
  {Sachrajda},\ and\ \citenamefont {Grutter}}]{Cockins2010}%
  \BibitemOpen
  \bibfield  {author} {\bibinfo {author} {\bibfnamefont {L.}~\bibnamefont
  {Cockins}}, \bibinfo {author} {\bibfnamefont {Y.}~\bibnamefont {Miyahara}},
  \bibinfo {author} {\bibfnamefont {S.~D.}\ \bibnamefont {Bennett}}, \bibinfo
  {author} {\bibfnamefont {A.~A.}\ \bibnamefont {Clerk}}, \bibinfo {author}
  {\bibfnamefont {S.}~\bibnamefont {Studenikin}}, \bibinfo {author}
  {\bibfnamefont {P.}~\bibnamefont {Poole}}, \bibinfo {author} {\bibfnamefont
  {A.}~\bibnamefont {Sachrajda}},\ and\ \bibinfo {author} {\bibfnamefont
  {P.}~\bibnamefont {Grutter}},\ }\bibfield  {title} {\bibinfo {title} {Energy
  levels of few-electron quantum dots imaged and characterized by atomic force
  microscopy},\ }\href {https://doi.org/10.1073/pnas.0912716107} {\bibfield
  {journal} {\bibinfo  {journal} {Proc. Natl. Acad. Sci. U S A}\ }\textbf
  {\bibinfo {volume} {107}},\ \bibinfo {pages} {9496} (\bibinfo {year}
  {2010})}\BibitemShut {NoStop}%
\bibitem [{\citenamefont {Franceschetti}\ and\ \citenamefont
  {Zunger}(2000)}]{Franceschetti2000}%
  \BibitemOpen
  \bibfield  {author} {\bibinfo {author} {\bibfnamefont {A.}~\bibnamefont
  {Franceschetti}}\ and\ \bibinfo {author} {\bibfnamefont {A.}~\bibnamefont
  {Zunger}},\ }\bibfield  {title} {\bibinfo {title} {Pseudopotential
  calculations of electron and hole addition spectra of {{InAs}}, {{InP}}, and
  {{Si}} quantum dots},\ }\href {https://doi.org/10.1103/PhysRevB.62.2614}
  {\bibfield  {journal} {\bibinfo  {journal} {Phys. Rev. B}\ }\textbf {\bibinfo
  {volume} {62}},\ \bibinfo {pages} {2614} (\bibinfo {year}
  {2000})}\BibitemShut {NoStop}%
\bibitem [{\citenamefont {Vurgaftman}\ \emph {et~al.}(2001)\citenamefont
  {Vurgaftman}, \citenamefont {Meyer},\ and\ \citenamefont
  {Ram-Mohan}}]{Vurgaftman2001}%
  \BibitemOpen
  \bibfield  {author} {\bibinfo {author} {\bibfnamefont {I.}~\bibnamefont
  {Vurgaftman}}, \bibinfo {author} {\bibfnamefont {J.~R.}\ \bibnamefont
  {Meyer}},\ and\ \bibinfo {author} {\bibfnamefont {L.~R.}\ \bibnamefont
  {Ram-Mohan}},\ }\bibfield  {title} {\bibinfo {title} {Band parameters for
  {{III-V}} compound semiconductors and their alloys},\ }\href
  {https://doi.org/10.1063/1.1368156} {\bibfield  {journal} {\bibinfo
  {journal} {J. Appl. Phys.}\ }\textbf {\bibinfo {volume} {89}},\ \bibinfo
  {pages} {5815} (\bibinfo {year} {2001})}\BibitemShut {NoStop}%
\bibitem [{\citenamefont {M\"uller}\ and\ \citenamefont
  {Koonin}(1996)}]{Muller1996}%
  \BibitemOpen
  \bibfield  {author} {\bibinfo {author} {\bibfnamefont {H.-M.}\ \bibnamefont
  {M\"uller}}\ and\ \bibinfo {author} {\bibfnamefont {S.~E.}\ \bibnamefont
  {Koonin}},\ }\bibfield  {title} {\bibinfo {title} {Phase transitions in
  quantum dots},\ }\href {https://doi.org/10.1103/PhysRevB.54.14532} {\bibfield
   {journal} {\bibinfo  {journal} {Phys. Rev. B}\ }\textbf {\bibinfo {volume}
  {54}},\ \bibinfo {pages} {14532} (\bibinfo {year} {1996})}\BibitemShut
  {NoStop}%
\bibitem [{\citenamefont {Tavernier}\ \emph {et~al.}(2003)\citenamefont
  {Tavernier}, \citenamefont {Anisimovas}, \citenamefont {Peeters},
  \citenamefont {Szafran}, \citenamefont {Adamowski},\ and\ \citenamefont
  {Bednarek}}]{Tavernier2003}%
  \BibitemOpen
  \bibfield  {author} {\bibinfo {author} {\bibfnamefont {M.~B.}\ \bibnamefont
  {Tavernier}}, \bibinfo {author} {\bibfnamefont {E.}~\bibnamefont
  {Anisimovas}}, \bibinfo {author} {\bibfnamefont {F.~M.}\ \bibnamefont
  {Peeters}}, \bibinfo {author} {\bibfnamefont {B.}~\bibnamefont {Szafran}},
  \bibinfo {author} {\bibfnamefont {J.}~\bibnamefont {Adamowski}},\ and\
  \bibinfo {author} {\bibfnamefont {S.}~\bibnamefont {Bednarek}},\ }\bibfield
  {title} {\bibinfo {title} {Four-electron quantum dot in a magnetic field},\
  }\href {https://doi.org/10.1103/PhysRevB.68.205305} {\bibfield  {journal}
  {\bibinfo  {journal} {Phys. Rev. B}\ }\textbf {\bibinfo {volume} {68}},\
  \bibinfo {pages} {205305} (\bibinfo {year} {2003})}\BibitemShut {NoStop}%
\bibitem [{\citenamefont {Nishi}\ \emph {et~al.}(2007)\citenamefont {Nishi},
  \citenamefont {Tokura}, \citenamefont {Gupta}, \citenamefont {Austing},\ and\
  \citenamefont {Tarucha}}]{Nishi2007}%
  \BibitemOpen
  \bibfield  {author} {\bibinfo {author} {\bibfnamefont {Y.}~\bibnamefont
  {Nishi}}, \bibinfo {author} {\bibfnamefont {Y.}~\bibnamefont {Tokura}},
  \bibinfo {author} {\bibfnamefont {J.}~\bibnamefont {Gupta}}, \bibinfo
  {author} {\bibfnamefont {G.}~\bibnamefont {Austing}},\ and\ \bibinfo {author}
  {\bibfnamefont {S.}~\bibnamefont {Tarucha}},\ }\bibfield  {title} {\bibinfo
  {title} {Ground-state transitions beyond the singlet-triplet transition for a
  two-electron quantum dot},\ }\href
  {https://doi.org/10.1103/PhysRevB.75.121301} {\bibfield  {journal} {\bibinfo
  {journal} {Phys. Rev. B}\ }\textbf {\bibinfo {volume} {75}},\ \bibinfo
  {pages} {121301} (\bibinfo {year} {2007})}\BibitemShut {NoStop}%
\bibitem [{\citenamefont {Klenovsk{\'y}}\ \emph {et~al.}(2017)\citenamefont
  {Klenovsk{\'y}}, \citenamefont {Steindl},\ and\ \citenamefont
  {Geffroy}}]{Klenovsky2017}%
  \BibitemOpen
  \bibfield  {author} {\bibinfo {author} {\bibfnamefont {P.}~\bibnamefont
  {Klenovsk{\'y}}}, \bibinfo {author} {\bibfnamefont {P.}~\bibnamefont
  {Steindl}},\ and\ \bibinfo {author} {\bibfnamefont {D.}~\bibnamefont
  {Geffroy}},\ }\bibfield  {title} {\bibinfo {title} {Excitonic structure and
  pumping power dependent emission blue-shift of {{type-II}} quantum dots},\
  }\href {https://doi.org/10.1038/srep45568} {\bibfield  {journal} {\bibinfo
  {journal} {Sci. Rep.}\ }\textbf {\bibinfo {volume} {7}},\ \bibinfo {pages}
  {45568} (\bibinfo {year} {2017})}\BibitemShut {NoStop}%
\bibitem [{\citenamefont {Klenovsky}(2025)}]{Klenovsky2025}%
  \BibitemOpen
  \bibfield  {author} {\bibinfo {author} {\bibfnamefont {P.}~\bibnamefont
  {Klenovsky}},\ }\href {https://arxiv.org/abs/2502.00776} {\bibinfo {title}
  {Coulomb correlated multi-particle polarons}} (\bibinfo {year} {2025}),\
  \Eprint {https://arxiv.org/abs/2502.00776} {arXiv:2502.00776} \BibitemShut
  {NoStop}%
\bibitem [{\citenamefont {Dyte}\ \emph {et~al.}(2025)\citenamefont {Dyte},
  \citenamefont {Manna}, \citenamefont {da~Silva}, \citenamefont {Rastelli},\
  and\ \citenamefont {Chekhovich}}]{Dyte2025}%
  \BibitemOpen
  \bibfield  {author} {\bibinfo {author} {\bibfnamefont {H.~E.}\ \bibnamefont
  {Dyte}}, \bibinfo {author} {\bibfnamefont {S.}~\bibnamefont {Manna}},
  \bibinfo {author} {\bibfnamefont {S.~F.~C.}\ \bibnamefont {da~Silva}},
  \bibinfo {author} {\bibfnamefont {A.}~\bibnamefont {Rastelli}},\ and\
  \bibinfo {author} {\bibfnamefont {E.~A.}\ \bibnamefont {Chekhovich}},\ }\href
  {https://arxiv.org/abs/2502.11092} {\bibinfo {title} {Storing quantum
  coherence in a quantum dot nuclear spin ensemble for over 100 milliseconds}}
  (\bibinfo {year} {2025}),\ \Eprint {https://arxiv.org/abs/2502.11092}
  {arXiv:2502.11092 [cond-mat.mes-hall]} \BibitemShut {NoStop}%
\bibitem [{\citenamefont {MacDonald}\ \emph {et~al.}(1993)\citenamefont
  {MacDonald}, \citenamefont {Eric~Yang},\ and\ \citenamefont
  {Johnson}}]{MacDonald1993}%
  \BibitemOpen
  \bibfield  {author} {\bibinfo {author} {\bibfnamefont {A.~H.}\ \bibnamefont
  {MacDonald}}, \bibinfo {author} {\bibfnamefont {S.~R.}\ \bibnamefont
  {Eric~Yang}},\ and\ \bibinfo {author} {\bibfnamefont {M.~D.}\ \bibnamefont
  {Johnson}},\ }\bibfield  {title} {\bibinfo {title} {Quantum dots in strong
  magnetic fields: Stability criteria for the maximum density droplet},\ }\href
  {https://doi.org/10.1071/PH930345} {\bibfield  {journal} {\bibinfo  {journal}
  {Aust. J. Phys.}\ }\textbf {\bibinfo {volume} {46}},\ \bibinfo {pages} {345}
  (\bibinfo {year} {1993})}\BibitemShut {NoStop}%
\bibitem [{\citenamefont {Huang}\ and\ \citenamefont {Hu}(2010)}]{Huang2010}%
  \BibitemOpen
  \bibfield  {author} {\bibinfo {author} {\bibfnamefont {C.-W.}\ \bibnamefont
  {Huang}}\ and\ \bibinfo {author} {\bibfnamefont {X.}~\bibnamefont {Hu}},\
  }\bibfield  {title} {\bibinfo {title} {Theoretical study of nuclear spin
  polarization and depolarization in self-assembled quantum dots},\ }\href
  {https://doi.org/10.1103/PhysRevB.81.205304} {\bibfield  {journal} {\bibinfo
  {journal} {Phys. Rev. B}\ }\textbf {\bibinfo {volume} {81}},\ \bibinfo
  {pages} {205304} (\bibinfo {year} {2010})}\BibitemShut {NoStop}%
\bibitem [{\citenamefont {Bulutay}(2012)}]{Bulutay2012}%
  \BibitemOpen
  \bibfield  {author} {\bibinfo {author} {\bibfnamefont {C.}~\bibnamefont
  {Bulutay}},\ }\bibfield  {title} {\bibinfo {title} {{Quadrupolar spectra of
  nuclear spins in strained In${}_{x}$Ga${}_{1\ensuremath{-}x}$As quantum
  dots}},\ }\href {https://doi.org/10.1103/PhysRevB.85.115313} {\bibfield
  {journal} {\bibinfo  {journal} {Phys. Rev. B}\ }\textbf {\bibinfo {volume}
  {85}},\ \bibinfo {pages} {115313} (\bibinfo {year} {2012})}\BibitemShut
  {NoStop}%
\bibitem [{\citenamefont {Coish}\ \emph {et~al.}(2008)\citenamefont {Coish},
  \citenamefont {Fischer},\ and\ \citenamefont {Loss}}]{Coish2008}%
  \BibitemOpen
  \bibfield  {author} {\bibinfo {author} {\bibfnamefont {W.~A.}\ \bibnamefont
  {Coish}}, \bibinfo {author} {\bibfnamefont {J.}~\bibnamefont {Fischer}},\
  and\ \bibinfo {author} {\bibfnamefont {D.}~\bibnamefont {Loss}},\ }\bibfield
  {title} {\bibinfo {title} {Exponential decay in a spin bath},\ }\href
  {https://doi.org/10.1103/PhysRevB.77.125329} {\bibfield  {journal} {\bibinfo
  {journal} {Phys. Rev. B}\ }\textbf {\bibinfo {volume} {77}},\ \bibinfo
  {pages} {125329} (\bibinfo {year} {2008})}\BibitemShut {NoStop}%
\bibitem [{\citenamefont {Klauser}\ \emph {et~al.}(2008)\citenamefont
  {Klauser}, \citenamefont {Coish},\ and\ \citenamefont {Loss}}]{Klauser2008}%
  \BibitemOpen
  \bibfield  {author} {\bibinfo {author} {\bibfnamefont {D.}~\bibnamefont
  {Klauser}}, \bibinfo {author} {\bibfnamefont {W.~A.}\ \bibnamefont {Coish}},\
  and\ \bibinfo {author} {\bibfnamefont {D.}~\bibnamefont {Loss}},\ }\bibfield
  {title} {\bibinfo {title} {Nuclear spin dynamics and {{Zeno}} effect in
  quantum dots and defect centers},\ }\href
  {https://doi.org/10.1103/PhysRevB.78.205301} {\bibfield  {journal} {\bibinfo
  {journal} {Phys. Rev. B}\ }\textbf {\bibinfo {volume} {78}},\ \bibinfo
  {pages} {205301} (\bibinfo {year} {2008})}\BibitemShut {NoStop}%
\bibitem [{\citenamefont {Reilly}\ \emph {et~al.}(2010)\citenamefont {Reilly},
  \citenamefont {Taylor}, \citenamefont {Petta}, \citenamefont {Marcus},
  \citenamefont {Hanson},\ and\ \citenamefont {Gossard}}]{Reilly2010}%
  \BibitemOpen
  \bibfield  {author} {\bibinfo {author} {\bibfnamefont {D.~J.}\ \bibnamefont
  {Reilly}}, \bibinfo {author} {\bibfnamefont {J.~M.}\ \bibnamefont {Taylor}},
  \bibinfo {author} {\bibfnamefont {J.~R.}\ \bibnamefont {Petta}}, \bibinfo
  {author} {\bibfnamefont {C.~M.}\ \bibnamefont {Marcus}}, \bibinfo {author}
  {\bibfnamefont {M.~P.}\ \bibnamefont {Hanson}},\ and\ \bibinfo {author}
  {\bibfnamefont {A.~C.}\ \bibnamefont {Gossard}},\ }\bibfield  {title}
  {\bibinfo {title} {Exchange control of nuclear spin diffusion in a double
  quantum dot},\ }\href {https://doi.org/10.1103/PhysRevLett.104.236802}
  {\bibfield  {journal} {\bibinfo  {journal} {Phys. Rev. Lett.}\ }\textbf
  {\bibinfo {volume} {104}},\ \bibinfo {pages} {236802} (\bibinfo {year}
  {2010})}\BibitemShut {NoStop}%
\bibitem [{\citenamefont {Gong}\ \emph {et~al.}(2011)\citenamefont {Gong},
  \citenamefont {qi~Yin},\ and\ \citenamefont {Duan}}]{Gong2011}%
  \BibitemOpen
  \bibfield  {author} {\bibinfo {author} {\bibfnamefont {Z.-X.}\ \bibnamefont
  {Gong}}, \bibinfo {author} {\bibfnamefont {Z.}~\bibnamefont {qi~Yin}},\ and\
  \bibinfo {author} {\bibfnamefont {L.-M.}\ \bibnamefont {Duan}},\ }\bibfield
  {title} {\bibinfo {title} {Dynamics of the {{Overhauser}} field under nuclear
  spin diffusion in a quantum dot},\ }\href
  {https://doi.org/10.1088/1367-2630/13/3/033036} {\bibfield  {journal}
  {\bibinfo  {journal} {New J. Phys.}\ }\textbf {\bibinfo {volume} {13}},\
  \bibinfo {pages} {033036} (\bibinfo {year} {2011})}\BibitemShut {NoStop}%
\bibitem [{\citenamefont {Lyanda-Geller}\ \emph {et~al.}(2002)\citenamefont
  {Lyanda-Geller}, \citenamefont {Aleiner},\ and\ \citenamefont
  {Altshuler}}]{LyandaGeller2002}%
  \BibitemOpen
  \bibfield  {author} {\bibinfo {author} {\bibfnamefont {Y.~B.}\ \bibnamefont
  {Lyanda-Geller}}, \bibinfo {author} {\bibfnamefont {I.~L.}\ \bibnamefont
  {Aleiner}},\ and\ \bibinfo {author} {\bibfnamefont {B.~L.}\ \bibnamefont
  {Altshuler}},\ }\bibfield  {title} {\bibinfo {title} {Coulomb ``blockade'' of
  nuclear spin relaxation in quantum dots},\ }\href
  {https://doi.org/10.1103/PhysRevLett.89.107602} {\bibfield  {journal}
  {\bibinfo  {journal} {Phys. Rev. Lett.}\ }\textbf {\bibinfo {volume} {89}},\
  \bibinfo {pages} {107602} (\bibinfo {year} {2002})}\BibitemShut {NoStop}%
\bibitem [{\citenamefont {Lyanda-Geller}\ \emph {et~al.}(2003)\citenamefont
  {Lyanda-Geller}, \citenamefont {Aleiner},\ and\ \citenamefont
  {Altshuler}}]{LyandaGeller2003}%
  \BibitemOpen
  \bibfield  {author} {\bibinfo {author} {\bibfnamefont {Y.~B.}\ \bibnamefont
  {Lyanda-Geller}}, \bibinfo {author} {\bibfnamefont {I.~L.}\ \bibnamefont
  {Aleiner}},\ and\ \bibinfo {author} {\bibfnamefont {B.~L.}\ \bibnamefont
  {Altshuler}},\ }\bibfield  {title} {\bibinfo {title} {Charging effects on
  nuclear spin relaxation in quantum dots},\ }\href
  {https://doi.org/10.1023/A:1025374125559} {\bibfield  {journal} {\bibinfo
  {journal} {J. Supercond.}\ }\textbf {\bibinfo {volume} {16}},\ \bibinfo
  {pages} {751} (\bibinfo {year} {2003})}\BibitemShut {NoStop}%
\bibitem [{\citenamefont {Baruffa}\ \emph {et~al.}(2010)\citenamefont
  {Baruffa}, \citenamefont {Stano},\ and\ \citenamefont
  {Fabian}}]{Baruffa2010}%
  \BibitemOpen
  \bibfield  {author} {\bibinfo {author} {\bibfnamefont {F.}~\bibnamefont
  {Baruffa}}, \bibinfo {author} {\bibfnamefont {P.}~\bibnamefont {Stano}},\
  and\ \bibinfo {author} {\bibfnamefont {J.}~\bibnamefont {Fabian}},\
  }\bibfield  {title} {\bibinfo {title} {Theory of anisotropic exchange in
  laterally coupled quantum dots},\ }\href
  {https://doi.org/10.1103/PhysRevLett.104.126401} {\bibfield  {journal}
  {\bibinfo  {journal} {Phys. Rev. Lett.}\ }\textbf {\bibinfo {volume} {104}},\
  \bibinfo {pages} {126401} (\bibinfo {year} {2010})}\BibitemShut {NoStop}%
\bibitem [{\citenamefont {Camenzind}\ \emph {et~al.}(2018)\citenamefont
  {Camenzind}, \citenamefont {Yu}, \citenamefont {Stano}, \citenamefont
  {Zimmerman}, \citenamefont {Gossard}, \citenamefont {Loss},\ and\
  \citenamefont {Zumb{\"u}hl}}]{Camenzind2018}%
  \BibitemOpen
  \bibfield  {author} {\bibinfo {author} {\bibfnamefont {L.~C.}\ \bibnamefont
  {Camenzind}}, \bibinfo {author} {\bibfnamefont {L.}~\bibnamefont {Yu}},
  \bibinfo {author} {\bibfnamefont {P.}~\bibnamefont {Stano}}, \bibinfo
  {author} {\bibfnamefont {J.~D.}\ \bibnamefont {Zimmerman}}, \bibinfo {author}
  {\bibfnamefont {A.~C.}\ \bibnamefont {Gossard}}, \bibinfo {author}
  {\bibfnamefont {D.}~\bibnamefont {Loss}},\ and\ \bibinfo {author}
  {\bibfnamefont {D.~M.}\ \bibnamefont {Zumb{\"u}hl}},\ }\bibfield  {title}
  {\bibinfo {title} {Hyperfine-phonon spin relaxation in a single-electron
  {{GaAs}} quantum dot},\ }\href {https://doi.org/10.1038/s41467-018-05879-x}
  {\bibfield  {journal} {\bibinfo  {journal} {Nat. Commun.}\ }\textbf {\bibinfo
  {volume} {9}},\ \bibinfo {pages} {3454} (\bibinfo {year} {2018})}\BibitemShut
  {NoStop}%
\bibitem [{\citenamefont {Gillard}\ \emph {et~al.}(2022)\citenamefont
  {Gillard}, \citenamefont {Clarke},\ and\ \citenamefont
  {Chekhovich}}]{Gillard2022}%
  \BibitemOpen
  \bibfield  {author} {\bibinfo {author} {\bibfnamefont {G.}~\bibnamefont
  {Gillard}}, \bibinfo {author} {\bibfnamefont {E.}~\bibnamefont {Clarke}},\
  and\ \bibinfo {author} {\bibfnamefont {E.~A.}\ \bibnamefont {Chekhovich}},\
  }\bibfield  {title} {\bibinfo {title} {Harnessing many-body spin environment
  for long coherence storage and high-fidelity single-shot qubit readout},\
  }\href {https://doi.org/10.1038/s41467-022-31618-4} {\bibfield  {journal}
  {\bibinfo  {journal} {Nat. Commun.}\ }\textbf {\bibinfo {volume} {13}},\
  \bibinfo {pages} {4048} (\bibinfo {year} {2022})}\BibitemShut {NoStop}%
\bibitem [{\citenamefont {Atature}(2006)}]{Atature2006}%
  \BibitemOpen
  \bibfield  {author} {\bibinfo {author} {\bibfnamefont {M.}~\bibnamefont
  {Atature}},\ }\bibfield  {title} {\bibinfo {title} {Quantum-dot spin-state
  preparation with near-unity fidelity},\ }\href
  {https://doi.org/10.1126/science.1126074} {\bibfield  {journal} {\bibinfo
  {journal} {Science}\ }\textbf {\bibinfo {volume} {312}},\ \bibinfo {pages}
  {551} (\bibinfo {year} {2006})}\BibitemShut {NoStop}%
\bibitem [{\citenamefont {Khaetskii}\ and\ \citenamefont
  {Nazarov}(2001)}]{Khaetskii2000}%
  \BibitemOpen
  \bibfield  {author} {\bibinfo {author} {\bibfnamefont {A.~V.}\ \bibnamefont
  {Khaetskii}}\ and\ \bibinfo {author} {\bibfnamefont {Y.~V.}\ \bibnamefont
  {Nazarov}},\ }\bibfield  {title} {\bibinfo {title} {{Spin-flip transitions
  between Zeeman sublevels in semiconductor quantum dots}},\ }\href
  {https://doi.org/10.1103/PhysRevB.64.125316} {\bibfield  {journal} {\bibinfo
  {journal} {Phys. Rev. B}\ }\textbf {\bibinfo {volume} {64}},\ \bibinfo
  {pages} {125316} (\bibinfo {year} {2001})}\BibitemShut {NoStop}%
\bibitem [{\citenamefont {Kroutvar}\ \emph {et~al.}(2004)\citenamefont
  {Kroutvar}, \citenamefont {Ducommun}, \citenamefont {Heiss}, \citenamefont
  {Bichler}, \citenamefont {Schuh}, \citenamefont {Abstreiter},\ and\
  \citenamefont {Finley}}]{Kroutvar2004}%
  \BibitemOpen
  \bibfield  {author} {\bibinfo {author} {\bibfnamefont {M.}~\bibnamefont
  {Kroutvar}}, \bibinfo {author} {\bibfnamefont {Y.}~\bibnamefont {Ducommun}},
  \bibinfo {author} {\bibfnamefont {D.}~\bibnamefont {Heiss}}, \bibinfo
  {author} {\bibfnamefont {M.}~\bibnamefont {Bichler}}, \bibinfo {author}
  {\bibfnamefont {D.}~\bibnamefont {Schuh}}, \bibinfo {author} {\bibfnamefont
  {G.}~\bibnamefont {Abstreiter}},\ and\ \bibinfo {author} {\bibfnamefont
  {J.~J.}\ \bibnamefont {Finley}},\ }\bibfield  {title} {\bibinfo {title}
  {Optically programmable electron spin memory using semiconductor quantum
  dots},\ }\href {https://doi.org/10.1038/nature03008} {\bibfield  {journal}
  {\bibinfo  {journal} {Nature}\ }\textbf {\bibinfo {volume} {432}},\ \bibinfo
  {pages} {81} (\bibinfo {year} {2004})}\BibitemShut {NoStop}%
\bibitem [{\citenamefont {Lu}\ \emph {et~al.}(2010)\citenamefont {Lu},
  \citenamefont {Zhao}, \citenamefont {Vamivakas}, \citenamefont {Matthiesen},
  \citenamefont {F\"alt}, \citenamefont {Badolato},\ and\ \citenamefont
  {Atat\"ure}}]{Lu2010}%
  \BibitemOpen
  \bibfield  {author} {\bibinfo {author} {\bibfnamefont {C.-Y.}\ \bibnamefont
  {Lu}}, \bibinfo {author} {\bibfnamefont {Y.}~\bibnamefont {Zhao}}, \bibinfo
  {author} {\bibfnamefont {A.~N.}\ \bibnamefont {Vamivakas}}, \bibinfo {author}
  {\bibfnamefont {C.}~\bibnamefont {Matthiesen}}, \bibinfo {author}
  {\bibfnamefont {S.}~\bibnamefont {F\"alt}}, \bibinfo {author} {\bibfnamefont
  {A.}~\bibnamefont {Badolato}},\ and\ \bibinfo {author} {\bibfnamefont
  {M.}~\bibnamefont {Atat\"ure}},\ }\bibfield  {title} {\bibinfo {title}
  {Direct measurement of spin dynamics in {{InAs/GaAs}} quantum dots using
  time-resolved resonance fluorescence},\ }\href
  {https://doi.org/10.1103/PhysRevB.81.035332} {\bibfield  {journal} {\bibinfo
  {journal} {Phys. Rev. B}\ }\textbf {\bibinfo {volume} {81}},\ \bibinfo
  {pages} {035332} (\bibinfo {year} {2010})}\BibitemShut {NoStop}%
\bibitem [{\citenamefont {Ulhaq}\ \emph {et~al.}(2016)\citenamefont {Ulhaq},
  \citenamefont {Duan}, \citenamefont {Zallo}, \citenamefont {Ding},
  \citenamefont {Schmidt}, \citenamefont {Tartakovskii}, \citenamefont
  {Skolnick},\ and\ \citenamefont {Chekhovich}}]{Ulhaq2016}%
  \BibitemOpen
  \bibfield  {author} {\bibinfo {author} {\bibfnamefont {A.}~\bibnamefont
  {Ulhaq}}, \bibinfo {author} {\bibfnamefont {Q.}~\bibnamefont {Duan}},
  \bibinfo {author} {\bibfnamefont {E.}~\bibnamefont {Zallo}}, \bibinfo
  {author} {\bibfnamefont {F.}~\bibnamefont {Ding}}, \bibinfo {author}
  {\bibfnamefont {O.~G.}\ \bibnamefont {Schmidt}}, \bibinfo {author}
  {\bibfnamefont {A.~I.}\ \bibnamefont {Tartakovskii}}, \bibinfo {author}
  {\bibfnamefont {M.~S.}\ \bibnamefont {Skolnick}},\ and\ \bibinfo {author}
  {\bibfnamefont {E.~A.}\ \bibnamefont {Chekhovich}},\ }\bibfield  {title}
  {\bibinfo {title} {Vanishing electron $g$ factor and long-lived nuclear spin
  polarization in weakly strained nanohole-filled {{GaAs/AlGaAs}} quantum
  dots},\ }\href {https://doi.org/10.1103/PhysRevB.93.165306} {\bibfield
  {journal} {\bibinfo  {journal} {Phys. Rev. B}\ }\textbf {\bibinfo {volume}
  {93}},\ \bibinfo {pages} {165306} (\bibinfo {year} {2016})}\BibitemShut
  {NoStop}%
\bibitem [{\citenamefont {Koppens}\ \emph {et~al.}(2006)\citenamefont
  {Koppens}, \citenamefont {Buizert}, \citenamefont {Tielrooij}, \citenamefont
  {Vink}, \citenamefont {Nowack}, \citenamefont {Meunier}, \citenamefont
  {Kouwenhoven},\ and\ \citenamefont {Vandersypen}}]{Koppens2006}%
  \BibitemOpen
  \bibfield  {author} {\bibinfo {author} {\bibfnamefont {F.~H.~L.}\
  \bibnamefont {Koppens}}, \bibinfo {author} {\bibfnamefont {C.}~\bibnamefont
  {Buizert}}, \bibinfo {author} {\bibfnamefont {K.~J.}\ \bibnamefont
  {Tielrooij}}, \bibinfo {author} {\bibfnamefont {I.~T.}\ \bibnamefont {Vink}},
  \bibinfo {author} {\bibfnamefont {K.~C.}\ \bibnamefont {Nowack}}, \bibinfo
  {author} {\bibfnamefont {T.}~\bibnamefont {Meunier}}, \bibinfo {author}
  {\bibfnamefont {L.~P.}\ \bibnamefont {Kouwenhoven}},\ and\ \bibinfo {author}
  {\bibfnamefont {L.~M.~K.}\ \bibnamefont {Vandersypen}},\ }\bibfield  {title}
  {\bibinfo {title} {Driven coherent oscillations of a single electron spin in
  a quantum dot},\ }\href {https://doi.org/10.1038/nature05065} {\bibfield
  {journal} {\bibinfo  {journal} {Nature}\ }\textbf {\bibinfo {volume} {442}},\
  \bibinfo {pages} {766} (\bibinfo {year} {2006})}\BibitemShut {NoStop}%
\bibitem [{\citenamefont {Shafiei}\ \emph {et~al.}(2013)\citenamefont
  {Shafiei}, \citenamefont {Nowack}, \citenamefont {Reichl}, \citenamefont
  {Wegscheider},\ and\ \citenamefont {Vandersypen}}]{Shafiei2013}%
  \BibitemOpen
  \bibfield  {author} {\bibinfo {author} {\bibfnamefont {M.}~\bibnamefont
  {Shafiei}}, \bibinfo {author} {\bibfnamefont {K.~C.}\ \bibnamefont {Nowack}},
  \bibinfo {author} {\bibfnamefont {C.}~\bibnamefont {Reichl}}, \bibinfo
  {author} {\bibfnamefont {W.}~\bibnamefont {Wegscheider}},\ and\ \bibinfo
  {author} {\bibfnamefont {L.~M.~K.}\ \bibnamefont {Vandersypen}},\ }\bibfield
  {title} {\bibinfo {title} {Resolving spin-orbit- and hyperfine-mediated
  electric dipole spin resonance in a quantum dot},\ }\href
  {https://doi.org/10.1103/PhysRevLett.110.107601} {\bibfield  {journal}
  {\bibinfo  {journal} {Phys. Rev. Lett.}\ }\textbf {\bibinfo {volume} {110}},\
  \bibinfo {pages} {107601} (\bibinfo {year} {2013})}\BibitemShut {NoStop}%

\setcounter{firstbib}{\value{NAT@ctr}}
\end{thebibliography}

\begin{thebibliography}{34}%
\makeatletter
\providecommand \@ifxundefined [1]{%
 \@ifx{#1\undefined}
}%
\providecommand \@ifnum [1]{%
 \ifnum #1\expandafter \@firstoftwo
 \else \expandafter \@secondoftwo
 \fi
}%
\providecommand \@ifx [1]{%
 \ifx #1\expandafter \@firstoftwo
 \else \expandafter \@secondoftwo
 \fi
}%
\providecommand \natexlab [1]{#1}%
\providecommand \enquote  [1]{``#1''}%
\providecommand \bibnamefont  [1]{#1}%
\providecommand \bibfnamefont [1]{#1}%
\providecommand \citenamefont [1]{#1}%
\providecommand \href@noop [0]{\@secondoftwo}%
\providecommand \href [0]{\begingroup \@sanitize@url \@href}%
\providecommand \@href[1]{\@@startlink{#1}\@@href}%
\providecommand \@@href[1]{\endgroup#1\@@endlink}%
\providecommand \@sanitize@url [0]{\catcode `\\12\catcode `\$12\catcode
  `\&12\catcode `\#12\catcode `\^12\catcode `\_12\catcode `\%12\relax}%
\providecommand \@@startlink[1]{}%
\providecommand \@@endlink[0]{}%
\providecommand \url  [0]{\begingroup\@sanitize@url \@url }%
\providecommand \@url [1]{\endgroup\@href {#1}{\urlprefix }}%
\providecommand \urlprefix  [0]{URL }%
\providecommand \Eprint [0]{\href }%
\providecommand \doibase [0]{https://doi.org/}%
\providecommand \selectlanguage [0]{\@gobble}%
\providecommand \bibinfo  [0]{\@secondoftwo}%
\providecommand \bibfield  [0]{\@secondoftwo}%
\providecommand \translation [1]{[#1]}%
\providecommand \BibitemOpen [0]{}%
\providecommand \bibitemStop [0]{}%
\providecommand \bibitemNoStop [0]{.\EOS\space}%
\providecommand \EOS [0]{\spacefactor3000\relax}%
\providecommand \BibitemShut  [1]{\csname bibitem#1\endcsname}%
\let\auto@bib@innerbib\@empty


\setcounter{NAT@ctr}{\value{firstbib}}


\bibitem [{\citenamefont {Oshiyama}\ and\ \citenamefont
  {Ohnishi}(1986)}]{Oshiyama1986}%
  \BibitemOpen
  \bibfield  {author} {\bibinfo {author} {\bibfnamefont {A.}~\bibnamefont
  {Oshiyama}}\ and\ \bibinfo {author} {\bibfnamefont {S.}~\bibnamefont
  {Ohnishi}},\ }\bibfield  {title} {\bibinfo {title} {{DX center: Crossover of
  deep and shallow states in Si-doped
  ${\mathrm{Al}}_{\mathrm{x}}$ ${\mathrm{Ga}}_{1\mathrm{\ensuremath{-}}\mathrm{x}}$As}},\
  }\href {https://doi.org/10.1103/PhysRevB.33.4320} {\bibfield  {journal}
  {\bibinfo  {journal} {Phys. Rev. B}\ }\textbf {\bibinfo {volume} {33}},\
  \bibinfo {pages} {4320} (\bibinfo {year} {1986})}\BibitemShut {NoStop}%
\bibitem [{\citenamefont {Mooney}(1990)}]{Mooney1990}%
  \BibitemOpen
  \bibfield  {author} {\bibinfo {author} {\bibfnamefont {P.~M.}\ \bibnamefont
  {Mooney}},\ }\bibfield  {title} {\bibinfo {title} {{Deep donor levels (DX
  centers) in III-V semiconductors}},\ }\href
  {https://doi.org/10.1063/1.345628} {\bibfield  {journal} {\bibinfo  {journal}
  {J. Appl. Phys.}\ }\textbf {\bibinfo {volume} {67}},\ \bibinfo {pages} {R1}
  (\bibinfo {year} {1990})}\BibitemShut {NoStop}%
\bibitem [{\citenamefont {Bloch}(1946)}]{Bloch1946}%
  \BibitemOpen
  \bibfield  {author} {\bibinfo {author} {\bibfnamefont {F.}~\bibnamefont
  {Bloch}},\ }\bibfield  {title} {\bibinfo {title} {Nuclear induction},\ }\href
  {https://doi.org/10.1103/PhysRev.70.460} {\bibfield  {journal} {\bibinfo
  {journal} {Phys. Rev.}\ }\textbf {\bibinfo {volume} {70}},\ \bibinfo {pages}
  {460} (\bibinfo {year} {1946})}\BibitemShut {NoStop}%
\bibitem [{\citenamefont {Checkhovich}\ \emph {et~al.}(2013)\citenamefont
  {Checkhovich}, \citenamefont {Glazov}, \citenamefont {Krysa}, \citenamefont
  {Hopkinson}, \citenamefont {Senellart}, \citenamefont {Lema\^{i}tre},
  \citenamefont {Skolnick},\ and\ \citenamefont
  {Tartakovskii}}]{Chekhovich2013NatPhys}%
  \BibitemOpen
  \bibfield  {author} {\bibinfo {author} {\bibfnamefont {E.~A.}\ \bibnamefont
  {Checkhovich}}, \bibinfo {author} {\bibfnamefont {M.~M.}\ \bibnamefont
  {Glazov}}, \bibinfo {author} {\bibfnamefont {A.~B.}\ \bibnamefont {Krysa}},
  \bibinfo {author} {\bibfnamefont {M.}~\bibnamefont {Hopkinson}}, \bibinfo
  {author} {\bibfnamefont {P.}~\bibnamefont {Senellart}}, \bibinfo {author}
  {\bibfnamefont {A.}~\bibnamefont {Lema\^{i}tre}}, \bibinfo {author}
  {\bibfnamefont {M.~S.}\ \bibnamefont {Skolnick}},\ and\ \bibinfo {author}
  {\bibfnamefont {A.~I.}\ \bibnamefont {Tartakovskii}},\ }\bibfield  {title}
  {\bibinfo {title} {Element-sensitive measurement of the hole-nuclear spin
  interaction in quantum dots},\ }\href {https://doi.org/10.1038/nphys2514}
  {\bibfield  {journal} {\bibinfo  {journal} {Nat. Phys.}\ }\textbf {\bibinfo
  {volume} {9}},\ \bibinfo {pages} {74} (\bibinfo {year} {2013})}\BibitemShut
  {NoStop}%
\bibitem [{\citenamefont {Gammon}\ \emph {et~al.}(2001)\citenamefont {Gammon},
  \citenamefont {Efros}, \citenamefont {Kennedy}, \citenamefont {Rosen},
  \citenamefont {Katzer}, \citenamefont {Park}, \citenamefont {Brown},
  \citenamefont {Korenev},\ and\ \citenamefont {Merkulov}}]{Gammon2001}%
  \BibitemOpen
  \bibfield  {author} {\bibinfo {author} {\bibfnamefont {D.}~\bibnamefont
  {Gammon}}, \bibinfo {author} {\bibfnamefont {A.~L.}\ \bibnamefont {Efros}},
  \bibinfo {author} {\bibfnamefont {T.~A.}\ \bibnamefont {Kennedy}}, \bibinfo
  {author} {\bibfnamefont {M.}~\bibnamefont {Rosen}}, \bibinfo {author}
  {\bibfnamefont {D.~S.}\ \bibnamefont {Katzer}}, \bibinfo {author}
  {\bibfnamefont {D.}~\bibnamefont {Park}}, \bibinfo {author} {\bibfnamefont
  {S.~W.}\ \bibnamefont {Brown}}, \bibinfo {author} {\bibfnamefont {V.~L.}\
  \bibnamefont {Korenev}},\ and\ \bibinfo {author} {\bibfnamefont {I.~A.}\
  \bibnamefont {Merkulov}},\ }\bibfield  {title} {\bibinfo {title} {Electron
  and nuclear spin interactions in the optical spectra of single {{GaAs}}
  quantum dots},\ }\href {https://doi.org/10.1103/PhysRevLett.86.5176}
  {\bibfield  {journal} {\bibinfo  {journal} {Phys. Rev. Lett.}\ }\textbf
  {\bibinfo {volume} {86}},\ \bibinfo {pages} {5176} (\bibinfo {year}
  {2001})}\BibitemShut {NoStop}%
\bibitem [{\citenamefont {Eble}\ \emph {et~al.}(2006)\citenamefont {Eble},
  \citenamefont {Krebs}, \citenamefont {Lema{\^i}tre}, \citenamefont {Kowalik},
  \citenamefont {Kudelski}, \citenamefont {Voisin}, \citenamefont {Urbaszek},
  \citenamefont {Marie},\ and\ \citenamefont {Amand}}]{Eble2006}%
  \BibitemOpen
  \bibfield  {author} {\bibinfo {author} {\bibfnamefont {B.}~\bibnamefont
  {Eble}}, \bibinfo {author} {\bibfnamefont {O.}~\bibnamefont {Krebs}},
  \bibinfo {author} {\bibfnamefont {A.}~\bibnamefont {Lema{\^i}tre}}, \bibinfo
  {author} {\bibfnamefont {K.}~\bibnamefont {Kowalik}}, \bibinfo {author}
  {\bibfnamefont {A.}~\bibnamefont {Kudelski}}, \bibinfo {author}
  {\bibfnamefont {P.}~\bibnamefont {Voisin}}, \bibinfo {author} {\bibfnamefont
  {B.}~\bibnamefont {Urbaszek}}, \bibinfo {author} {\bibfnamefont
  {X.}~\bibnamefont {Marie}},\ and\ \bibinfo {author} {\bibfnamefont
  {T.}~\bibnamefont {Amand}},\ }\bibfield  {title} {\bibinfo {title} {{Dynamic
  nuclear polarization of a single charge-tunable InAs/GaAs quantum dot}},\
  }\href {https://doi.org/10.1103/PhysRevB.74.081306} {\bibfield  {journal}
  {\bibinfo  {journal} {Phys. Rev. B}\ }\textbf {\bibinfo {volume} {74}},\
  \bibinfo {pages} {081306} (\bibinfo {year} {2006})}\BibitemShut {NoStop}%
\bibitem [{\citenamefont {Skiba-Szymanska}\ \emph {et~al.}(2008)\citenamefont
  {Skiba-Szymanska}, \citenamefont {Chekhovich}, \citenamefont {Nikolaenko},
  \citenamefont {Tartakovskii}, \citenamefont {Makhonin}, \citenamefont
  {Drouzas}, \citenamefont {Skolnick},\ and\ \citenamefont
  {Krysa}}]{Skiba2008}%
  \BibitemOpen
  \bibfield  {author} {\bibinfo {author} {\bibfnamefont {J.}~\bibnamefont
  {Skiba-Szymanska}}, \bibinfo {author} {\bibfnamefont {E.~A.}\ \bibnamefont
  {Chekhovich}}, \bibinfo {author} {\bibfnamefont {A.~E.}\ \bibnamefont
  {Nikolaenko}}, \bibinfo {author} {\bibfnamefont {A.~I.}\ \bibnamefont
  {Tartakovskii}}, \bibinfo {author} {\bibfnamefont {M.~N.}\ \bibnamefont
  {Makhonin}}, \bibinfo {author} {\bibfnamefont {I.}~\bibnamefont {Drouzas}},
  \bibinfo {author} {\bibfnamefont {M.~S.}\ \bibnamefont {Skolnick}},\ and\
  \bibinfo {author} {\bibfnamefont {A.~B.}\ \bibnamefont {Krysa}},\ }\bibfield
  {title} {\bibinfo {title} {Overhauser effect in individual
  {{$\mathrm{In}\mathrm{P}/{\mathrm{Ga}}_{x}{\mathrm{In}}_{1\ensuremath{-}x}\mathrm{P}$}}
  dots},\ }\href {https://doi.org/10.1103/PhysRevB.77.165338} {\bibfield
  {journal} {\bibinfo  {journal} {Phys. Rev. B}\ }\textbf {\bibinfo {volume}
  {77}},\ \bibinfo {pages} {165338} (\bibinfo {year} {2008})}\BibitemShut
  {NoStop}%
\bibitem [{\citenamefont {Ragunathan}\ \emph {et~al.}(2019)\citenamefont
  {Ragunathan}, \citenamefont {Kobak}, \citenamefont {Gillard}, \citenamefont
  {Pacuski}, \citenamefont {Sobczak}, \citenamefont {Borysiuk}, \citenamefont
  {Skolnick},\ and\ \citenamefont {Chekhovich}}]{Ragunathan2019}%
  \BibitemOpen
  \bibfield  {author} {\bibinfo {author} {\bibfnamefont {G.}~\bibnamefont
  {Ragunathan}}, \bibinfo {author} {\bibfnamefont {J.}~\bibnamefont {Kobak}},
  \bibinfo {author} {\bibfnamefont {G.}~\bibnamefont {Gillard}}, \bibinfo
  {author} {\bibfnamefont {W.}~\bibnamefont {Pacuski}}, \bibinfo {author}
  {\bibfnamefont {K.}~\bibnamefont {Sobczak}}, \bibinfo {author} {\bibfnamefont
  {J.}~\bibnamefont {Borysiuk}}, \bibinfo {author} {\bibfnamefont {M.~S.}\
  \bibnamefont {Skolnick}},\ and\ \bibinfo {author} {\bibfnamefont {E.~A.}\
  \bibnamefont {Chekhovich}},\ }\bibfield  {title} {\bibinfo {title} {Direct
  measurement of hyperfine shifts and radio frequency manipulation of nuclear
  spins in individual $\mathrm{CdTe}/\mathrm{ZnTe}$ quantum dots},\ }\href
  {https://doi.org/10.1103/PhysRevLett.122.096801} {\bibfield  {journal}
  {\bibinfo  {journal} {Phys. Rev. Lett.}\ }\textbf {\bibinfo {volume} {122}},\
  \bibinfo {pages} {096801} (\bibinfo {year} {2019})}\BibitemShut {NoStop}%
\bibitem [{\citenamefont {Chekhovich}\ \emph {et~al.}(2017)\citenamefont
  {Chekhovich}, \citenamefont {Ulhaq}, \citenamefont {Zallo}, \citenamefont
  {Ding}, \citenamefont {Schmidt},\ and\ \citenamefont
  {Skolnick}}]{Chekhovich2017}%
  \BibitemOpen
  \bibfield  {author} {\bibinfo {author} {\bibfnamefont {E.~A.}\ \bibnamefont
  {Chekhovich}}, \bibinfo {author} {\bibfnamefont {A.}~\bibnamefont {Ulhaq}},
  \bibinfo {author} {\bibfnamefont {E.}~\bibnamefont {Zallo}}, \bibinfo
  {author} {\bibfnamefont {F.}~\bibnamefont {Ding}}, \bibinfo {author}
  {\bibfnamefont {O.~G.}\ \bibnamefont {Schmidt}},\ and\ \bibinfo {author}
  {\bibfnamefont {M.~S.}\ \bibnamefont {Skolnick}},\ }\bibfield  {title}
  {\bibinfo {title} {Measurement of the spin temperature of optically cooled
  nuclei and {{GaAs}} hyperfine constants in {{GaAs/AlGaAs}} quantum dots},\
  }\href {https://doi.org/10.1038/nmat4959} {\bibfield  {journal} {\bibinfo
  {journal} {Nat. Mater.}\ }\textbf {\bibinfo {volume} {16}},\ \bibinfo {pages}
  {982} (\bibinfo {year} {2017})}\BibitemShut {NoStop}%
\bibitem [{\citenamefont {Wang}\ and\ \citenamefont {Takeda}(2021)}]{Wang2021}%
  \BibitemOpen
  \bibfield  {author} {\bibinfo {author} {\bibfnamefont {Y.}~\bibnamefont
  {Wang}}\ and\ \bibinfo {author} {\bibfnamefont {K.}~\bibnamefont {Takeda}},\
  }\bibfield  {title} {\bibinfo {title} {Speedup of nuclear spin diffusion in
  hyperpolarized solids},\ }\href {https://doi.org/10.1088/1367-2630/ac0d6e}
  {\bibfield  {journal} {\bibinfo  {journal} {New J. Phys.}\ }\textbf {\bibinfo
  {volume} {23}},\ \bibinfo {pages} {073015} (\bibinfo {year}
  {2021})}\BibitemShut {NoStop}%
\bibitem [{\citenamefont {Chekhovich}\ \emph {et~al.}(2012)\citenamefont
  {Chekhovich}, \citenamefont {Kavokin}, \citenamefont {Puebla}, \citenamefont
  {Krysa}, \citenamefont {Hopkinson}, \citenamefont {Andreev}, \citenamefont
  {Sanchez}, \citenamefont {Beanland}, \citenamefont {Skolnick},\ and\
  \citenamefont {Tartakovskii}}]{Chekhovich2012}%
  \BibitemOpen
  \bibfield  {author} {\bibinfo {author} {\bibfnamefont {E.~A.}\ \bibnamefont
  {Chekhovich}}, \bibinfo {author} {\bibfnamefont {K.~V.}\ \bibnamefont
  {Kavokin}}, \bibinfo {author} {\bibfnamefont {J.}~\bibnamefont {Puebla}},
  \bibinfo {author} {\bibfnamefont {A.~B.}\ \bibnamefont {Krysa}}, \bibinfo
  {author} {\bibfnamefont {M.}~\bibnamefont {Hopkinson}}, \bibinfo {author}
  {\bibfnamefont {A.~D.}\ \bibnamefont {Andreev}}, \bibinfo {author}
  {\bibfnamefont {A.~M.}\ \bibnamefont {Sanchez}}, \bibinfo {author}
  {\bibfnamefont {R.}~\bibnamefont {Beanland}}, \bibinfo {author}
  {\bibfnamefont {M.~S.}\ \bibnamefont {Skolnick}},\ and\ \bibinfo {author}
  {\bibfnamefont {A.~I.}\ \bibnamefont {Tartakovskii}},\ }\bibfield  {title}
  {\bibinfo {title} {{Structural analysis of strained quantum dots using
  nuclear magnetic resonance}},\ }\href
  {https://doi.org/10.1038/nnano.2012.142} {\bibfield  {journal} {\bibinfo
  {journal} {Nat. Nanotechnol.}\ }\textbf {\bibinfo {volume} {7}},\ \bibinfo
  {pages} {646} (\bibinfo {year} {2012})}\BibitemShut {NoStop}%
\bibitem [{\citenamefont {Chekhovich}\ \emph {et~al.}(2018)\citenamefont
  {Chekhovich}, \citenamefont {Griffiths}, \citenamefont {Skolnick},
  \citenamefont {Huang}, \citenamefont {Covre~da Silva}, \citenamefont {Yuan},\
  and\ \citenamefont {Rastelli}}]{Chekhovich2018}%
  \BibitemOpen
  \bibfield  {author} {\bibinfo {author} {\bibfnamefont {E.~A.}\ \bibnamefont
  {Chekhovich}}, \bibinfo {author} {\bibfnamefont {I.~M.}\ \bibnamefont
  {Griffiths}}, \bibinfo {author} {\bibfnamefont {M.~S.}\ \bibnamefont
  {Skolnick}}, \bibinfo {author} {\bibfnamefont {H.}~\bibnamefont {Huang}},
  \bibinfo {author} {\bibfnamefont {S.~F.}\ \bibnamefont {Covre~da Silva}},
  \bibinfo {author} {\bibfnamefont {X.}~\bibnamefont {Yuan}},\ and\ \bibinfo
  {author} {\bibfnamefont {A.}~\bibnamefont {Rastelli}},\ }\bibfield  {title}
  {\bibinfo {title} {Cross calibration of deformation potentials and
  gradient-elastic tensors of {{GaAs}} using photoluminescence and nuclear
  magnetic resonance spectroscopy in {{GaAs/AlGaAs}} quantum dot structures},\
  }\href {https://doi.org/10.1103/PhysRevB.97.235311} {\bibfield  {journal}
  {\bibinfo  {journal} {Phys. Rev. B}\ }\textbf {\bibinfo {volume} {97}},\
  \bibinfo {pages} {235311} (\bibinfo {year} {2018})}\BibitemShut {NoStop}%
\bibitem [{\citenamefont {Birner}\ \emph {et~al.}(2007)\citenamefont {Birner},
  \citenamefont {Zibold}, \citenamefont {Andlauer}, \citenamefont {Kubis},
  \citenamefont {Sabathil}, \citenamefont {Trellakis},\ and\ \citenamefont
  {Vogl}}]{Birner2007}%
  \BibitemOpen
  \bibfield  {author} {\bibinfo {author} {\bibfnamefont {S.}~\bibnamefont
  {Birner}}, \bibinfo {author} {\bibfnamefont {T.}~\bibnamefont {Zibold}},
  \bibinfo {author} {\bibfnamefont {T.}~\bibnamefont {Andlauer}}, \bibinfo
  {author} {\bibfnamefont {T.}~\bibnamefont {Kubis}}, \bibinfo {author}
  {\bibfnamefont {M.}~\bibnamefont {Sabathil}}, \bibinfo {author}
  {\bibfnamefont {A.}~\bibnamefont {Trellakis}},\ and\ \bibinfo {author}
  {\bibfnamefont {P.}~\bibnamefont {Vogl}},\ }\bibfield  {title} {\bibinfo
  {title} {{Nextnano: General purpose 3-D simulations}},\ }\href
  {https://doi.org/10.1109/TED.2007.902871} {\bibfield  {journal} {\bibinfo
  {journal} {IEEE Transactions on Electron Devices}\ }\textbf {\bibinfo
  {volume} {54}},\ \bibinfo {pages} {2137} (\bibinfo {year}
  {2007})}\BibitemShut {NoStop}%
\bibitem [{\citenamefont {Schliwa}\ \emph {et~al.}(2009)\citenamefont
  {Schliwa}, \citenamefont {Winkelnkemper},\ and\ \citenamefont
  {Bimberg}}]{Schliwa:09}%
  \BibitemOpen
  \bibfield  {author} {\bibinfo {author} {\bibfnamefont {A.}~\bibnamefont
  {Schliwa}}, \bibinfo {author} {\bibfnamefont {M.}~\bibnamefont
  {Winkelnkemper}},\ and\ \bibinfo {author} {\bibfnamefont {D.}~\bibnamefont
  {Bimberg}},\ }\bibfield  {title} {\bibinfo {title} {Few-particle energies
  versus geometry and composition of
  {{${\text{In}}_{x}{\text{Ga}}_{1\ensuremath{-}x}\text{As}/\text{GaAs}$}}
  self-organized quantum dots},\ }\href
  {https://doi.org/10.1103/PhysRevB.79.075443} {\bibfield  {journal} {\bibinfo
  {journal} {Phys. Rev. B}\ }\textbf {\bibinfo {volume} {79}},\ \bibinfo
  {pages} {075443} (\bibinfo {year} {2009})}\BibitemShut {NoStop}%
\bibitem [{\citenamefont {Stier}(2000)}]{Stier2000}%
  \BibitemOpen
  \bibfield  {author} {\bibinfo {author} {\bibfnamefont {O.}~\bibnamefont
  {Stier}},\ }\href@noop {} {Ph.D. thesis},\ \bibinfo  {school} {Technische
  Universit{\"a}t Berlin,} (\bibinfo {year} {2000})\BibitemShut {NoStop}%
\end{thebibliography}

\pagebreak


\renewcommand{\thesection}{Supplementary Section \arabic{section}}
\setcounter{section}{0}
\renewcommand{\thefigure}{\arabic{figure}}
\renewcommand{\figurename}{Supplementary Figure}
\setcounter{figure}{0}
\renewcommand{\theequation}{S\arabic{equation}}
\setcounter{equation}{0}
\renewcommand{\thetable}{\arabic{table}}
\renewcommand{\tablename}{Supplementary Table}
\setcounter{table}{0}

\makeatletter
\def\l@subsubsection#1#2{}
\makeatother

\pagebreak \pagenumbering{arabic}
\newpage


\section*{Supplemental Material}

\section{Sample structure}
\label{sec:Sample}

The sample structure used in this work is the same semiconductor wafer that was used previously in Refs.~\cite{MillingtonHotze2022,Zaporski2022,MillingtonHotze2023,Dyte2023,Dyte2025}. The sample is grown using molecular beam epitaxy (MBE) on a semi-insulating GaAs (001) substrate. The layer sequence of the semiconductor structure is shown in Supplementary Fig.~\ref{Fig:SSample}. The growth starts with a layer of Al$_{0.95}$Ga$_{0.05}$As followed by a single pair of Al$_{0.2}$Ga$_{0.8}$As and Al$_{0.95}$Ga$_{0.05}$As layers acting as a Bragg reflector in optical experiments. Then, a 95~nm thick layer of Al$_{0.15}$Ga$_{0.85}$As is grown, followed by a 95~nm thick layer of Al$_{0.15}$Ga$_{0.85}$As doped with Si at a volume concentration of $1.0\times10^{18}$~cm$^{-3}$. The low Al concentration of $0.15$ in the Si doped layer mitigates the issues caused by the deep DX centers \cite{Oshiyama1986,Mooney1990,Zhai2020}. The $n$-type doped layer is followed by the electron tunnel barrier layers: first, a 5~nm thick Al$_{0.15}$Ga$_{0.85}$As layer is grown at a reduced temperature of $560$~$^{\circ}$C to suppress Si segregation, followed by a 10~nm thick Al$_{0.15}$Ga$_{0.85}$As layer and then a 15~nm thick Al$_{0.33}$Ga$_{0.67}$As layer grown at $600$~$^{\circ}$C. Aluminium droplets are grown on the surface of the Al$_{0.33}$Ga$_{0.67}$As layer and are used to etch the nanoholes \cite{Heyn2009,Atkinson2012}. Atomic force microscopy shows that typical nanoholes have a depth of $\approx6.5$~nm and are $\approx70$~nm in diameter \cite{MillingtonHotze2022}. Next, a 2.1~nm thick layer of GaAs is grown to form QDs by infilling the nanoholes as well as to form the quantum well (QW) layer. Thus, the maximum height of the QDs in the growth $z$ direction is $\approx9$~nm. The GaAs layer is followed by a 268~nm thick Al$_{0.33}$Ga$_{0.67}$As barrier layer. Finally, the $p$-type contact layers doped with C are grown: a 65~nm thick layer of Al$_{0.15}$Ga$_{0.85}$As with a $5\times10^{18}$~cm$^{-3}$ doping concentration, followed by a 5~nm thick layer of Al$_{0.15}$Ga$_{0.85}$As with a $9\times10^{18}$~cm$^{-3}$ concentration, and a 10~nm thick layer of GaAs with a $9\times10^{18}$~cm$^{-3}$ concentration.

\begin{figure}
\centering
\includegraphics[width=0.53\linewidth]{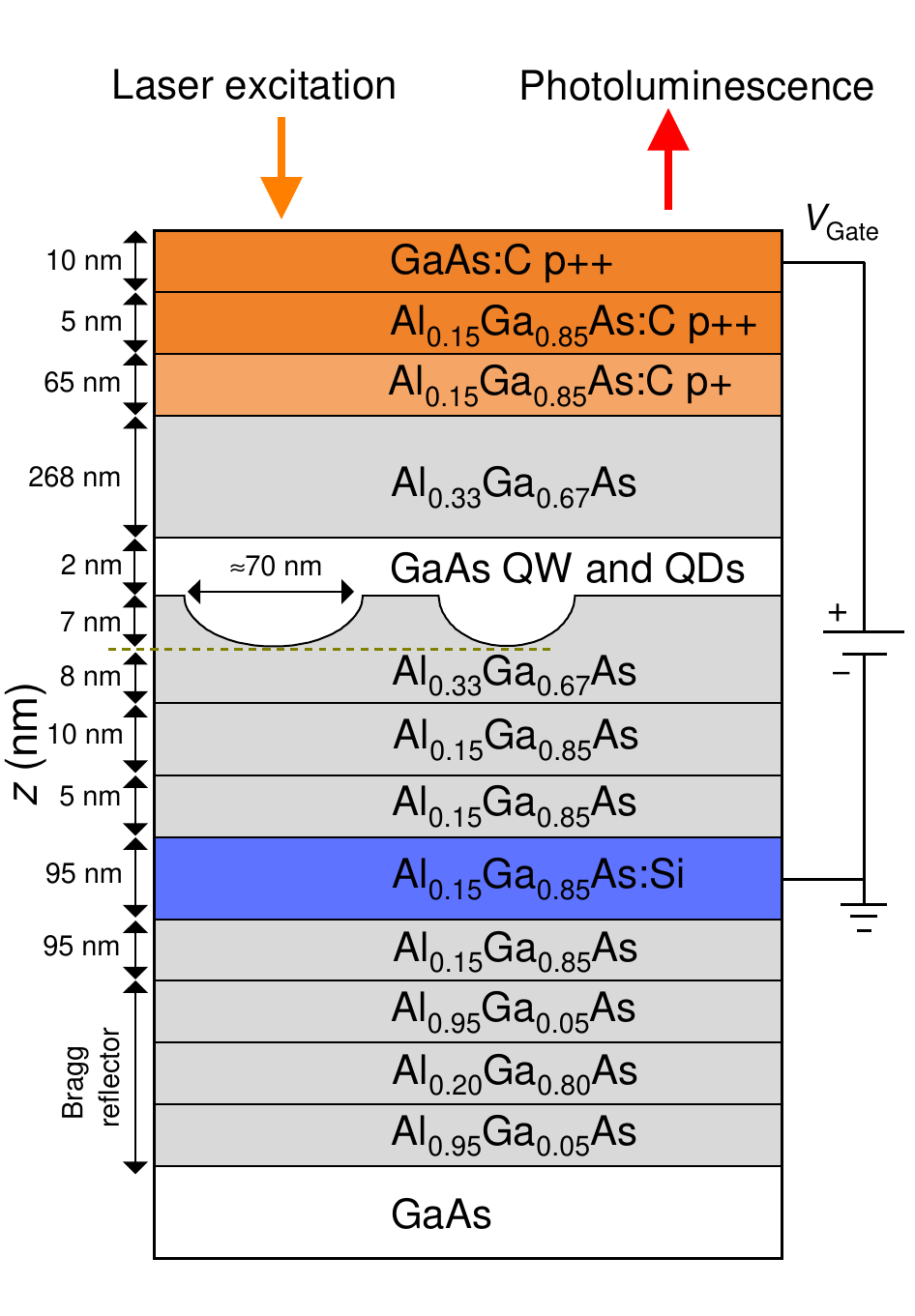}
\caption{\label{Fig:SSample} {Schematic of the quantum dot sample structure.} }
\end{figure}

The sample is processed into a $p$-$i$-$n$ diode structure. Mesa
structures with a height of 250~nm are formed by etching away the
$p$-doped layers and depositing layers of metal on the etched areas in the following order: Ni(10~nm), AuGe(150~nm), Ni(40~nm), and Au(100~nm). The sample is then annealed to
enable diffusion of metal atoms down to the $n$-doped layer to form the ohmic
back contact. The top gate contact is formed by depositing Ti(15~nm)/Au(100~nm) on to the $p$-type surface of the mesa areas. QD photoluminescence (PL) is excited and collected through the top of the sample. The
sample gate bias $V_{\rm{Gate}}$ is the bias of the $p$-type top
contact with respect to the grounded $n$-type back contact. Due to the
large thickness of the top Al$_{0.33}$Ga$_{0.67}$As layer, the tunneling of holes is suppressed, whereas tunnel
coupling to the $n$-type layer enables deterministic charging of the QDs with electrons by changing $V_{\rm{Gate}}$. 

Most of experimental work is conducted on a sample fabricated as discussed above. In this sample, QDs are subject to natural elastic strain on the order of $\approx0.03\%$ arising from GaAs and AlGaAs lattice mismatch. The nuclear quadrupolar shifts arising from such strain are on the order of tens of kHz, leading to an overlap of the nuclear magnetic resonance (NMR) spectral components. In order to resolve the quadrupolar NMR components, we use a separate piece of the same semiconductor sample subject to a uniaxial mechanical stress. To this end, the semiconductor wafer is first cleaved into a small piece with a rectangular surface area of 0.7~mm $\times$ 2.35~mm. The edges of the rectangular profile are aligned along the $[110]$ and $[1\bar{1}0]$ crystallographic directions. The thickness of the sample along the $[001]$ growth direction is 0.35~mm. Thus, the sample is shaped as a parallelepiped. The sample is then inserted into a home-made stress cell. This is done in such a way that the two 0.7~mm $\times$ 0.35~mm surfaces of the sample are contacted to the flat titanium surfaces of the stress cell bracket. A titanium screw is then used to apply compressive stress, directed along the 2.35~mm long edge of the sample.

\section{Experimental techniques}
\label{sec:ExpTechn}

\subsection{Experimental setup}

The sample is placed in a liquid helium bath cryostat. A
superconducting coil is used to apply magnetic field up to
$B_{\rm{z}}=10$~T. The field is parallel to the sample growth
direction and the optical axis $z$ (Faraday geometry). We use
confocal microscopy configuration. An aspheric lens with a focal
distance of 1.45~mm and NA=0.58 is used as an objective for
optical excitation of the QD and for PL
collection. The excitation laser is focused into a spot
with a diameter of $\approx1~\mu$m. The collected PL is dispersed
in a two-stage grating spectrometer, each stage with a 0.85~m
focal length, and recorded with a charge-coupled device (CCD)
camera. The changes in the spectral splitting of a negatively
charged trion $X^-$, derived from the PL spectra, are used to
measure the hyperfine shifts $E_{\rm{hf}}$ proportional to the
nuclear spin polarization degree.

\begin{figure}
\centering
\includegraphics[width=0.95\linewidth]{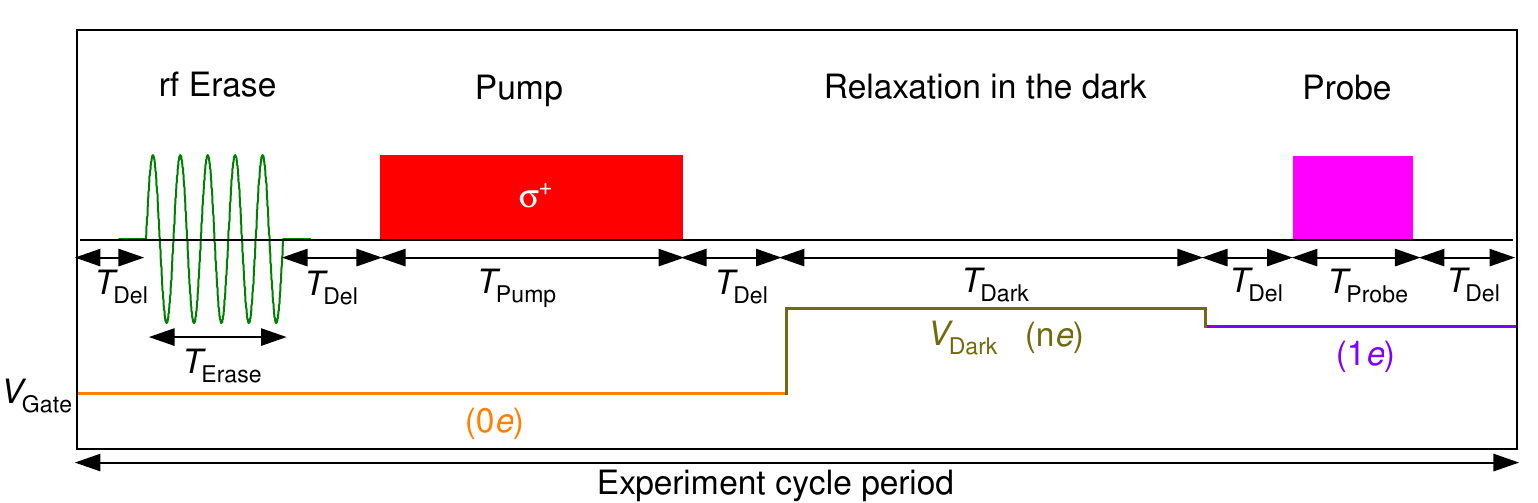}
\caption{\label{Fig:STDiag} {Detailed timing diagram of the nuclear spin relaxation measurement cycle.} }
\end{figure}

Supplementary Fig.~\ref{Fig:STDiag} is a detailed version of Fig.~1(c) of the main text and shows the timing of the nuclear spin relaxation (NSR) measurement. In what follows we describe the individual elements of the timing sequence. While this discussion is specific to the NSR measurement, the same principles apply to other time-resolved measurements.

\subsection{Radio frequency depolarization of nuclear spins}

Investigation of NSR and spin diffusion relies on the ability to prepare a reproducible state of the nuclear spin
polarization. This is achieved with a radio frequency (rf) erase
pulse (Supplementary Fig.~\ref{Fig:STDiag}) which effectively resets the nuclear spin polarization to zero in the entire sample. This is achieved by
saturating the nuclear magnetic resonance of the As and Ga
isotopes. When subject to an oscillating magnetic field, resonant with the nuclear Larmor frequency, the nuclear spins undergo Rabi rotation, periodically transitioning between the spin states parallel and antiparallel to the external magnetic field \citep{Bloch1946}. Due to the nuclear-nuclear dipolar interactions each nuclear spin is subject to a local field. The randomness of these local fields perturbs the Rabi precession frequencies, resulting in ensemble dephasing. Consequently, the nuclei become randomly oriented (depolarized) after a long resonant rf saturation pulse. The required oscillating magnetic field $B_{\rm{x}}\perp
z$ is produced by a copper wire coil placed at a distance of $\approx0.5$~mm from the QD sample. The coil is made of 10 turns of a 0.1~mm diameter enameled copper wire wound on a $\approx0.4$~mm diameter spool in 5 layers, with 2 turns in each layer. The axis of the spiral winding of the coil is perpendicular to the $z$ axis. The coil is driven by a class-A rf amplifier (rated up to 20~W) which is fed by the output of an arbitrary waveform generator. The spectrum of the rf
excitation consists of three bands, each 340~kHz wide and centered
on the NMR frequency of the corresponding As or Ga isotope. For each magnetic field the frequencies are adjusted based on NMR spectroscopy. To give a specific example, the central frequencies at 10~T are 73.079, 102.471 and 130.199~MHz for $^{75}$As,
$^{69}$Ga and $^{71}$Ga, respectively.  Each
rf band is generated as a frequency comb
\citep{Chekhovich2013NatPhys} with a mode spacing of 120~Hz, much
smaller than the homogeneous NMR linewidth. The rf power density
in the comb is chosen to be low enough and the rf pulse duration $T_{\rm{Erase}}$
long enough (ranging between 0.1 and 10~s depending on magnetic
field) to achieve noncoherent exponential depolarization of the
nuclear spin ensemble.

\subsection{Optical pumping of the quantum dot nuclear spins}

Optical pumping of the QD nuclear spin polarization (labeled Pump in Supplementary Fig.~\ref{Fig:STDiag}) is achieved using the emission of a 690~nm circularly polarized diode laser, which is resonant with the GaAs QW states. Optical dynamical nuclear spin polarization is a well known process, that has been observed in many types of QDs \citep{Gammon2001,Eble2006,Skiba2008,Ulhaq2016,Ragunathan2019}, see Ref. \citep{Urbaszek2013} for a review. In brief, dynamic nuclear polarization is a three-stage cyclic process. At the first stage a spin-polarized electron is created optically. This is made possible by the selection rules, which allow conversion of the circularly polarized photons into spin-polarized electron-hole pairs in group III-V semiconductors. At the second stage, the electron exchanges its spin with one of the nuclei through the flip-flop term of the electron-nuclear hyperfine Hamiltonian. The third stage is the electron-hole optical recombination, which removes the flipped electron. This final step is required in order to let the QD accept new spin-polarized electrons and continue polarizing the ensemble of $\approx 10^5$ nuclear spins of the QD. Given that optical pumping is resonant with the QW, it is possible that dynamical nuclear polarization takes place not only in the QDs but also in the adjacent parts of the QW. On the other
hand, the pump laser photon energy is well below the bandgap of
the AlGaAs barriers. For that reason we assume that dynamic
nuclear polarization in AlGaAs is induced only through spin
diffusion from the GaAs layer of the QW and QDs. During the optical pump
the sample gate is set to a large reverse bias, typically
$V_{\rm{Gate}}=-2$~V. The pump power is $\approx300~\mu$W, which
is two orders of magnitude higher than the ground-state PL
saturation power. The resulting hyperfine shifts do not exceed
$|E_{\rm{hf}}|<50~\mu$eV, corresponding to initial nuclear spin
polarization degree within $|P_{\rm{N},0}|\lesssim0.45$. While
polarization as high as $P_{\rm{N},0}\approx0.95$ is possible
\citep{Chekhovich2017,MillingtonHotze2023}, we deliberately use lower values to ensure linear regime of NSR and spin diffusion, free from hyperpolarization regime corrections \citep{Wang2021,MillingtonHotze2023}. Moreover, high nuclear spin polarization degrees require resonant optical pumping of the QD. The use of a non-resonant pump laser, greatly simplifies the experiment by removing the need to search for the optical resonance for each individual QD and at each value of the external magnetic field.

\subsection{Optical probing of the quantum dot nuclear spins}

\begin{figure}
\includegraphics[width=0.8\linewidth]{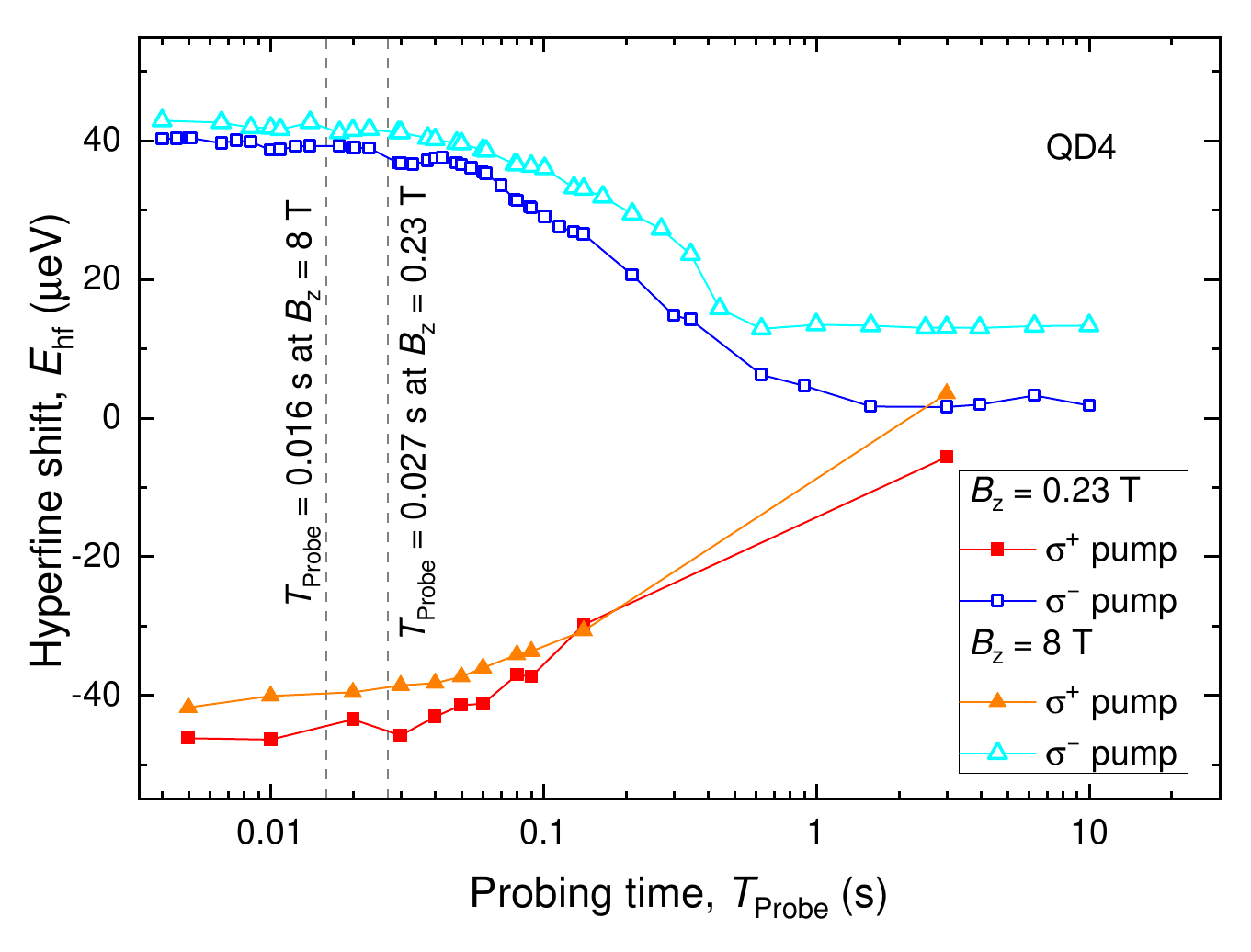}
\caption{\label{Fig:SProbe} {\bf{Calibration of the optical probing of the QD nuclear spin polarization.}} Hyperfine shift measured as a function of the probing time $T_{\rm{Probe}}$ following a $\sigma^+$ or $\sigma^-$ pumping of the nuclear spin polarization in a QD. Results are shown for $B_{\rm{z}}=0.23$~T (squares) and $B_{\rm{z}}=8$~T (circles) Vertical dashed lines shows the $T_{\rm{Probe}}$ value chosen for the NSR measurements on this individual QD at the particular magnetic fields.}
\end{figure}

For optical probing of the nuclear spin polarization we use a diode laser emitting at 640~nm. For most magnetic fields the sample forward bias, typically $V_{\rm{Probe}}=+0.5$~V, and the probe power are
chosen to maximize (saturate) PL intensity of the ground state
$X^-$ trion. At low magnetic fields $B_{\rm{z}}\lesssim 0.4$~T the PL spectral splitting of $X^-$ is not resolved, and we use PL of the $X^{3-}$ trion charged with 3 electrons. Fig.~1(d) of the main text shows $X^-$ PL probe spectra measured at $B_{\rm{z}}=0.5$~T. The
difference in spectral splitting of the $X^-$ trion doublet
reveals the hyperfine shifts $E_{\rm{hf}}$. These shifts are
used to monitor the average QD nuclear spin polarization in NSR
experiments such as shown in Fig.~1(e) of the main text. Illumination with a probe laser inevitably acts back on the nuclear spin polarization. An example of the probe pulse calibration is shown in Supplementary Fig.~\ref{Fig:SProbe}. In this experiment the QD is first pumped with a $\sigma^+$ or $\sigma^-$ polarized laser in order to create large initial nuclear polarization. Then a probe laser pulse is applied. The hyperfine shift $E_{\rm{hf}}$ is measured from PL spectroscopy at the end of this probe. Such calibration is carried out for each individual QD at each magnetic field. It can be seen that the probe induces decay of the nuclear spin polarization on a timescale of a few hundred milliseconds. The probe time $T_{\rm{Probe}}$ used in the NSR experiments is chosen to ensure minimal distortion of the measured $E_{\rm{hf}}$. For example, for the data shown in Supplementary Fig.~\ref{Fig:SProbe} we choose $T_{\rm{Probe}}=0.027$~s at $B_{\rm{z}}=0.23$~T and $T_{\rm{Probe}}=0.016$~s at $B_{\rm{z}}=8$~T. Typical $T_{\rm{Probe}}$ values range between 10 and 80 ms, depending on individual QD and magnetic field.

\subsection{Pump probe experiment implementation}

Optical pump and probe pulses are formed by mechanical shutters.
The switching time and the uncertainty in that switching time are both in the range of a few milliseconds. In order to accommodate these shutter transients, small delays $T_{\rm{Del}}=10~$ms are introduced in the timing sequences as shown in Supplementary Fig.~\ref{Fig:STDiag}. Under certain regimes in $B_{\rm{z}}$ and $V_{\rm{Gate}}$ (e.g. resonant cotunnelling with the Fermi reservoir) this $T_{\rm{Del}}$ is comparable to the
nuclear spin relaxation times $T_{\rm{1,N}}$. However, the
relaxation time in an empty (0$e$) or singly charged (1$e$) QD is always considerably
longer. Thus, during the shutter switching delay the QD is kept under either the 0$e$ bias (after the pump) or the 1$e$ bias (prior to the probe). The dark time $T_{\rm{Dark}}$ is implemented by closing both the pump and probe shutters and pulsing the gate bias to the chosen dark-state value
$V_{\rm{Gate}}$ for a duration $T_{\rm{Dark}}$. The QD device responds to the bias on a sub-microsecond scale. This way we ensure that the switching delays $T_{\rm{Del}}$ have minimal effect on the measured NSR dynamics and are also able to implement an effective $T_{\rm{Dark}}$ shorter than the response time of the shutter.

\subsection{Model fitting of the NSR bias spectra}

The NSR peaks, such as observed in Fig.~2(a) of the main text, can be modeled in the same way that transport through QDs is treated \citep{Kouwenhoven2001}. The key difference is that here it is the ``spin current'' that gives rise to NSR, rather than actual charge current through QD. Resonant electron cotunnelling requires that the electron Fermi reservoir of the $n$-type doped GaAs has both the filled states at energy $\mu_n$, which can provide electrons to tunnel into the QD, and unfilled states at the same energy $\mu_n$, which can accept electrons tunneling back from the QD. Thus, for each individual peak, the rate of the nuclear spin transfer can be modelled as $\propto
f(1-f)$, where $f=1/(1+e^{(\mu_n-E_{\rm{F}})/k_{\rm{B}}T})$ is the
Fermi-Dirac distribution of the electronic energies in the Fermi reservoir, $T$ is the temperature, $E_{\rm{F}}$ is the Fermi energy, and $k_{\rm{B}}$ is the Boltzmann constant \citep{Cockins2010,Beenakker1991}.

The electrochemical potentials $\mu_n$ of the $n$-electron states in the QD are tuned with respect to $E_{\rm{F}}$ by varying the gate bias $V_{\rm{G}}$. Thus, the cotunnelling contribution of each NSR peak can be modelled as $\frac{2\Gamma_{\rm{R}}^{(ne)}}{1+\cosh{\left(\frac{V_{\rm{G}}-V_{\rm{R}}^{(ne)}}{w_{\rm{n}}}\right)}}$, where $w_{\rm{n}}$ is the width, $V_{\rm{R}}$ is the resonant bias, and $\Gamma_{\rm{R}}^{(ne)}$ is the maximum NSR rate of the resonant peak between the $n-1$ and $n$ electron states of the QD. 

Cotunnelling coupling between the QD few-electron state and the electron Fermi reservoir is not the only NSR mechanism. Between the resonant cotunelling peaks (i.e. at the plateaus) NSR is dominated by other mechanisms, such as nuclear spin diffusion. We denote these NSR rates as $\Gamma_{\rm{N}}^{(ne)}$ and treat them as bias-independent. The overall dependence of the NSR rate on the sample gate bias is then modelled as:
\begin{equation}
\sum_{n=1}\frac{2\Gamma_{\rm{R}}^{(ne)}}{1+\cosh{\left(\frac{V_{\rm{G}}-V_{\rm{R}}^{(ne)}}{w_{\rm{n}}}\right)}}+\sum_{n=0}\Gamma_{\rm{N}}^{(ne)}p_n,    
\end{equation}
where $p_n$ is the probability for the QD to be occupied by $n$ resonant electrons and is calculated from the same Fermi-Dirac distribution as follows. The probability for a QD to be occupied with $n=0$ electrons is
\begin{equation}
p_0=\left(1-\frac{1}{1+e^{-(V_{\rm{G}}-V_{\rm{R}}^{(1e)})/w_{\rm{1}}}}\right)
\end{equation}
The probability for $n=1$ electron is
\begin{equation}
p_1=\left(\frac{1}{1+e^{-(V_{\rm{G}}-V_{\rm{R}}^{(1e)})/w_{\rm{1}}})}\right)\left(1-\frac{1}{1+e^{-(V_{\rm{G}}-V_{\rm{R}}^{(2e)})/w_{\rm{2}}}}\right)
\end{equation}
For $n=2$ electrons the probability is
\begin{equation}
p_2=\left(\frac{1}{1+e^{-(V_{\rm{G}}-V_{\rm{R}}^{(1e)})/w_{\rm{1}}})}\right)\left(\frac{1}{1+e^{-(V_{\rm{G}}-V_{\rm{R}}^{(2e)})/w_{\rm{2}}})}\right)\left(1-\frac{1}{1+e^{-(V_{\rm{G}}-V_{\rm{R}}^{(3e)})/w_{\rm{3}}}}\right),
\end{equation}
and so on.

The values of $\Gamma_{\rm{N}}^{(ne)}$, $\Gamma_{\rm{R}}^{(ne)}$, and $w_n$ are used as fitting parameters. Examples of the fitted curves are shown by the solid lines in Figs.~2(a) and 3(a) of the main text.

\section{Nuclear magnetic resonance of individual quantum dots}
\label{sec:NMR}

\subsection{Inverse NMR spectroscopy}

Nuclear magnetic resonance characterization
is conducted using the inverse NMR method
\citep{Chekhovich2012} which enhances the signal for $I>1/2$ spins
and improves the signal to noise ratio. In this method the nuclear spins are first polarized with a pump laser ($T_{\rm{Pump}}=6.5$~s)
and are then depolarized by a weak rf field, whose spectral profile is a
broadband frequency comb with a narrow gap of width
$w_{\rm{gap}}$ in the center. The frequency comb has a total spectral width of 600~kHz and its mode spacing is 125~Hz. The value of $w_{\rm{gap}}$ controls the balance
between the measured NMR signal and the spectral resolution. In an empty QD
(0$e$), NMR spectra of As and Ga measured with
$w_{\rm{gap}}=6$~kHz consist of well-resolved quadrupolar-split
triplets, consistent with previous observations for these \citep{MillingtonHotze2022} and similar QD
structures \citep{Ulhaq2016,Chekhovich2018}. The unpaired spin of a few electron configuration ($ne$) leads to inhomogeneous Knight shifts. The electron spin lifetime, which is on the order of
milliseconds, is typically shorter than the rf burst (typically 0.18~s) used in inverse NMR. The dynamic spectral Knight shifts disrupt the enhancement of
the inverse NMR method, as described previously \citep{MillingtonHotze2022}. For that reason, we employ pulsed NMR spectroscopy techniques described next.

\subsection{Pulsed NMR spectroscopy}

Pulsed NMR measurements were conducted to probe the distributions of the hyperfine Knight shifts for various electron number configurations (see Fig.~5 of the main text). The oscillating magnetic field $B_{\rm{x}}~\perp~z$ required to perform pulsed NMR is produced by a similar copper wire coil, but the driving power is provided by a class-AB high-power rf amplifier (Tomco BT01000-AlphaSA rated up to 1000~W). The pulsed NMR measurements were conducted on the uniaxially stressed sample, where the quadrupolar-split NMR transitions are well resolved.

The measurement cycle used to obtain the data in Fig.~5 of the main text is shown in its basic form in the inset of that figure. Here we discuss additional technical details. First, the nuclei are polarized by a circularly polarized optical pump ($T_{\rm{Pump}}~=~3.5-5.5~$s). After this we perform transfer of nuclear spin state populations in order to improve the optically detected NMR signal. Consider the case where the nuclei are pumped using $\sigma^{+}$ polarized light then the resulting population distribution leaves $I_{\rm{z}}=-3/2$ as the most populated state, while the other states are depopulated. This creates a problem if the actual measurement is to be performed on the -1/2$\leftrightarrow$+1/2 NMR transition. As the NMR signal is related to relative population change, we wish to maximize the population difference between the $I_{\rm{z}}=-1/2,+1/2$ states. To achieve this we apply rf $\pi$ pulses to two of the NMR transitions. First, a $\pi$ pulse is applied to the $-3/2\leftrightarrow-1/2$ transition leaving $I_{\rm{z}}=-1/2$ as the most populated state, then a $\pi$ pulse is applied to the $+1/2\leftrightarrow-+3/2$ transition leaving $I_{\rm{z}}=+1/2$ as the least populated state. This way we achieve the largest possible population difference for the states with $I_{\rm{z}}=-1/2,+1/2$. The same process can be applied to maximize the population difference for any pair of the nuclear spin states. Both the optical pumping and population transfer are conducted with a large reverse gate bias $V_{\rm{G}}~=~-2.4$~V, which empties the QD. This ensures that the NMR transitions are not distorted by the electron knight field, maximizing the rotation fidelity. A delay of 1~ms is added between the population transfer pulses to allow time for dephasing of any unwanted nuclear spin coherence, which may arise from imperfect pulse rotations.

Once the nuclear spin state is prepared through optical pumping and population transfer, the gate bias is changed to match the charging plateau of the desired charge state, causing a set number of electrons to tunnel into the QD from the Fermi reservoir. The electrons are then left to equilibrate for a time $T_{\rm{Load}}=30-90$~ms to ensure that any transient effects decay before the measurement. A single rf pulse with a raised-cosine envelope or a sequence of such rf pulses is then applied, depending on the desired type of the NMR measurement. For the measurements shown in Fig.~5(a) of the main text we use free induction decay (FID) spectroscopy, where two short ($T_{\rm{RF}}=10~\mu$s, full width at half maximum (FWHM) is 5~$\mu$s) $\pi/2$ pulses are applied, separated by a free evolution delay $T_{\rm{FreeEvol}}$. The decay of the nuclear polarization is measured as a function of $T_{\rm{FreeEvol}}$ with the second pulse set either in phase or in quadrature with the first pulse. A phase-sensitive Fourier transform is then performed to derive the NMR spectral lineshape. The exception in Fig.~5(a) of the main text is the 4$e$ spectrum at $B_{\rm{z}}=0.98$~T, recorded by varying the frequency offset of a single rf pulse. In this case the pulse is $T_{\rm{RF}}=80~\mu$s long (FWHM is 40~$\mu$s) and its amplitude is reduced to perform a $\pi/2$ rotation, which reduces the distortion of the measured NMR lineshape. In case of Fig.~5(b) of the main text, all measurements are performed by scanning the frequency offset of a single $\pi$-rotation rf pulse with duration $T_{\rm{RF}}=40~\mu$s (FWHM is 20~$\mu$s). In these experiments with $1e$ and $3e$ configurations the NMR measurements are performed in single-shot regime, and the histograms are built to reconstruct the NMR lineshapes \citep{Dyte2025}. Once the rf pulse sequence is finished, the electron and nuclei are left to evolve for time $T_{\rm{Dwell}}=20-100~$ms, which is much longer than the nuclear spin dephasing time $T_{2,\rm{N}}^{*} \approx50-250~\mu$s \cite{Chekhovich2020}. This ensures dissipation of any unwanted coherence before the readout of the longitudinal nuclear spin polarization. 

Following the rf pulse sequence, the sample gate bias is changed to unload the electron(s) from the dot. A reverse transfer of nuclear spin population is then performed to enhance the optically detected NMR signal. Using the example from above, if the $-1/2\leftrightarrow +1/2$ NMR transition is measured, $\pi$ pulses are applied to the $-3/2\leftrightarrow-1/2$ and $+1/2\leftrightarrow-+3/2$ transitions. This moves the population difference from the $I_{\rm{z}}=-1/2,+1/2$ states to the $I_{\rm{z}}=-3/2,+3/2$, which triples the NMR signal measured via the hyperfine coupling to the optically excited electron spins.

After the reverse population transfer a short $T_{\rm{Probe}}=40-50~$ms optical probe pulse is applied to readout the final nuclear spin polarization. During the optical readout the bias across the sample is set to obtain the desired exciton and to maximize the PL signal. The QD PL induced by the probe is collected and the change in the nuclear spin longitudinal polarization is measured through changes in the PL spectral doublet splitting (see Fig.~1(d) of the main text). In most NMR experiments the measurements are differential, where we compare the PL spectral splitting to that measured without the rf pulses.

\subsection{Electron spin readout via single-shot NMR}

The readout of the electron spin $z$-projection, needed to obtain the $T_{1,e}$ times, is carried out using detuned rf $\pi$ pulses applied to the nuclear spins. Here we give a brief summary of the measurement principles, a full description can be found in Ref.~\cite{Dyte2023}. As seen from Fig.~5(b) in the main text, introducing one or three electrons to the dot causes a Knight shift in the NMR frequencies, with the direction of this shift dependent on the electron spin state. When detuned rf pulses are applied to a nuclear spin transition, the detected values of the final nuclear spin polarization (the NMR signals) display a clear bimodal distribution \cite{Dyte2023}. In one of the modes there is a substantial NMR signal, which corresponds to the case where the electron spin projection (e.g. $s_{\rm{z}}=+1/2$) Knight shifts the nuclear spins into resonance with the detuned rf pulse. The other mode is characterized by a very small NMR signal, which corresponds to the opposite electron spin projection (e.g. $s_{\rm{z}}=-1/2$) where the nuclei are detuned even further from the rf pulse by the Knight shifts. By counting the number of readouts in each mode, the average electron spin polarization can be obtained from the ratio of the two values. Electron spin polarizations measured in this way are shown in Fig.~6(f) of the main text.


\section{Supplementary results}
\label{sec:SupRes}

\subsection{NSR at high magnetic fields in the 5$e$ configuration}

\begin{figure}[h]
\includegraphics[width=0.75\linewidth]{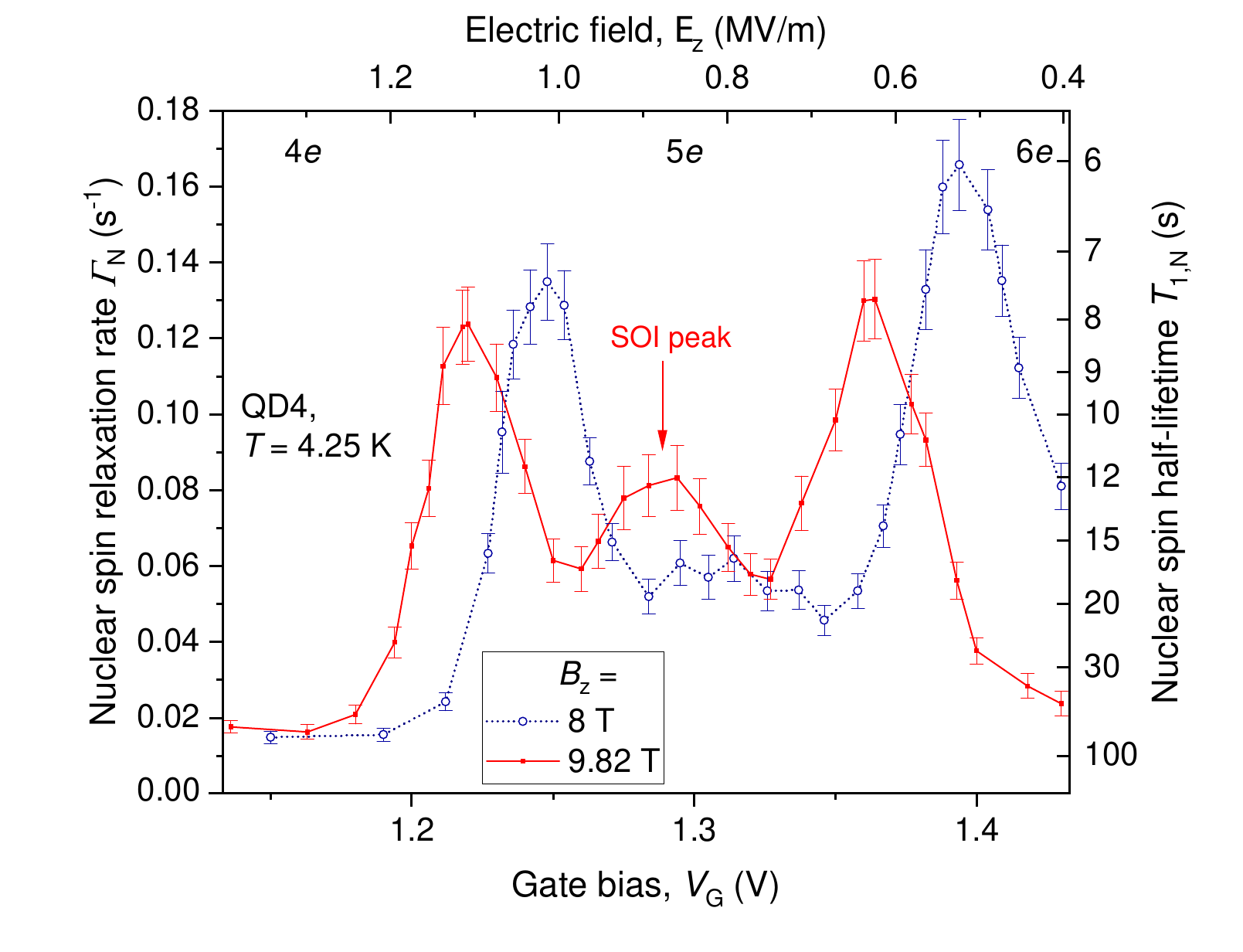}
\caption{\label{Fig:5eSOI} {\bf{Nuclear spin relaxation in the 5$e$ configuration.}} NSR rate as a function of the gate bias measured at $B_{\rm{z}}=8$~T (open symbols) and $B_{\rm{z}}=9.82$~T (solid symbols).}
\end{figure}

Supplementary Fig.~\ref{Fig:5eSOI} shows the NSR rates measured around the 5$e$ Coulomb blockade regime. At $B_{\rm{z}}=8$~T (open symbols), the NSR dependence consists of the two resonant tunneling peaks and a relatively flat plateau in the middle. This is a typical structure observed for different Coulomb blockade plateaus in different individual QDs. When the field is increased to the maximum possible of $B_{\rm{z}}=9.82$~T (solid symbols) an extra peak, of the same width as the resonant tunneling peaks emerges in the center of the plateau. Such a peculiar structure has been predicted theoretically \citep{LyandaGeller2002,LyandaGeller2003}, and ascribed to the spin-orbit interaction (SOI). Experimentally, we observe this additional peak only at one value of the external field, so that our attribution to SOI is tentative. Further experimental studies would benefit from stronger magnetic fields, lower temperatures, and higher electron addition energies, which would facilitate observation of the peak-in-the-plateau structure.

\subsection{Absolute nuclear spin relaxation rates}

\begin{figure}[ht!]
\includegraphics[width=1.0\linewidth]{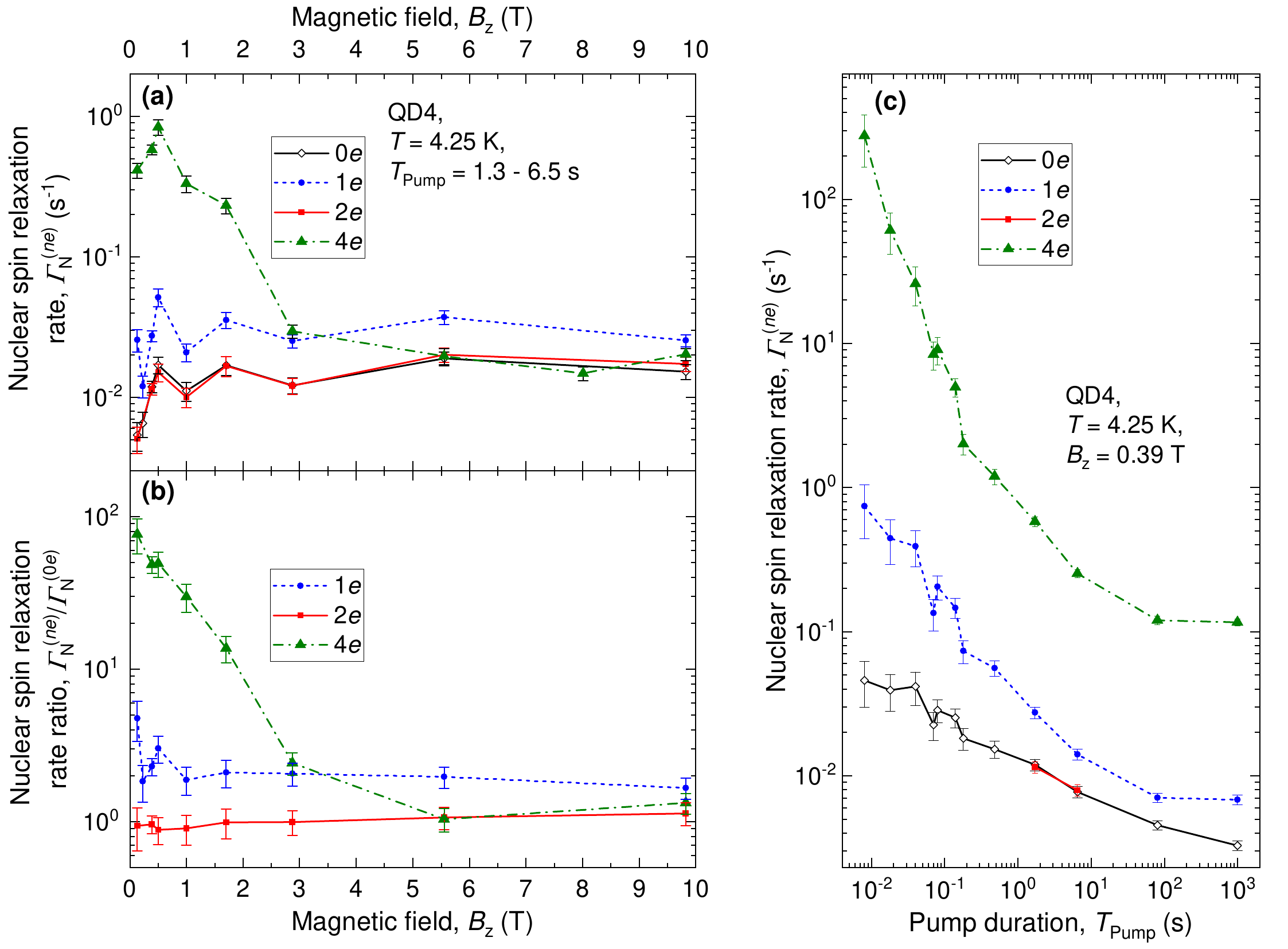}
\caption{\label{Fig:GammaNAbs} {\bf{Absolute nuclear spin relaxation rates.}} (a) Magnetic field dependence of the NSR rates measured in a Coulomb blockade regime where the QD is charged with $n$ electrons or is empty ($n=0$). (b) NSR rates
$\varGamma_{\rm{N}}^{(ne)}$ normalized by the rate
$\varGamma_{\rm{N}}^{(0e)}$ measured in an empty QD. (c) NSR rates measured as a function of the optical pump duration $T_{\rm{Pump}}$ in different charge states at $B_{\rm{z}}=0.39$~T.}
\end{figure}

Supplementary Fig.~\ref{Fig:GammaNAbs}(a) shows the absolute NSR rates measured as a function of magnetic field $B_{\rm{z}}$ at different charge states, characterized by the number $n$ of electrons in the QD. The measurements are conducted in the Coulomb blockade regime for $n\geq1$ or under large reverse bias for $n=0$. Supplementary Fig.~\ref{Fig:GammaNAbs}(b) is derived from the data of Supplementary Fig.~\ref{Fig:GammaNAbs}(a) and shows the NSR rates $\varGamma_{\rm{N}}^{(ne)}$ for $n\geq1$ normalized by the NSR rate $\varGamma_{\rm{N}}^{(0e)}$ in an empty QD ($n=0$). Supplementary Fig.~\ref{Fig:GammaNAbs}(b) shows the same data as Fig.~4(a) of the main text.

It can be seen from Supplementary Fig.~\ref{Fig:GammaNAbs}(a) that for all charge states the NSR rates depend on magnetic field. This includes the empty QD state $n=0$ where there is no electron spin that could affect the NSR. The reason for this dependence is the nuclear spin diffusion, a process where a localized nuclear spin polarization decays through spin-exchange flip-flops between the adjacent nuclear spins, which carries away the nuclear spin polarization from its original location (the optically polarized QD) to the distant parts of the sample. The nuclear spin diffusion process is thus very sensitive to the initial spatial distribution of the optically-pumped nuclear polarization. When the magnetic field is changed, it is impossible to recreate the exact same conditions for optical nuclear spin pumping, which results in an indirect dependence of $\varGamma_{\rm{N}}^{(ne)}$ on $B_{\rm{z}}$, as observed in Supplementary Fig.~\ref{Fig:GammaNAbs}(a). By taking the ratios $\varGamma_{\rm{N}}^{(ne)}/\varGamma_{\rm{N}}^{(0e)}$, shown in Supplementary Fig.~\ref{Fig:GammaNAbs}(b), this unwanted dependence on the optical pumping conditions can be largely eliminated, allowing us to study the effect of the QD few-electron state on the nuclear spin dynamics.

The role of the initial optically pumped nuclear spin state on the NSR rate is demonstrated in Supplementary Fig.~\ref{Fig:GammaNAbs}(c), where the absolute NSR rates $\varGamma_{\rm{N}}^{(ne)}$ are shown as a function of the optical pump duration $T_{\rm{Pump}}$ used to polarize the nuclear spins before they are allowed to depolarize in the dark in presence of the $n$-electron configuration in a QD. Longer optical pumping of nuclear spins is seen to reduce the rate of the subsequent NSR. This is due to the spin diffusion during optical pumping, which results in a spatially broad distribution of nuclear spin polarization at the end of pumping. Such distribution suppresses further spin diffusion, observed as a reduced $\varGamma_{\rm{N}}^{(ne)}$. By contrast, shorter $T_{\rm{Pump}}$ results in a sharp spatial profile of nuclear polarization, localized in a QD, which accelerates subsequent spin diffusion. As can be seen from Supplementary Fig.~\ref{Fig:GammaNAbs}(c) such acceleration is disproportionately faster for 1$e$ and 4$e$ states, when compared to the neutral 0$e$ QD state. This effect has been examined previously for the 1$e$ state \cite{MillingtonHotze2022} and ascribed to electron-mediated nuclear-nuclear spin interactions. A similar effect, but even more pronounced, is observed here for the 4$e$ state below the phase-transition critical magnetic field: the $\varGamma_{\rm{N}}^{(4e)}$ rate changes by a factor of $\approx300$ over the studied range of $T_{\rm{Pump}}$, whereas the change in $\varGamma_{\rm{N}}^{(0e)}$ is only by a factor of $\approx15$. We therefore conclude that the polarized 4$e$ ground state significantly enhances NSR both through non-diffusion and spin diffusion mechanisms, as discussed in the main text. 

As discussed above, the duration of the optical pumping $T_{\rm{Pump}}$ is not the only factor that affects the NSR rate. The optical pumping power and the matching of the pump laser wavelength to a particular transition in the optical absorption spectrum of the QD also affect the initial nuclear spin state, and then indirectly affect the rate of the subsequent NSR. The combination of these factors is what makes it difficult to reproduce the nuclear spin pumping conditions and requires the use of the normalized ratios $\varGamma_{\rm{N}}^{(ne)}/\varGamma_{\rm{N}}^{(0e)}$. As a target, at each magnetic field we chose $T_{\rm{Pump}}$ that yields nuclear spin polarization close to 85\% of the nuclear spin polarization achieved under steady-state nuclear spin pumping (i.e. in the limit of very long $T_{\rm{Pump}}$). The $T_{\rm{Pump}}$ values chosen in this way vary between 1.3~s and 6.5~s for the dataset used in Supplementary Figs.~\ref{Fig:GammaNAbs}(a) and (b), as well as Figs.~2(a), 3(a), and 4(a) of the main text.

\section{Configuration interaction calculations}
\label{sec:CI}

\subsection{Theory}

In eight-band ${\bf k}\cdot{\bf p}$, the single-particle electron states are considered as linear combination of $s$-orbital~like and $x$,~$y$,~$z$~$p$-orbital~like Bloch waves~\cite{Klenovsky2017} at $\Gamma$ point of the Brillouin zone,~i.e.,
\begin{equation}
\label{eq:wfkp}
    \Psi_{e_n}(\mathbf{r}) = \sum_{\nu\in\{s,x,y,z\}\otimes \{\uparrow,\downarrow\}} \chi_{e_n,\nu}(\mathbf{r})u^{\Gamma}_{\nu}\,,
\end{equation}
where $u^{\Gamma}_{\nu}$ is the Bloch wave-function of $s$- and $p$-like conduction and valence bands at $\Gamma$ point, respectively, $\uparrow$/$\downarrow$ marks the spin, and $\chi_{e_n,\nu}$ is~the~envelope function for electron of the $n$-th single-particle state.
Thereafter, the following envelope-function $\mathbf{k}\!\cdot\!\mathbf{p}$ Schr\"{o}dinger equation is solved
\begin{equation}
\label{eq:EAkp}
\begin{split}    
    &\sum_{\nu\in\{s,x,y,z\}\otimes \{\uparrow,\downarrow\}}\left(\left[E_\nu^{\Gamma}-\frac{\hbar^2{\bf \nabla}^2}{2m_e}+V_{0}({\bf r})\right]\delta_{\nu'\nu}+\frac{{\hbar}{\bf \nabla}\cdot{\bf p}_{\nu'\nu}}{m_e}\right.\\
    &\left. + \hat{H}^{\rm str}_{\nu'\nu}({\bf r})+\hat{H}^{\rm so}_{\nu'\nu}({\bf r})\right)\chi_{e_n,\nu}({\bf r})=E^{k\cdot p}_n\cdot \chi_{e_n,\nu'}({\bf r}),
\end{split}    
\end{equation}
where the term in round brackets on left side of the equation is the envelope function $\mathbf{k}\cdot\mathbf{p}$ Hamiltonian $\hat{H}_0^{k\cdot p}$, and $E^{k\cdot p}_n$ on the right side is the $n$-th single-particle eigenenergy. Furthermore, $E_\nu^{\Gamma}$ is the energy of the bulk $\Gamma$-point Bloch band $\nu$, $V_0({\bf r})$ is the scalar potential (e.g. due to piezoelectricity or externally applied electric field), $\hat{H}^{\rm str}_{\nu'\nu}({\bf r})$ is the Pikus-Bir Hamiltonian introducing the effect of elastic strain, and $\hat{H}^{\rm so}_{\nu'\nu}({\bf r})$ is the spin-orbit Hamiltonian.~\cite{Birner2007} Further, $\hbar$ is the reduced Planck's constant, $m_e$ the free electron mass, $\delta$ the Kronecker delta, and $\nabla := \left( \frac{\partial}{\partial x}, \frac{\partial}{\partial y}, \frac{\partial}{\partial z} \right)^T$.

We use single-particle states computed by the aforementioned ${\bf k}\cdot{\bf p}$ as basis states for our configuration interaction (CI). In CI we consider the multi-electron states as linear combinations of the Slater determinants
%
\begin{equation}
    \psi_i^{\rm CI}(\mathbf{r}) = \sum_{\mathit m=1}^{n_{\rm SD}} \mathit \eta_{i,m} \left|D_m^{\rm CI}\right>, \label{eq:CIwfSD}
\end{equation}
where $n_{\rm SD}$ is the number of Slater determinants $\left|D_m^{\rm CI}\right>$, and $\eta_{i,m}$ is the $i$-th CI coefficient which is found along with the eigenenergy using the variational method by solving the Schr\"{o}dinger equation 
\begin{equation}
\label{CISchrEq}
\hat{H}^{\rm{CI}} \psi_i^{\rm CI}(\mathbf{r}) = E_i^{\rm{CI}} \psi_i^{\rm CI}(\mathbf{r}),
\end{equation}
where $E_i^{\rm{CI}}$ is the $i$-th eigenenergy of multi-electron state $\psi_i^{\rm CI}(\mathbf{r})$, and~$\hat{H}^{\rm{CI}}$ is the CI Hamiltonian which reads
\begin{equation}
\label{CIHamiltonian}
\hat{H}^{\rm{CI}}=E^{k\cdot p}+\hat{V}^{\rm{CI}},
\end{equation}
with $E^{k\cdot p}$ being the eigenenergies of single-particle states and $\hat{V}^{\rm{CI}}$ representing the Coulomb interaction between Slater determinants constructed from these single-particle states. The Slater determinant $\left|D_m^{\rm CI}\right>$ is defined as:
\begin{equation}
\label{SlaterDeterminant}
\left|D_m^{\rm CI}\right> = \frac{1}{\sqrt{N!}} \sum_{\tau \in S_N} \rm sgn \mathit(\tau) \Psi_{\tau\{i_1\}}(\mathbf{r}_1) \Psi_{\tau\{i_2\}}(\mathbf{r}_2) \dots \Psi_{\tau\{i_N\}}(\mathbf{r}_N),
\end{equation}
where $N$ is the number of interacting electrons. Further, the matrix element of $\hat{V}^{\rm{CI}}$ in the basis of the Slater determinants $\left|D_m^{\rm CI}\right>$ is~\cite{Klenovsky2017}
\begin{equation}
\begin{split}
    &V^{\rm CI}_{n,m}=\bra{D_n^{\rm CI}}\hat{J}\ket{D_m^{\rm CI}} = \sum_{ijkl} (1-\delta_{ij})(1-\delta_{kl})\iint {\rm d}\mathbf{r} {\rm d}\mathbf{r}^{\prime} \frac{e^2}{4\pi\epsilon_0\,\epsilon(\mathbf{r},\mathbf{r}^{\prime})|\mathbf{r}-\mathbf{r}^{\prime}|} \\
    &\times \{ \Psi^*_i(\mathbf{r})\Psi^*_j(\mathbf{r}^{\prime})\Psi_k(\mathbf{r})\Psi_l(\mathbf{r}^{\prime}) - \Psi^*_i(\mathbf{r})\Psi^*_j(\mathbf{r}^{\prime})\Psi_l(\mathbf{r})\Psi_k(\mathbf{r}^{\prime})\}
    \\
    &= \sum_{ijkl}(1-\delta_{ij})(1-\delta_{kl})\left(V^{\rm CI}_{ij,kl} - V^{\rm CI}_{ij,lk}\right).\\
\end{split}
\label{eq:CoulombMatrElem}
\end{equation}
Here $\hat{J}$ marks the Coulomb operator, $e$ labels the elementary charge and $\epsilon(\mathbf{r},\mathbf{r}^{\prime})$ is the spatially dependent relative dielectric function, $\epsilon_0$ is the vacuum permittivity. Note that for $\epsilon(\mathbf{r},\mathbf{r}^{\prime})$ in Eq.~\eqref{eq:CoulombMatrElem} we use the position-dependent bulk dielectric constant. The Coulomb interaction in Eq.~\eqref{eq:CoulombMatrElem} described by $V^{\rm CI}_{ij,kl}$ ($V^{\rm CI}_{ij,lk}$) is called direct (exchange).

We note that because wavefunctions in Eq.~\eqref{eq:wfkp} are orthonormal, there exist only three different kinds of non-zero Coulomb integrals in Eq.~\eqref{eq:CoulombMatrElem} depending on the number of different indices in Eq.~\eqref{eq:CoulombMatrElem}, i.e.,
\begin{align}
    \left<D_n^M\right|\hat{J}\left|D_n^M\right>&=\frac{1}{2}\sum_{ij}(1-\delta_{ij})\left(V^{\rm CI}_{ij,ij}-V^{\rm CI}_{ij,ji}\right)\,\,{\rm for\,2\,different\,indices}\,,\label{eq:CI1}\\
    \left<D_n^M\right|\hat{J}\left|D_m^M\right>&=\frac{1}{4}\sum_{ijk}(1-\delta_{ij})(1-\delta_{kj})\left(V^{\rm CI}_{ij,kj}-V^{\rm CI}_{ij,jk}\right)\,\,{\rm for\,3\,different\,indices}\,,\label{eq:CI2}\\
    \left<D_n^M\right|\hat{J}\left|D_m^M\right>&=\frac{1}{4}\sum_{ijkl}(1-\delta_{ij})(1-\delta_{kl})\left(V^{\rm CI}_{ij,kl}-V^{\rm CI}_{ij,lk}\right)\,\,{\rm for\,4\,different\,indices}\,,\label{eq:CI3}
\end{align}
if we further emphasize that $i<j\,\&\,k<l$, which stems from the CI matrix Hermiticity, and consider further also $m\geq n$ only (the lower triangular matrix is then obtained by Hermitian transpose). 
%

The sixfold integral in Eq.~\eqref{eq:CoulombMatrElem} is evaluated using the~Green's function method.~\cite{Schliwa:09,Stier2000,Klenovsky2017} The integral in Eq.~\eqref{eq:CoulombMatrElem} is split into solution of the Poisson's equation for one electron $a$ only, followed by a three-fold integral for another electron $b$ in the electrostatic potential generated by $a$ and resulting from the previous step. That procedure, thus, makes the whole solution numerically more feasible and is described by
\begin{equation}
\begin{split}
    \nabla \left[ \epsilon(\mathbf{r}) \nabla \hat{U}_{ajl}(\mathbf{r}) \right] &= \frac{4\pi e^2}{\epsilon_0}\Psi^*_{aj}(\mathbf{r})\Psi_{al}(\mathbf{r}),\\
    V^{\rm{\rm CI}}_{ij,kl} &= \int {\rm d}\mathbf{r}'\,\hat{U}_{ajl}(\mathbf{r}')\Psi^*_{bi}(\mathbf{r}')\Psi_{bk}(\mathbf{r}'),
\end{split}
\label{eq:GreenPoisson}
\end{equation}
where $a,b$ mark the electrons.

\subsection{Implementation}


\begin{figure}[ht!]
\includegraphics[width=0.98\linewidth]{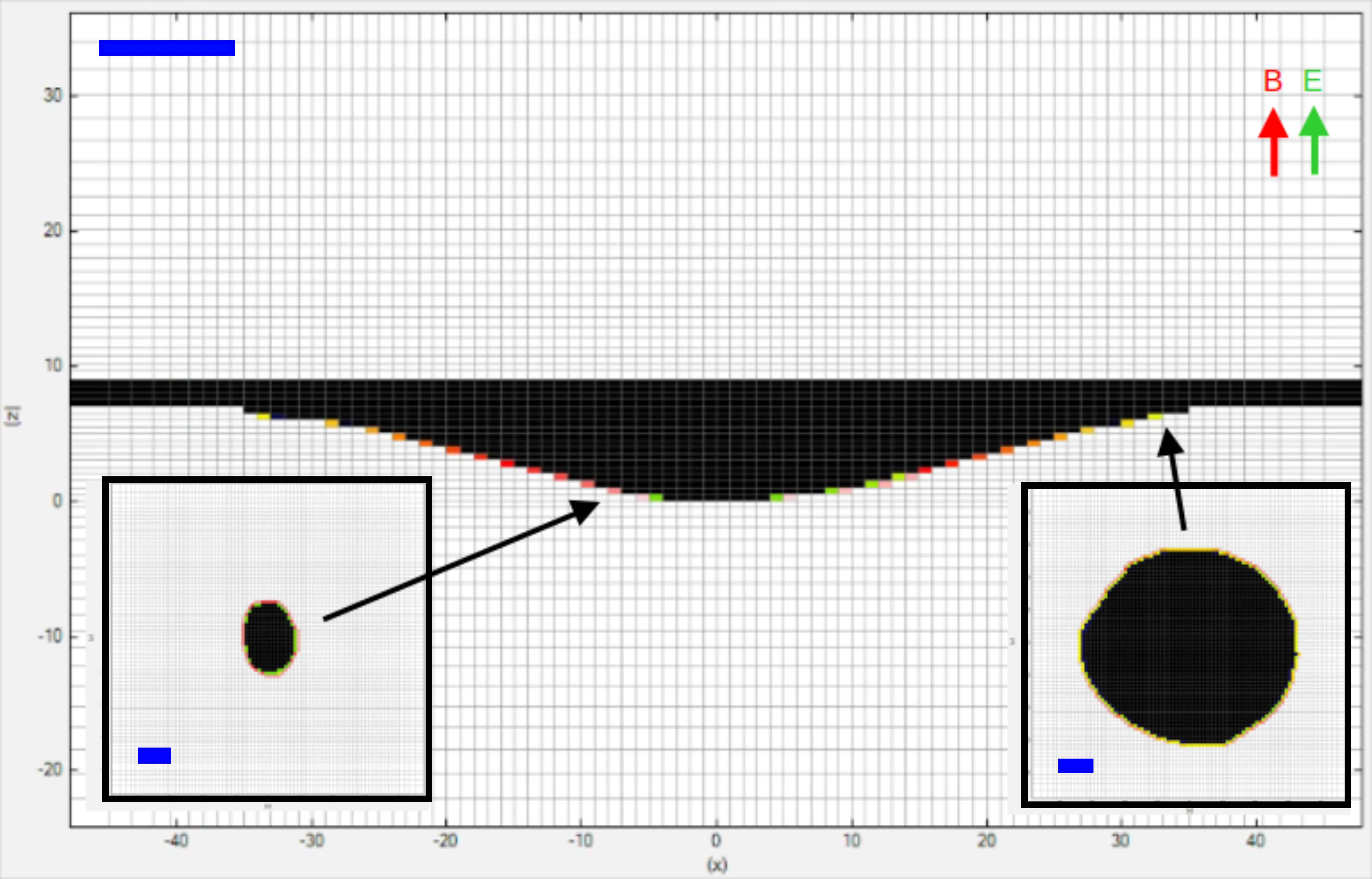}
\caption{\label{Fig:SimQDAFM} {\bf{Side view of the GaAs QD model used in the calculations.}} QD shape was derived from AFM measurement of the AlGaAs nanohole. Overgrowth of the Al$_{0.4}$Ga$_{0.6}$As surface with 2~nm GaAs results in nanohole filling and in a quantum well. The insets show cuts parallel to growth plane close to the apex and base of the nanohole. The blue scale bars correspond to 10~nm. The red and green arrows mark the positive values of applied magnetic and electric fields, respectively. The mesh in the background marks the simulation grid.}
\end{figure}
Fig.~\ref{Fig:SimQDAFM} shows a cross section through the center of the simulated QD structure, along with the indicated directions of the positive increase of the applied magnetic flux density and electric field.

\begin{figure}[ht!]
\includegraphics[width=0.4\linewidth]{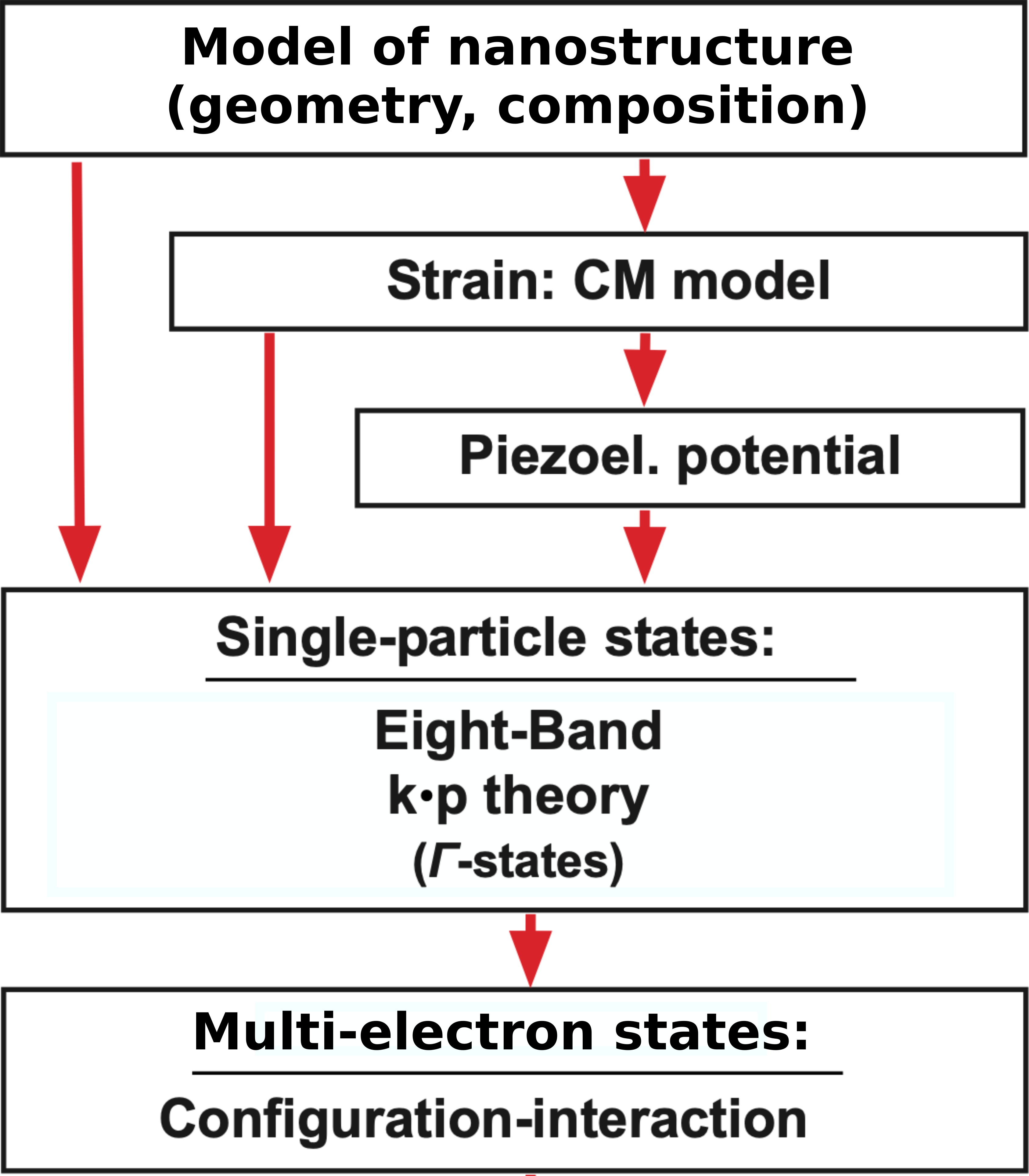}
\caption{\label{Fig:CalcSheme} {\bf{Flow chart of the modeling procedure applied in this work.}}}
\end{figure}

The flowchart of the whole calculation is depicted in Fig.~\ref{Fig:CalcSheme}. In the calculation we first implement the 3D QD model structure (size, shape, chemical composition), see Fig.~\ref{Fig:SimQDAFM}. This is followed by the calculation of elastic strain by minimizing the total strain energy in the structure and subsequent evaluation of piezoelectricity. The resulting strain and polarization fields then enter the eight-band $\mathbf{k}\cdot\mathbf{p}$ Hamiltonian~\cite{Birner2007}. The single-particle states obtained that way are thereafter used as basis states for CI calculations using our own implementation~\cite{Klenovsky2017}.

\subsection{Extended results of configuration interaction modeling}

Here we present the extended results of CI calculations for charge configurations ranging from 2 to 5 electrons in a QD. For these calculations we use the QD model based on AFM data. The results are shown in Supplementary Fig.~\ref{Fig:CISpin}. We start with the 2$e$ configuration. The spectrum of the 2$e$ eigenenergies as a function of the external magnetic field $B_{\rm{z}}$ is shown in Supplementary Fig.~\ref{Fig:CISpin}(a). For each eigenstate with wavefunction $\psi$ we take the expectation values of the spin projections to calculate the magnitude of the total spin projection: $\vert S \vert=\sqrt{\langle \psi \vert \hat{S}_{\rm{x}} \vert \psi \rangle^2+\langle \psi \vert \hat{S}_{\rm{y}} \vert \psi \rangle^2+\langle \psi \vert \hat{S}_{\rm{z}} \vert \psi \rangle^2}$. The components of the total spin projection operators $\hat{S}_{\rm{x}}$, $\hat{S}_{\rm{y}}$, and $\hat{S}_{\rm{z}}$ are defined as sums of the corresponding single-particle operators $\hat{s}_{{\rm{x}},i}$, $\hat{s}_{{\rm{y}},i}$, and $\hat{s}_{{\rm{z}},i}$ over all individual electrons $i\in [1,n]$. The total spin projections of three lowest energy 2$e$ states are shown in Supplementary Fig.~\ref{Fig:CISpin}(b). We further define the polar angle of the total spin projection: $\theta = \arctan{\left(\sqrt{\langle \psi \vert \hat{S}_{\rm{x}} \vert \psi \rangle^2+\langle \psi \vert \hat{S}_{\rm{y}} \vert \psi \rangle^2}\big/\langle \psi \vert \hat{S}_{\rm{z}} \vert \psi \rangle\right)}$, where $\arctan{}$ is defined to return values within $[0,\pi]$. If the electron eigenstate possesses a finite polarization, this polar angle describes its quantization axis with respect to the $z$ axis. The polar angles of the 2$e$ eigenstates are shown in Supplementary Fig.~\ref{Fig:CISpin}(c).

At low magnetic fields, the total spin projection of the ground 2$e$ state is very small $\vert S \vert <10^{-4}$, which allows this state to be identified as a singlet $S=0$, formed by two electrons occupying the lowest $s$-shell single-particle orbital. The first and second excited states are polarized along the negative ($\theta \approx 180^\circ$) and positive ($\theta \approx 0^\circ$) directions of the $z$ axis, respectively. The magnitude of the spin projection is reduced to $\vert S \vert \approx 0.1$ at zero field, which can be ascribed to the anisotropic exchange interaction mixing of these two excited states. Once the magnetic field applied along the $z$ axis reaches $B_{\rm{z}}\approx1$~T, the total spin projections approach unity $\vert S \vert \approx 1$, which allows these states to be described as $S_{\rm{z}}=\pm 1$ spin projection states of the triplet $S=1$. In a simplified shell-filling picture, this triplet is formed when one electron occupies the lowest $s$-shell orbital and the other electron is promoted to the next $p$-shell single-particle orbital. The next excited state, split off by exchange interaction, is ascribed to the $S_{\rm{z}}=0$ projection state of the triplet $S=1$. The CI model predicts phase transition of the 2$e$ ground state into a triplet for magnetic fields $B_{\rm{z}}\gtrsim5$~T. This transition is not observed in experiments. The most likely explanation is that such transition does take place in real QDs, but at magnetic fields exceeding the maximum available in our setup ($10$~T). As we show below for other charge configurations, the CI model predicts correctly the phase transitions, but tends to underestimate the critical fields.

\begin{figure}
\includegraphics[width=0.99\linewidth]{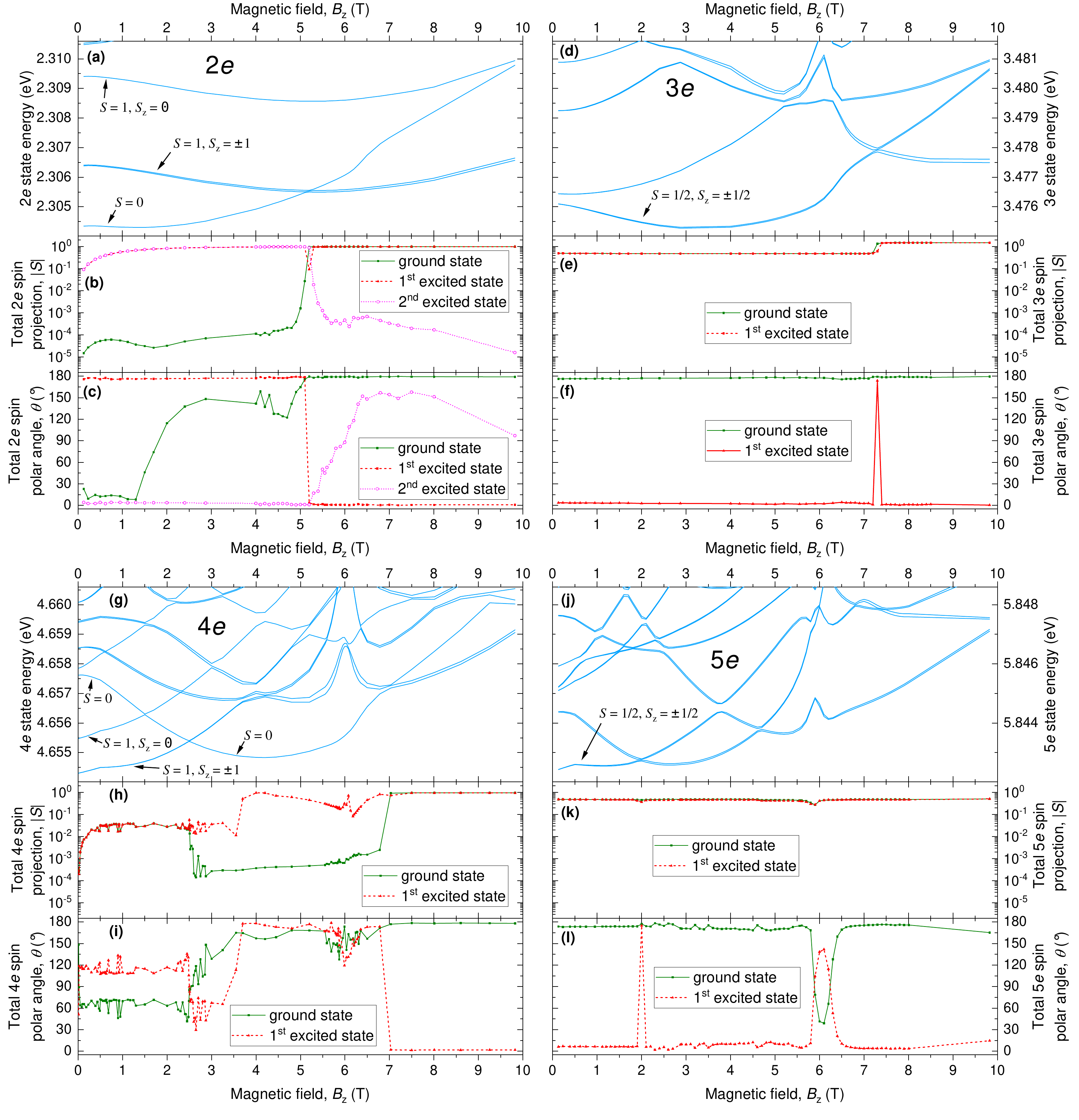}
\caption{\label{Fig:CISpin} {\bf{Configuration Interaction modeling of few-electron states in a GaAs/AlGaAs QD.}} (a) Energies of the 2$e$ states as a function of magnetic field. (b) Total spin projection of the ground and first excited states of the 2$e$ configuration. (c) Orientation of the spin projection for the ground and first excited states of the 2$e$ configuration, plotted as a polar angle between the spin projection and the $z$ axis. (d-f) Same as (a-c) but for the 3$e$ configuration. (g-i) Same as (a-c) but for 4$e$. (j-l) Same as (a-c) but for 5$e$.}
\end{figure}

We next consider the 3$e$ configuration. The spectrum of the 3$e$ eigenenergies as a function of the external magnetic field $B_{\rm{z}}$ is shown in Supplementary Figs.~\ref{Fig:CISpin}(d) - \ref{Fig:CISpin}(f). Here, the lowest energy manifold is the doublet of states with a total spin projection $\vert S \vert \approx 1/2$ and quantized close to the $z$ axis ($\theta \approx 0, 180^\circ$). This doublet can be ascribed to the unpaired spin-1/2 of the $p$-shell electron, added to the spin-singlet formed by the other two $s$-shell electrons of the 3$e$ configuration. This prediction of a $p$-shell spin qubit state is in good agreement with NMR and electron spin lifetime experimental data. At $B_{\rm{z}}\approx 7$~T the CI model predicts a phase transition to a ground state with $\vert S \vert \approx 1.5$, which can be interpreted as a $S=3/2$ configuration of the 3$e$ charge configuration, polarized along $+z$ or $-z$ axis. We do not observe any experimental signatures of such a phase transition, once again suggesting that the CI model underestimates the critical transition fields.

The case of four electrons (4$e$) is shown in Supplementary Figs.~\ref{Fig:CISpin}(g) - \ref{Fig:CISpin}(i), which reproduce Figs.~4(c) - 4(e) of the main text. The CI model predicts the triplet-singlet phase transition to occur at $B_{\rm{z,cr}}^{(4e)}\approx2.5$~T, observed as a drop of the total spin projection from $\vert S \vert \approx 3\times10^{-2}$ at $B_{\rm{z}}=0.5-2.5$~T to $\vert S \vert \approx 3\times10^{-4}$ at $B_{\rm{z}}>2.5$~T. Unlike in the 3$e$ configuration the low-field ($B_{\rm{z}}=0.5-2.5$~T) polarized ground-state doublet of the 4$e$ configuration has a small spin magnitude $\vert S \vert \approx 3\times10^{-2}$ and is polarized nearly orthogonal to the external field ($\theta \approx 70^\circ$ and $\theta \approx 110^\circ$) even when magnetic field is as large as $2.5$~T (just before the phase transition). At very small magnetic fields ($B_{\rm{z}}<0.5$~T) the spin magnitude $\vert S \vert$ is seen to drop, which can be explained by mixing of the two lowest $4e$ states in the anisotropic potential of the QD. The CI model predicts a second phase transition into a triplet ground state $S=1$ at $B_{\rm{z}}\approx 7$~T, which is not observed in experiments up to $B_{\rm{z}}\approx 10$~T. 

The results for five electrons (5$e$) in a QD are shown in Supplementary Figs.~\ref{Fig:CISpin}(j) - \ref{Fig:CISpin}(l). Similar to the 3$e$ case, the lowest-energy manifold is a doublet of states with $\vert S\vert=0.5$ and spin projections approximately along the $\pm z$ directions. In the limit of low magnetic fields, the ground state 5$e$ doublet can be understood as a combination of a spin-singlet electron pair in the $s$-shell, a spin-singlet pair in the $p$-shell, and another unpaired electron in the $p$-shell. In both cases of the 3$e$ and 5$e$ configurations, the character of the low-field ground state is defined by the unpaired spin of the odd-numbered electron spin configuration. However, the quantitative properties are different. For the 3$e$ configuration the CI model yields $\vert S \vert \approx 0.499$, with spin projection polar angles of $\theta \approx 3.8^\circ$ and $\theta \approx 176.2^\circ$ (at the lowest sampled field $B_{\rm{z}}\approx 0.13$~T). For the 5$e$ configuration at the same field we find a reduced $\vert S \vert \approx 0.482$, with spin projection polar angles $\theta \approx 6.3^\circ$ and $\theta \approx 173.7^\circ$, indicating further deviation from the $z$ axis. Interactions and correlations in a few-electron complex cause the departure from a simple orbital-filling model: the larger the number of the electrons, the stronger is the deviation from an ideal spin-1/2 picture with $S=1/2$ and spin projections $S_{\rm{z}}=\pm 1/2$ aligned along the $z$ axis (i.e. $\theta=0, 180^\circ$ in the ideal case). Yet, when compared to the 4$e$ ground state doublet, both the 3$e$ and 5$e$ ground state doublets are a good approximation to a spin-1/2 particle. The doublet state of the even-numbered 4$e$ configuration, with its order-of-magnitude smaller spin projection $\vert S \vert \approx 0.03$ and quantization nearly orthogonal to the external magnetic field, is very different from the odd-numbered 3$e$ and 5$e$ doublet states. 

Supplementary Fig.~\ref{Fig:CISpin}(k) shows that the 5$e$ ground state maintains the $\vert S \vert \approx 0.5$ character through the entire range of fields up to $B_{\rm{z}}\approx 10$~T. However, a closer look at the single-particle content of the 5$e$ wavefunctions shows that the ground state doublet in the $B_{\rm{z}}\approx 2 - 4$~T range has its unpaired electron promoted from the single-particle $p$-shell into the $d$-shell, indicating ground-state phase transitions. The critical fields are in principle within the experimentally accessible range of magnetic fields, however no experimental observations were made since the 5$e$ Coulomb blockade plateau is too narrow at the base temperature $T\approx4.2$~K in our setup. The rich structure of correlations and phase transitions in many-electron ($n\geq5$) states is a subject of future work, both experimental and computational.

Overall, we find that the configuration interaction numerical model accurately captures the qualitative picture of the many-body physics in the few-electron spin system of an epitaxial quantum dot. The quantitative match is more difficult to achieve, with clear discrepancies in numeric parameters such as phase transition critical fields. The exact spectrum of the few-electron states is sensitive to a large number of parameters such as size, shape, chemical composition, and strain distribution within the quantum dot. The limited knowledge of these parameters precludes exact numerical prediction of the few-electron spectra. The challenge for future work is to establish a closed loop algorithm where these rich spectra of the few-electron quantum states can be used as a fingerprint in deducing the structure and morphology of the semiconductor quantum dot.  



\end{document}